\documentclass[article,aps,preprint,floatfix,nofootinbib,showpacs]{revtex4-1}
\pdfoutput=1
\usepackage{dcolumn}
\usepackage{bm}
\usepackage{graphicx}
\usepackage{amssymb,amsmath}
\usepackage{siunitx}
\usepackage{multirow}
\usepackage{changes}
\usepackage{color,url}
\usepackage{tabu}
\usepackage{array}
\usepackage{slashed}
\usepackage{array}
\newcolumntype{L}[1]{>{\raggedright\arraybackslash}p{#1}}
\usepackage{float} 
\usepackage[colorlinks=true,urlcolor=blue,anchorcolor=blue
,citecolor=blue,filecolor=blue,linkcolor=blue,menucolor=blue
,linktocpage=true,pdfproducer=medialab,pdfa=true]{hyperref}
\usepackage{colordvi}
\usepackage{subcaption}
\def\beqa{\begin{eqnarray}}
\def\eeqa{\end{eqnarray}}

\usepackage{soul}
\usepackage{booktabs}
\usepackage{siunitx}    
\usepackage{array}      
\usepackage{caption}    
\usepackage{ragged2e}   

\begin{document} 
\preprint{CPTNP-2025-034}

\title{Searching for leptophilic composite asymmetric dark sector at $e^+e^-$ colliders}
\def\slash#1{#1\!\!\!/}

\author{Chih-Ting Lu$^{1,2}$, Changbin Xi$^1$}
\affiliation{
$^1$ Department of Physics and Institute of Theoretical Physics, Nanjing Normal University, Nanjing, 210023, P. R. China}  
\affiliation{$^2$ Nanjing Key Laboratory of Particle Physics and Astrophysics, Nanjing, 210023, China}


\begin{abstract}

Composite asymmetric dark matter (ADM) models provide a well-motivated paradigm that simultaneously explains dark matter (DM) relic density and matter-antimatter asymmetry. In these models, the mass of the DM candidate (the lightest dark baryon) is generated through the dark confinement scale dynamics. Although the leptophilic composite ADM model offers a viable framework, comprehensive studies of its collider phenomenology are absent. This work systematically explores novel signatures from the leptophilic composite asymmetric dark sector at both low-energy and high-energy $e^+e^-$ colliders as well as other existing collider constraints. We demonstrate detectability of TeV-scale mediators along with sub-GeV to GeV-scale lightest dark mesons at Belle II and its proposed far detector, GAZELLE, as well as {\color{black}Circular Electron Positron Collider}/{\color{black}Future Circular Collider-electron-positron} experiments. Moreover, these experiments exhibit complementary coverage of the model parameter space.

\end{abstract}

\maketitle

\section{Introduction}

The particle nature of dark matter (DM) and the origin of matter-antimatter asymmetry remain outstanding challenges in particle physics, astrophysics, and cosmology. The proposed DM models span mass scales from ultralight to tens of solar masses~\cite{Battaglieri:2017aum,Lin:2019uvt,Ferreira:2020fam,Belenchia:2021rfb}, with well-motivated candidates including weakly interacting massive particles (WIMPs)~\cite{Arcadi:2017kky,Arcadi:2024ukq}, strongly interacting massive particles~\cite{Hochberg:2014dra}, hidden sector DM~\cite{Battaglieri:2017aum}, WIMPless DM~\cite{Feng:2008ya}, asymmetric DM (ADM)~\cite{Petraki:2013wwa,Zurek:2013wia}, freeze-in DM~\cite{Bernal:2017kxu}, QCD axions~\cite{Baryakhtar:2025jwh}, and primordial black holes~\cite{Villanueva-Domingo:2021spv}. Despite extensive searches, no conclusive evidence supports any candidate to date. Consequently, diverse experimental strategies have been developed to probe their properties.

Dynamically generating the observed matter-antimatter asymmetry requires satisfying Sakharov's three conditions~\cite{Sakharov:1967dj}: 1) baryon number violation, 2) $C/CP$ violation, and 3) departure from thermal equilibrium. Successful mechanisms include electroweak baryogenesis~\cite{Bodeker:2020ghk}, leptogenesis~\cite{Davidson:2008bu}, and the Affleck-Dine mechanism~\cite{Affleck:1984fy}. The observed DM-to-baryon density ratio $\Omega_{\text{DM}}/\Omega_{\text{B}}\approx 5$ suggests a common origin for both abundances. The ADM paradigm~\cite{Petraki:2013wwa,Zurek:2013wia} achieves this by positing that DM abundance, like baryonic abundance, arises from particle-antiparticle asymmetry generated by shared physics connecting dark and {\color{black}standard model} (SM) sectors.

Composite ADM~\cite{Blinnikov:1982eh,Blinnikov:1983gh,Khlopov:1989fj,Alves:2009nf,SpierMoreiraAlves:2010err,Bai:2013xga,Ibe:2018juk,Ibe:2018tex,Ibe:2019ena,Hall:2019rld,Zhang:2021orr,Bottaro:2021aal,Ibe:2021gil,Ritter:2022opo} represents a prominent scenario in which the dark sector features a hidden $SU(N)$ gauge symmetry analogous to SM QCD. In such models, accidental dark baryon number conservation stabilizes the lightest dark baryon, making it a natural DM candidate. Crucially, the DM mass is directly related to the dark confinement scale $\Lambda_d$, thereby being dynamically determined rather than a free parameter. Mediators facilitate dark sector and SM sector coupling to generate shared particle-antiparticle asymmetry. We adopt the framework in Refs.~\cite{Bai:2013xga,Zhang:2021orr}{\color{black},} where bi-charged scalar mediator decays produce asymmetries in both sectors.

Two bi-charged scalar mediators exist in the literature: baryophilic~\cite{Bai:2013xga} and leptophilic~\cite{Zhang:2021orr} types, both capable of simultaneously explaining DM relic density and matter-antimatter asymmetry. The baryophilic mediator carries charges under both hidden $SU(N)$ and SM QCD symmetries, coupling to a dark quark and an SM quark. This enables copious dark quark production at the LHC where, given the dark quark mass is much smaller than the collision energy, the produced dark quarks become energetic and highly boosted. When the scale of $\Lambda_d$ is also much smaller than the collision energy, subsequent QCD-like showering and hadronization generate distinctive dark jets that facilitate novel search strategies at the {\color{black}LHC}~\cite{Schwaller:2015gea,Cohen:2015toa,Cohen:2017pzm,Park:2017rfb,Renner:2018fhh,Mies:2020mzw,Linthorne:2021oiz,Archer-Smith:2021ntx,Carrasco:2023loy,Carmona:2024tkg,Liu:2025bbc}. In contrast, the leptophilic mediator is coupled with a dark quark and an SM lepton, suppressing LHC production rates as noted in~\cite{Zhang:2021orr}. Therefore, lepton colliders consequently provide ideal environments for probing such models. Although Ref.~\cite{Zhang:2021orr} proposed preliminary studies, comprehensive {\color{black}analyses} remain lacking. This work{\color{black}, therefore,} systematically explores leptophilic composite ADM signatures on low-energy and high-energy $e^+e^-$ colliders.

Collider searches for leptophilic scalar mediators resemble searches of slepton in supersymmetry models when dark hadrons inside dark jets escape detection at {\color{black}the LHC and} LEP~\cite{ATLAS:2019lff,CMS:2020bfa,CMS:2022syk,CMS:2024gyw,ATLAS:2024fub,Fox:2011fx}. We{\color{black}, therefore,} rescale existing slepton constraints to these mediators. While collision energies exceeding the dark confinement scale $\Lambda_d$ can produce dark showers, the intensity of showering depends critically on energy: higher energies generate denser parton showers. At low-energy $e^+e^-$ colliders such as $B$ {\color{black}factories}, sparse dark showers enable tracking of individual displaced dark pions and kinematic reconstruction. Similar approaches have been studied for the dark photon mediator~\cite{Bernreuther:2022jlj}, though signatures differ fundamentally from this work. Conversely, at high-energy $e^+e^-$ colliders {\color{black}[Circular Electron Positron Collider(CEPC), Future Circular Collider-electron-positron(FCC-ee), and ILC]}, dense dark showers necessitate identification of displaced, collimated lepton jets rather than individual dark pions~\cite{Zhang:2021orr}. When dark jets become undetectable, {\color{black}monophoton} signatures~\cite{Liu:2019ogn} become relevant. This work establishes complementary coverage of parameter space through these distinct search strategies.

This paper is structured as follows: {\color{black}Sec.}~\ref{sec:model} reviews simplified leptophilic composite ADM models. Section~\ref{sec:constraint} examines existing collider constraints. Novel dark shower signatures at $B$ {\color{black}factories} and Higgs factories are explored in {\color{black}Secs.}~\ref{sec:Babar} and~\ref{sec:CEPC}, respectively. Section~\ref{sec:conclusion} presents our conclusions and future research directions.

\section{A brief review of QCD-like dark sector with bi-charged mediators}
\label{sec:model}
In the leptophilic composite ADM models proposed in Ref.~\cite{Zhang:2021orr}, new heavy particles, including Majorana fermions and $SU(3)_D$ charged Dirac fermions, generate out-of-equilibrium processes and explicitly break hidden baryon number conservation. However, these particles lie significantly above current collider energy scales and are consequently decoupled from our analysis. We{\color{black}, therefore,} restrict the Lagrangian to the relevant scalar mediator $X$ and dark quark $q_d$ interactions as follows\footnote{The possible Higgs portal coupling between the SM Higgs doublet field and the $X$ field is irrelevant to this study. Therefore, we assume that its contributions are negligible and can be neglected here.}:
\begin{equation}
\mathcal{L} \supset \overline{q_d} ( \slashed{D} - m_{q_d}) q_d 
+ (D_\mu X)^\dagger (D^\mu X) 
- m^2_X X^\dagger X 
- \frac{1}{4} G^{d,\mu\nu} G^d_{\mu\nu} 
- \left( \kappa X \overline{q_d}_L l_R + \text{H.c.} \right), 
\label{Eq:lag}
\end{equation}
where the covariant derivative \( D_{\mu} = \partial_{\mu} -ig_d G^d_{\mu} \) is defined with \( g_d \) and \( G^d_{\mu} \) representing the \( SU(3)_D \) gauge coupling and dark gluon field respectively. The field strength tensor \( G^d_{\mu\nu} \) corresponds to the \( SU(3)_D \) gauge field \( G^d_{\mu} \), while \( m_{q_d} \) and \( m_X \) denote the dark quark and mediator masses. The coupling matrix \( \kappa \) (a \( n \times 3 \) matrix) corresponds to interactions between \( n\)-flavor dark quarks and three-flavor charged leptons. Note that the mediator $X$ plays a crucial role in generating asymmetries in both SM and dark sectors through its decays, as it carries both SM lepton number and dark baryon number.  

\begin{figure}[t!]\centering
\includegraphics[width=0.4\textwidth]{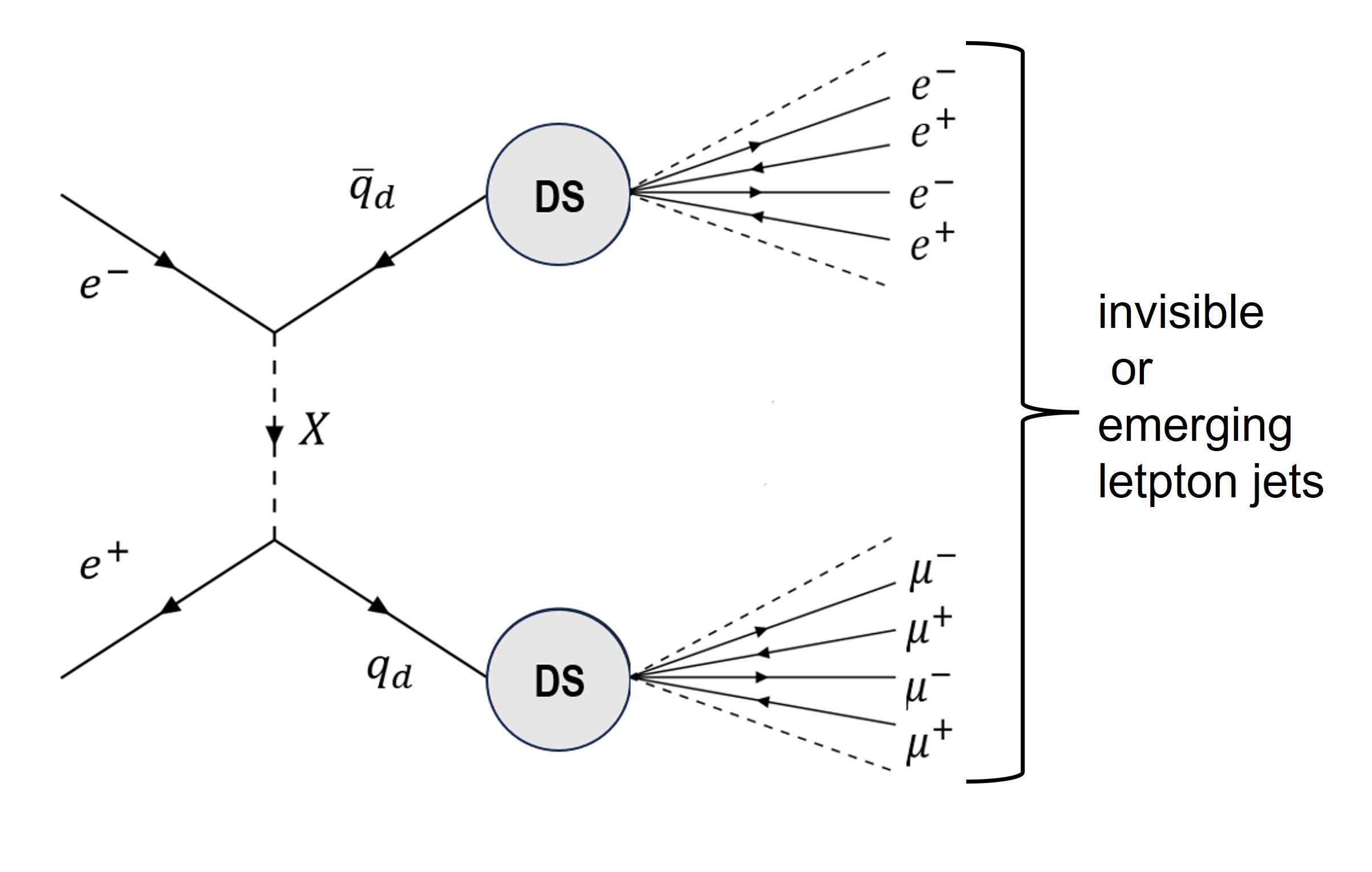}
\includegraphics[width=0.5\textwidth]{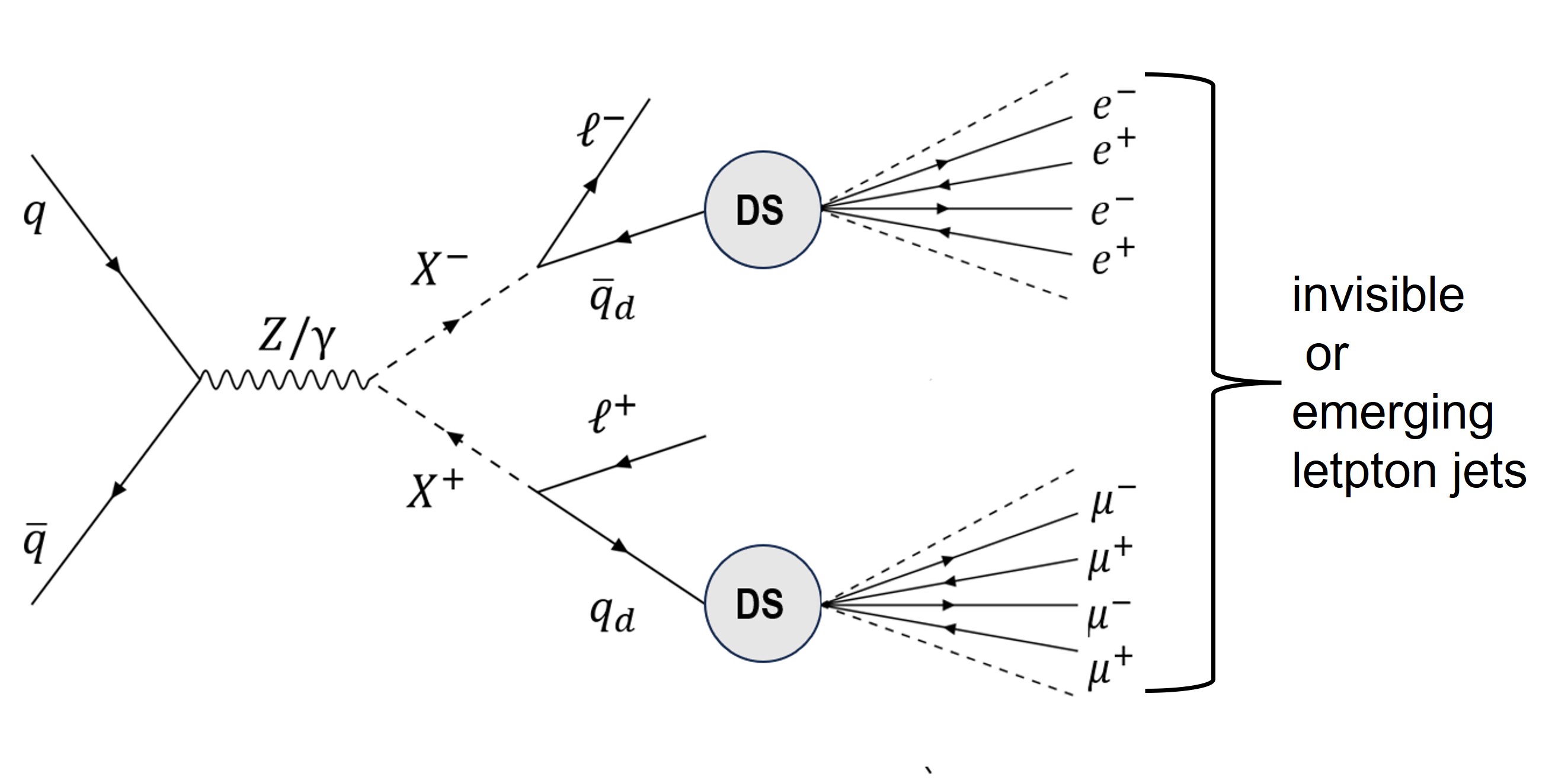}
\caption{Diagrams illustrating dark shower production evolving into invisible or emerging lepton jets. Left panel: $e^+ e^-$ annihilation via $t$-channel mediator $X$ producing a dark quark pair ($q_d\overline{q_d}$). Right panel: $q\overline{q}$ annihilation yielding mediator pairs, where each mediator decays to a dark quark and a SM lepton.}
\label{fig:Feynman}
\end{figure}

At lepton and hadron colliders, dark quarks are produced through \( t\)-channel processes and/or mediator decays, as shown in Fig.~\ref{fig:Feynman}. {\color{black}Postproduction}, they acquire significant kinetic energy (\( E \gg \Lambda_d \)), where \(\Lambda_d\) denotes the dark confinement scale. Assuming the dark QCD sector mirrors SM QCD dynamics, these dark quarks undergo dark parton shower evolution and hadronization, forming a collinear bunch of dark mesons.

In subsequent sections, we adopt the parameter settings for the QCD-like dark sector from Refs.~\cite{Schwaller:2015gea,Mies:2020mzw,Archer-Smith:2021ntx,Carrasco:2023loy} to enable concrete simulation predictions. We assume three dark colors ($N_{C_d}=3$) and seven dark flavors ($N_{f_d}=7$), with the following mass relations: $\Lambda_d = m_{q_d} = 2m_{\pi_d}$, and $m_{\rho_d} = 4m_{\pi_d}$. Note that the dark confinement scale $\Lambda_d$ is closely related to the mass of the lightest dark baryon, which serves as the DM candidate~\cite{Bai:2013xga,Schwaller:2015gea,Renner:2018fhh,Mies:2020mzw,Zhang:2021orr}. All dark quarks are taken to be mass-degenerate, while dark pions and dark rho mesons are similarly degenerate in both flavor diagonal and off-diagonal states. Other dark baryons and mesons, assumed substantially heavier than dark pions and rho mesons, exhibit negligible production rates from dark showers and hadronization due to mass suppression. We{\color{black}, therefore,} exclude them from this analysis.

Integrating out the heavy mediator {\color{black}[final terms in Eq.~(\ref{Eq:lag})]} yields a dimension-six operator, 
\begin{equation}
\mathcal{L}_{\text{dim-6}} \supset \frac{\kappa^2}{m^2_X}(\overline{q_{d}}_L \ell_R)(\overline{\ell_R}q_{dL}), 
\end{equation}
which enables dark pion (\(\pi_d\)) decays to charged lepton pairs. The decay width is~\cite{Schwaller:2015gea}
\begin{equation}
\Gamma(\pi_d \to \bar{\ell}\ell) = \frac{\kappa^4 f_{\pi_d}^2 m_{\ell}^2 m_{\pi_d}}{32 \pi m_X^4} \sqrt{1 - \frac{4m_\ell^2}{m_{\pi_d}^2}} , \quad \text{with} \quad \ell = e, \mu, \tau. 
\label{decay_width_of_dark_pion}
\end{equation}
Here \( f_{\pi_d} \) is the dark pion decay constant and \( m_\ell \) the lepton mass. {\color{black}Parametrizing} the proper lifetime as 
\begin{equation}
c\tau_0 = \frac{c\hbar}{\Gamma_{\pi_d}} \approx 1.92 \, \text{m} \times \left(\frac{1}{\kappa^4}\right) \left(\frac{1 \, \text{GeV}}{f_{\pi_d}}\right)^2 \left(\frac{0.1 \, \text{GeV}}{m_{\ell}}\right)^2 \left(\frac{1 \, \text{GeV}}{m_{\pi_d}}\right) \left(\frac{m_X}{1 \, \text{TeV}}\right)^4 \left({1 - \frac{4m_\ell^2}{m_{\pi_d}^2}}\right)^{-\frac{1}{2}}
\label{eq:proper_decay_length}
\end{equation} 
reveals that dark pions propagate measurable distances from the primary vertex before decaying into charged lepton pairs.

\begin{figure}[t!]\centering
\includegraphics[width=0.8\textwidth]{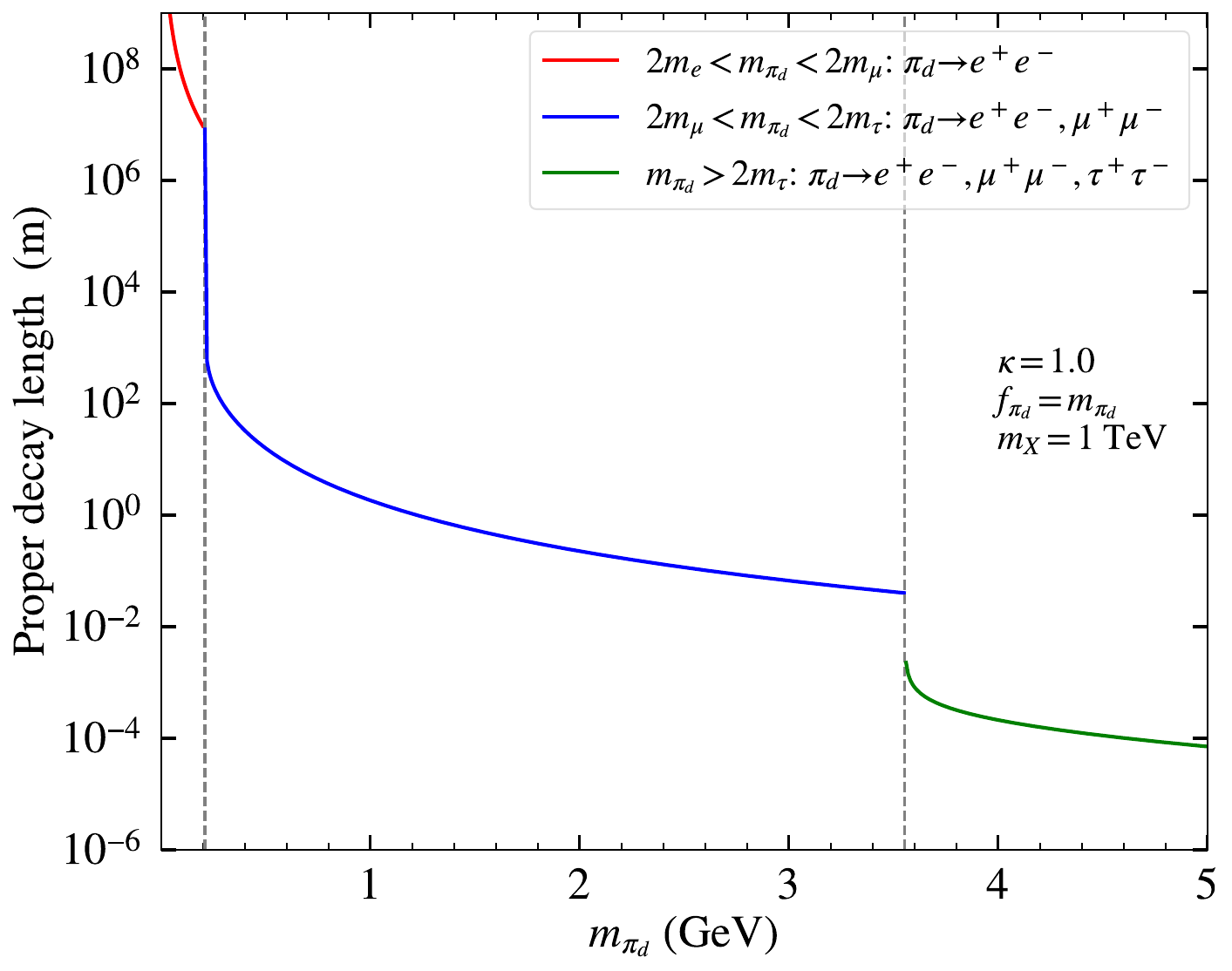}
\caption{The dark pion proper decay length with respect to $m_{\pi_d}$. Here we fixed the parameters $\kappa = 1.0$, $f_{\pi_d} = m_{\pi_d}$, and $m_X = 1$ TeV.}
\label{fig:proper_decay_length}
\end{figure}

To illustrate the typical proper decay length of $\pi_d$ versus $m_{\pi_d}$, we fix $\kappa = 1.0$, $f_{\pi_d} = m_{\pi_d}$, and $m_X = 1$ TeV as benchmark values. In order to simplify our discussions here, the lepton flavor {\color{black}universality} (LFU) is assumed such that the common coupling $\kappa$ applies to the interactions of mediator with a dark quark and a SM charged lepton for each lepton flavor. Figure~\ref{fig:proper_decay_length} shows that for $2m_e < m_{\pi_d} < 2m_{\mu}$, $\pi_d$ decays to electron-positron pairs. The proper decay length exceeds $10^7$ m, allowing electrons and positrons to escape detection. When $2m_{\mu} < m_{\pi_d} < 2m_{\tau}$, the muon decay channel opens, with $\pi_d$ predominantly decaying into muon pairs. The resulting decay length $\sim 1$ m makes these long-lived $\pi_d$ detectable. Finally, for $m_{\pi_d} > 2m_\tau$, the tau channel becomes accessible, and $\pi_d$ primarily decays to tau pairs, potentially exhibiting prompt decays. 


\section{Existing collider constraints on model parameters}
\label{sec:constraint}

Figure~\ref{fig:proper_decay_length} reveals a dramatic change in the proper decay length of $\pi_d$ at the $m_{\pi_d} \geq 2m_\mu$ threshold. We{\color{black}, therefore,} classify two dark pion mass ranges: $10\text{ MeV} \leq m_{\pi_d} < 2m_\mu$ and $2m_\mu \leq m_{\pi_d} < 2m_\tau$, and discuss their respective existing collider constraints.

\begin{figure}
    \centering
    \includegraphics[width=0.8\linewidth]{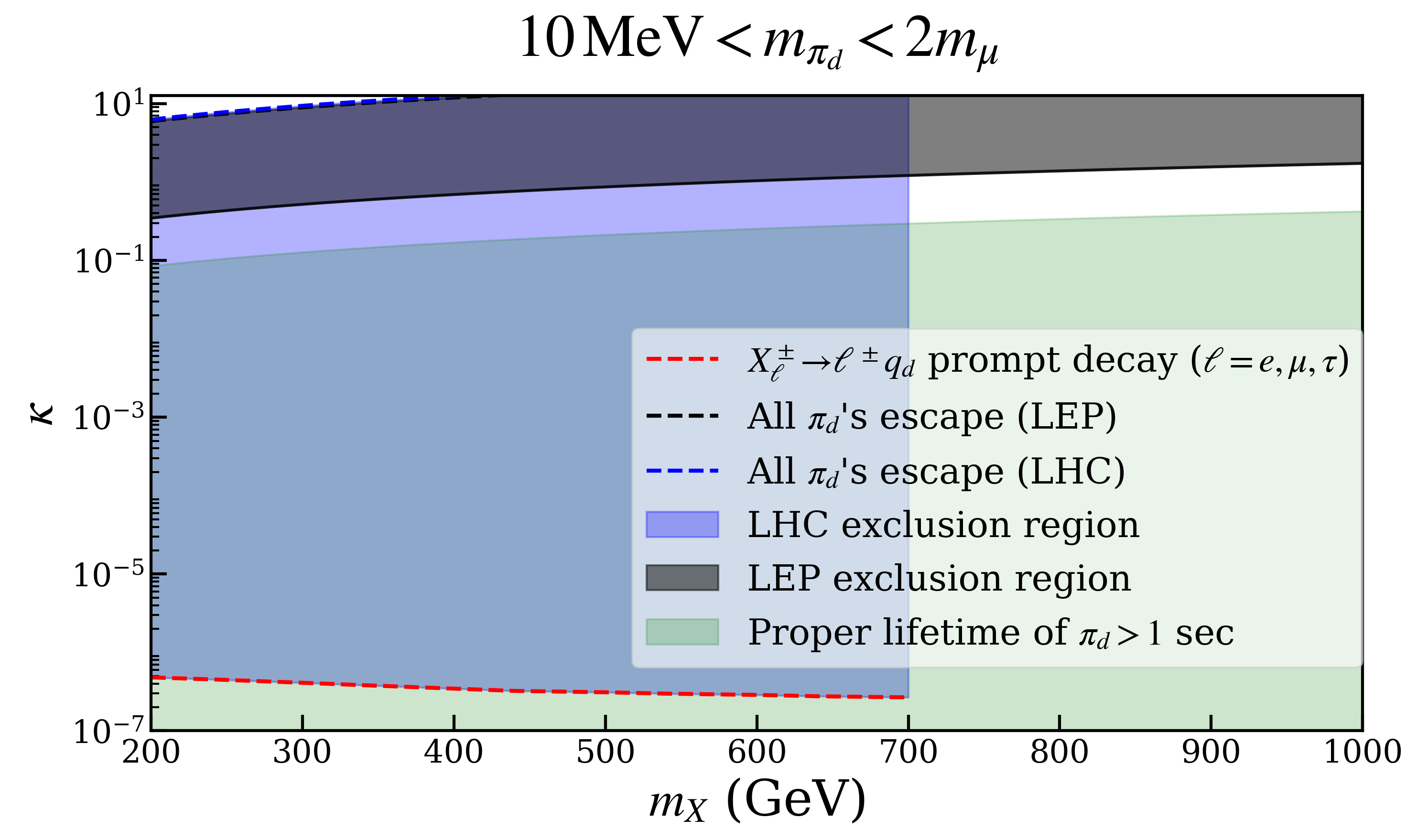}
    \caption{Exclusion region for dark pion masses $10\text{ MeV}\leq m_{\pi_d} < 2m_\mu$. The red dashed line separates long-lived (below) and prompt (above) $X^\pm$ decays. The blue and black dashed lines indicate where all $\pi_d$ particles escape detection at the LHC and LEP, respectively. The blue and black shaded regions correspond to slepton search constraints (LHC)~\cite{ATLAS:2019lff} and {\color{black}monophoton} search constraints (LEP)~\cite{Fox:2011fx, Liu:2019ogn}. The green region represents the parameter space where the proper lifetime of $\pi_d$ exceeds {\color{black}one} second, conservatively evaluated at $m_{\pi_d} = 0.2\text{ GeV}$. }
    \label{fig:light}
\end{figure}

\begin{figure}
    \centering
    \includegraphics[width=0.8\linewidth]{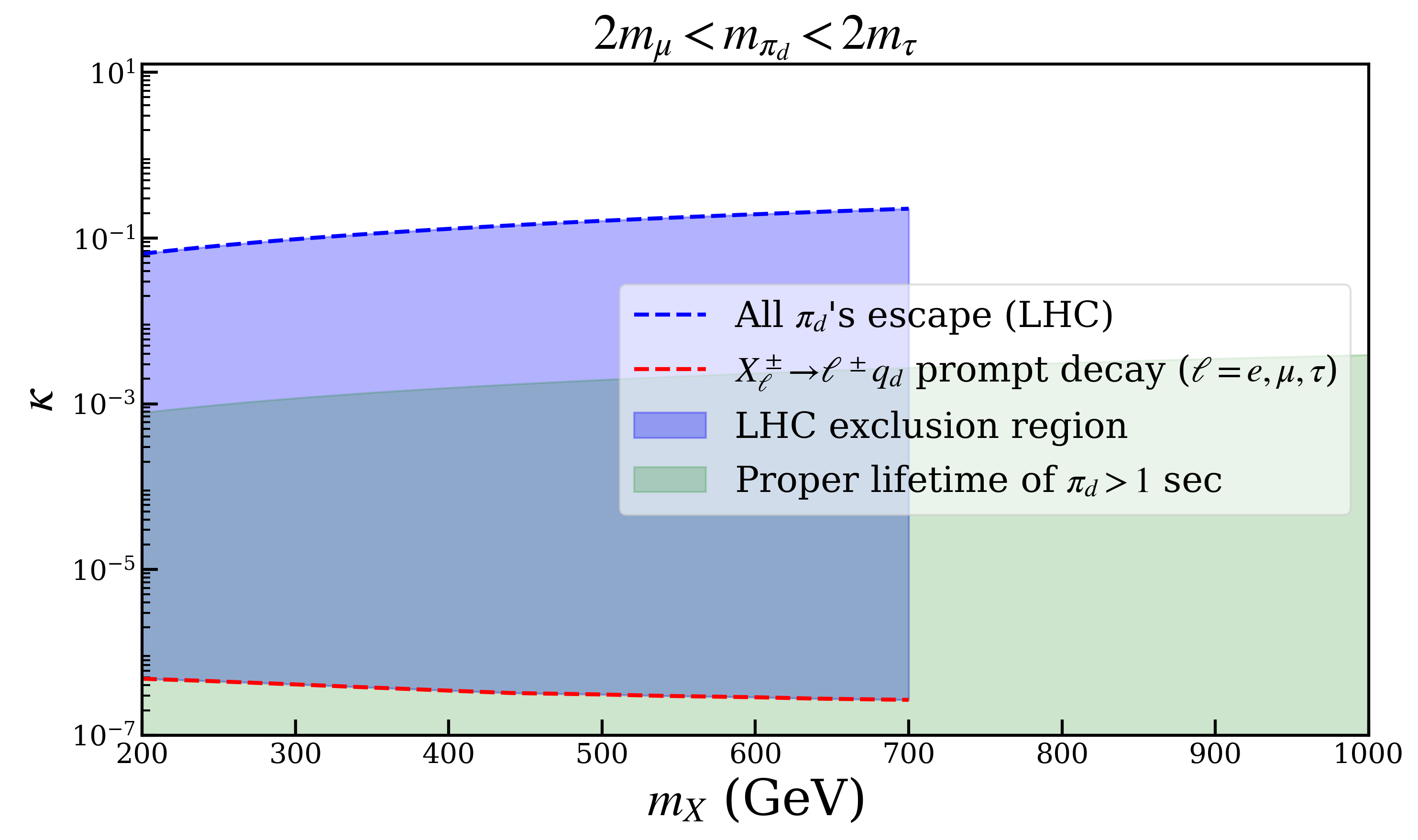}
    \caption{Exclusion region for dark pion masses $2m_\mu < m_{\pi_d} < 2m_\tau$. The blue dashed line indicates where all $\pi_d$ particles escape detection at the LHC, making this region subject to slepton search constraints. The red dashed line retains the same interpretation as in Fig.~\ref{fig:light}. The green region represents the parameter space where the proper lifetime of $\pi_d$ exceeds {\color{black}one} second, conservatively evaluated at $m_{\pi_d} = 3\text{ GeV}$. }
    \label{fig:heavy}
\end{figure}

First, we consider the production of scalar mediator pairs through $pp\to X^+ X^-$, with decays $X^+\to\ell^+ q_d$ and $X^-\to\ell^-\overline{q_d}$ ($\ell = e, \mu, \tau$). Here, $\kappa$ in Eq.~(\ref{Eq:lag}) affects only the $X^{\pm}$ lifetime, not the production {\color{black}cross section}. When $X^{\pm}$ decays promptly and dark parton showers are invisible at ATLAS and CMS detectors, signatures resemble slepton and stau searches in supersymmetry models~\cite{ATLAS:2019lff,ATLAS:2024fub}. Under the LFU assumption, we recast these constraints onto this model. We generate a UFO model file for Eq.~(\ref{Eq:lag}) via \texttt{FeynRules}~\cite{Alloul:2013bka} and simulate signal processes with \texttt{MadGraph5\_aMC@NLO}~\cite{Alwall:2014hca} (version 3.5.8). For each $m_X$, we scan $\kappa$ and define the mediator prompt decay region by $\gamma\beta c\tau_X < 0.1\text{ mm}$ in the lab frame. Figures~\ref{fig:light} and~\ref{fig:heavy} show regions above the red dashed line {\color{black}[$\kappa \gtrsim \mathcal{O}(10^{-7}-10^{-6})$]} for the mediator prompt decays recast from slepton and stau constraints\footnote{{\color{black}The signal production cross sections rescaled to match Fig.~3 of Ref.~\cite{Zhang:2021orr} with QCD NLO+NLL corrections at $13\text{ TeV}$ LHC~\cite{Fiaschi:2018xdm}.}}.  
Additionally, because there are plenty of dark pions produced after dark showers and hadronization, we require the $\pi_d$ proper decay length (without any boost) larger than $10\text{ m}$ to conservatively assume all dark pions escape ATLAS and CMS detectors as missing energy. Specifically, we set $m_{\pi_d} = f_{\pi_d} = 0.2\text{ GeV}$ in Fig.~\ref{fig:light} and $m_{\pi_d} = f_{\pi_d} = 3\text{ GeV}$ in Fig.~\ref{fig:heavy} to estimate the upper bounds of $\kappa$. Regions below blue dashed lines {\color{black}[$\kappa \lesssim \mathcal{O}(1-10)$ for $10\text{ MeV}\leq m_{\pi_d} < 2m_\mu$; $\kappa \lesssim \mathcal{O}(0.1-1)$ for $2m_\mu \leq m_{\pi_d} < 2m_\tau$]} correspond to invisible dark showers which are adapted for recasting. 

Under the LFU assumption, the mediator $X^{\pm}$ has approximately equal decay branching ratios of $1/3$ to each flavor channel ($e^{\pm}q_d$, $\mu^{\pm}q_d$, $\tau^{\pm}q_d$). Consequently, we include an extra factor of $1/9$ when applying stau pair production constraints. Comparison with ATLAS results from Table~{\color{black}1} of auxiliary material in Ref.~\cite{ATLAS:2024fub} shows that for $m_X > 200$ GeV and the dark quark mass $m_{q_d} = 1$ GeV, the effective {\color{black}cross section} is at least a factor of {\color{black}2} below the experimental bounds. Thus, no exclusion region exists for $m_X > 200$ GeV, though stau constraints exclude $m_X \lesssim 160$ GeV.

For slepton pair production constraints in Fig.~{\color{black}3(c)} of Ref.~\cite{ATLAS:2019lff}, the analysis involves all combinations of a pair of leptons from electrons and muons in the final states. Therefore, we apply a factor of $4/9$ in our process. Notably, while the constraints include contributions from both $\tilde{\ell}^+_L\tilde{\ell}^-_L$ and $\tilde{\ell}^+_R\tilde{\ell}^-_R$ production, our signal process specifically corresponds to the $\tilde{\ell}^+_R\tilde{\ell}^-_R$ channel. We{\color{black}, therefore,} isolate the right-handed slepton constraints using the fraction of $pp\to\tilde{\ell}^+_L\tilde{\ell}^-_L$ and $pp\to\tilde{\ell}^+_R\tilde{\ell}^-_R$ production {\color{black}cross sections} from Refs.~\cite{Fuks:2013lya,Fiaschi:2018xdm}. The resulting effective {\color{black}cross section} in our process exceeds ATLAS exclusion limits for $\tilde{\ell}^+_R\tilde{\ell}^-_R$ in the mass range $m_X \in [200, 700]$ GeV. Additionally, after accounting for the average Lorentz boost of the mediator ($\gamma\beta\simeq 1.5$), we estimate the prompt decay boundaries within this mass range.

Conversely, for $\pi_d$ with moderate decay lengths, the novel signatures of displaced lepton jets are expected to be observable at the LHC. However, Ref.~\cite{Zhang:2021orr} notes detection challenges for the $pp\to X^+ X^-$ production channel due to (1) insufficient production {\color{black}cross sections}, and (2) detection efficiencies after considering the whole cut-flow  below $\sim 0.1$. Moreover, significant irreducible cosmic-ray backgrounds further complicate the observation. While LHC Run 3 and HL-LHC may improve sensitivity, current ATLAS displaced lepton jet searches~\cite{ATLAS:2022izj} exclusively target Higgs exotic decays, distinct from this model's signature of two opposite-sign leptons plus two displaced lepton jets. Consequently, recasting strategies from Ref.~\cite{ATLAS:2022izj} are inapplicable. A dedicated analysis (beyond the scope of this work) would be required, though we establish that the LHC constraints remain weaker than those derived for $e^+e^-$ colliders in {\color{black}Secs}~\ref{sec:Babar} and~\ref{sec:CEPC}.

Dark quark pairs can also be produced via $t$-channel mediator exchange at the LEP. Following dark showers and hadronization, if all dark mesons escape the detector, they contribute to missing energy. With {\color{black}initial-state radiation}, this also yields a {\color{black}monophoton} signature. For $10\text{ MeV}\leq m_{\pi_d} < 2m_\mu$, we conservatively set $m_{\pi_d} = f_{\pi_d} = 0.2\text{ GeV}$ and require $\pi_d$ proper decay length greater than $12$ m to ensure that all dark mesons escape the detector volume of LEP. The black dashed line in Fig.~\ref{fig:light} gives upper bounds $\kappa\sim\mathcal{O}(1-10)$. We further recast {\color{black}monophoton} constraints at the LEP from the $t$-channel scalar mediator and Dirac fermionic DM model with the {\color{black}cutoff} scale $\Lambda$ {\color{black}[Eq.~(4)]} and the purple line in Fig.~4 ($\bar{\chi}e\bar{e}\chi$) of Ref.~\cite{Fox:2011fx} to the leptophilic composite ADM via
\begin{equation}
\frac{1}{\Lambda^2} \simeq \frac{1}{(333.86\text{ GeV})^2} = 3 \frac{\kappa^2}{m^2_{X}}, 
\end{equation}
where the factor of 3 originates from $SU(3)_D$ charges. Assuming light dark quarks and determining lower bounds of $\Lambda$, the black regions in Fig.~\ref{fig:light} indicate where predicted {\color{black}cross sections} exceed experimental bounds, consistent with the $1/\Lambda^2$ scaling.

Finally, a dark pion lifetime exceeding one second would violate observed light-element abundances and the {\color{black} cosmic microwave background (CMB)} spectrum, as constrained by {\color{black}big bang nucleosynthesis and CMB} measurements~\cite{Lindley:1984bg,Dev:2025pru}. We{\color{black}, therefore,} incorporate these constraints by excluding regions where the proper lifetime of $\pi_d$ is greater than {\color{black}one second}, indicated by the green shaded areas in Figs.~\ref{fig:light} and~\ref{fig:heavy}. Notably, only a narrow region of parameter space remains in Fig.~\ref{fig:light} for $m_X \lesssim 1$ TeV.

Although portions of the parameter space are excluded in Figs.~\ref{fig:light} and~\ref{fig:heavy}, significant regions remain accessible for exploration at both low-energy and high-energy $e^+e^-$ colliders. In subsequent sections, we derive exclusion limits for three benchmark points (BPs): $m_{\pi_d} = 0.32$, $0.7$, and $1$ GeV. 
\section{Sparse dark showers at BaBar, Belle II and GAZELLE} 
\label{sec:Babar}

As previously discussed, the intensity of dark showers depends critically on the energy of the produced dark quarks. This section focuses on low-energy $e^+e^-$ colliders capable of generating sparse dark showers, where individual displaced vertices from long-lived dark pion decays can be resolved. We specifically examine the BaBar and Belle II experiments, along with the proposed GAZELLE far detector for Belle II.

In this and the following section, we employ \texttt{FeynRules}~\cite{Alloul:2013bka} to generate a UFO model file for Eq.~(\ref{Eq:lag}), and then parton-level Monte Carlo events are generated using \texttt{MadGraph5\_aMC@NLO}~\cite{Alwall:2014hca}. Subsequently, the Hidden Valley (HV) module within \texttt{Pythia8.313}~\cite{Sjostrand:2014zea} simulates dark parton showering and hadronization processes. Parameter settings closely follow Refs.~\cite{Schwaller:2015gea,Mies:2020mzw,Archer-Smith:2021ntx,Carrasco:2023loy}, specifically{\color{black},} $N_{C_d} = 3$, $N_{f_d} = 7$, $\Lambda_d = m_{q_d} = 2m_{\pi_d}$, and $m_{\rho_d} = 4m_{\pi_d}$. In our simulation, diagonal and off-diagonal HV-mesons share identical decay patterns. We implement the running coupling $\alpha_d$ and apply a $p_T$ cutoff of $1.1\Lambda_d$ to dark showers. The fraction of vector and pseudoscalar HV-mesons is calculated by following Appendix A of Ref.~\cite{Knapen:2021eip}. Finally, the ratio of dark pion mass over dark pion decay constant, $m_{\pi_d}/f_{\pi_d}$, can range from $\mathcal{O}(1-10)$~\cite{Berlin:2018tvf,Choi:2018iit,Kuwahara:2023vfc,Bernreuther:2023kcg}. For this analysis, we fix $f_{\pi_d} = m_{\pi_d}$. This choice only affects the dark pion decay width, which can be easily rescaled using Eq.~(\ref{decay_width_of_dark_pion}). 



\subsection{Recasting LLP searches at the BaBar} 

The BaBar experiment utilized the PEP-II collider, colliding $9\,\text{GeV}$ electrons with $3.1\,\text{GeV}$ positrons at a center-of-mass energy $\sqrt{s}=10.58\,\text{GeV}$. Dark quark pairs are produced via the $t$-channel process (Fig.~\ref{fig:Feynman}, left panel). Following dark showering and hadronization, long-lived dark pions decay to muon pairs, creating multiple displaced vertices. Such novel signature is an icon of dark showers.

To recast the previous BaBar long-lived particle (LLP) searches to this model, we implement the event selection criteria from Refs.~\cite{BaBar:2015jvu,Bernreuther:2022jlj}: 
\begin{itemize} 
\item The transverse displacement of $\pi_d$ is required to be within the range 10 mm $< R_{xy} <$ 500 mm from the interaction point (IP). 
\item The final state must include at least three charged tracks with $p_T > 0.12$~GeV, among which at least two of them must have $p_T > 0.18$~GeV.
\item The invariant mass of the decay products must satisfy either $m_{\mu^+\mu^-} < 0.37$~GeV or $m_{\mu^+\mu^-} > 0.5$~GeV. 
\item Each muon is required to have $d_0/\sigma_{d_0} > 3$, where $d_0$ is the transverse impact parameter and $\sigma_{d_0}$ is its uncertainty.
\end{itemize}    
The detector resolution is performed by the smearing effect of the relevant uncertainties of observables. Since recasting efficiencies are only available for LLPs with $m > 0.5~\text{GeV}$ and {\color{black}$c\tau > 5~\text{mm}$,} as noted in Refs.~\cite{BaBar:2015jvu,Bernreuther:2022jlj}, we restrict analysis to two of our BPs: $m_{\pi_d} = 0.7~\text{GeV}$ and $1~\text{GeV}$, requiring $c\tau > 5~\text{mm}$ for dark pions.


Specifically, to calculate the impact parameter $d_0$, we first select the long-lived dark pion candidate and its daughter muons. We record their respective three-momenta and compute the sine of the angle $\alpha$ between them. The transverse displacement is defined as $R_{xy} = \sqrt{R_x^2 + R_y^2}$, where $R$ is the flight distance of the dark pion in the lab frame, and $R_x$ ($R_y$) is the $x$ ($y$) component of $R$. The impact parameter is then given by $d_0 = R_{xy} \sin\alpha$. The position resolution of displaced vertices and the momentum resolution of the final-state particles are given by~\cite{BaBar:1995bns}
\begin{equation}
\sigma_{xy} = \sigma_z = \left[ \frac{50\text{ GeV}}{p_T} + 15 \right]\mu\text{m}, 
\end{equation}

\begin{equation}
\frac{\sigma_{p_T}}{p_T} = \left[0.21 + 0.14\frac{p_T}{\text{GeV}}\right]\%.
\end{equation} 
Here, $p_T$ denotes the transverse momentum of the charged track. In the calculation of $d_0$, we use $\sigma_{xy}$ to account for the detector resolution in the transverse plane.

\begin{figure}[t!]
\centering
\includegraphics[width=0.45\textwidth]{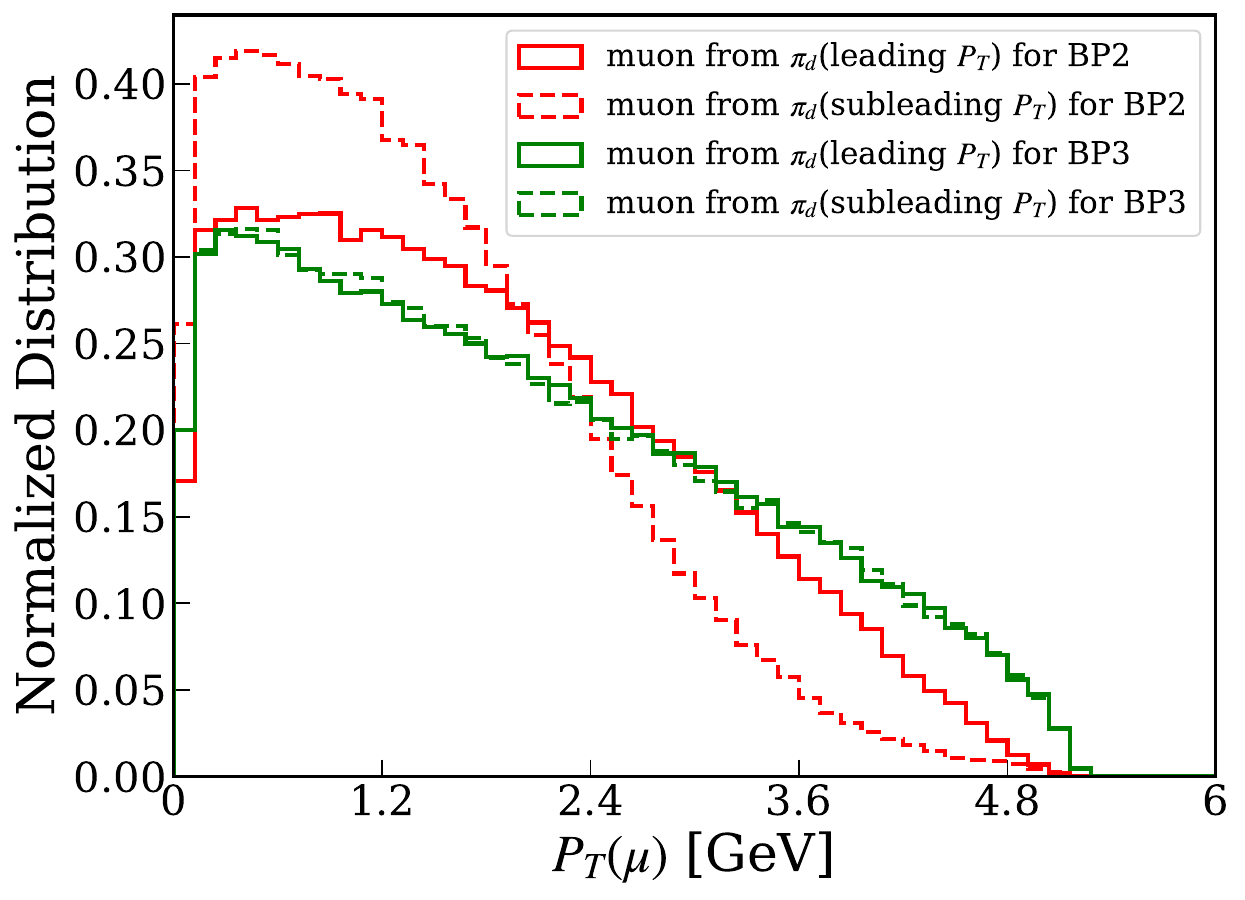}
\hfill
\includegraphics[width=0.45\textwidth]{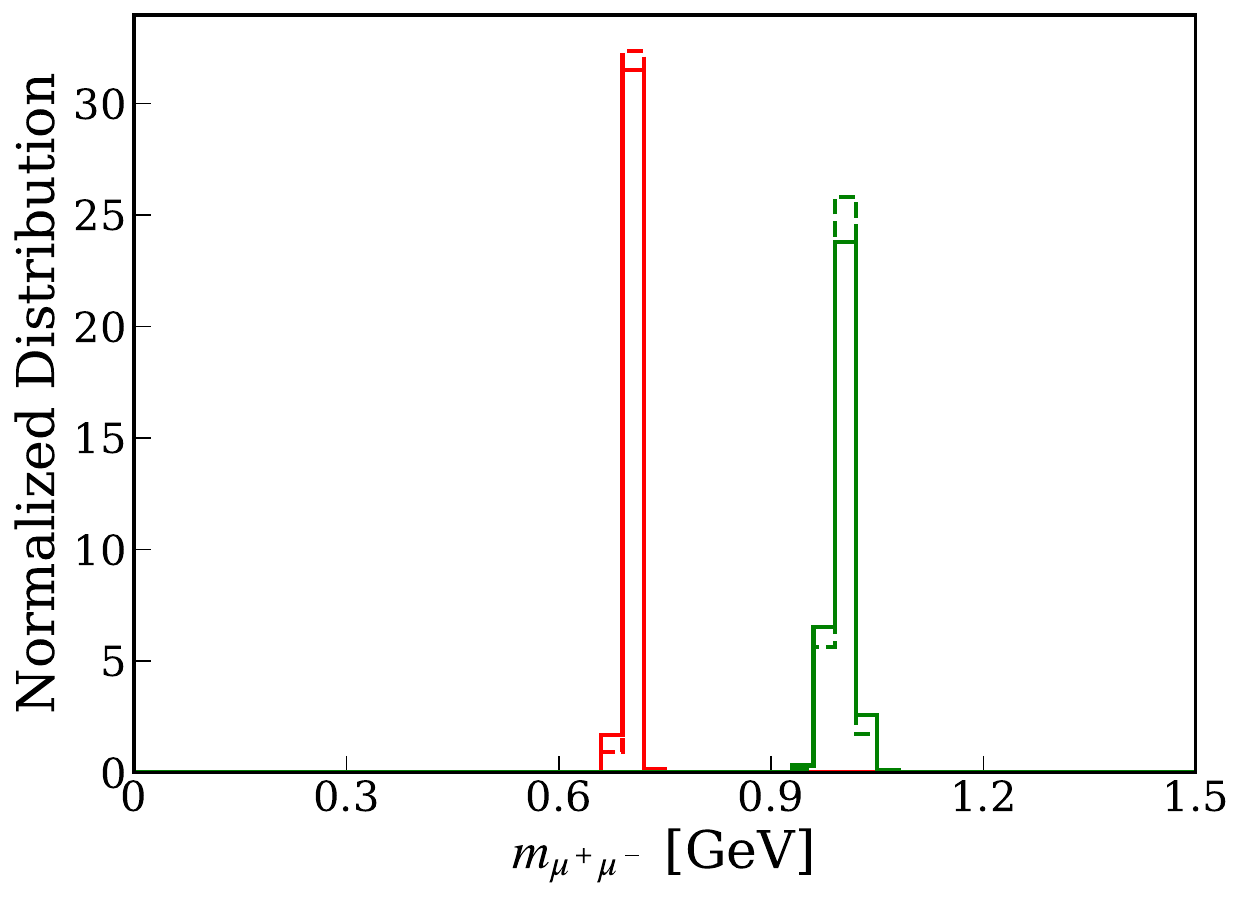}

\vspace{0.5cm}

\includegraphics[width=0.45\textwidth]{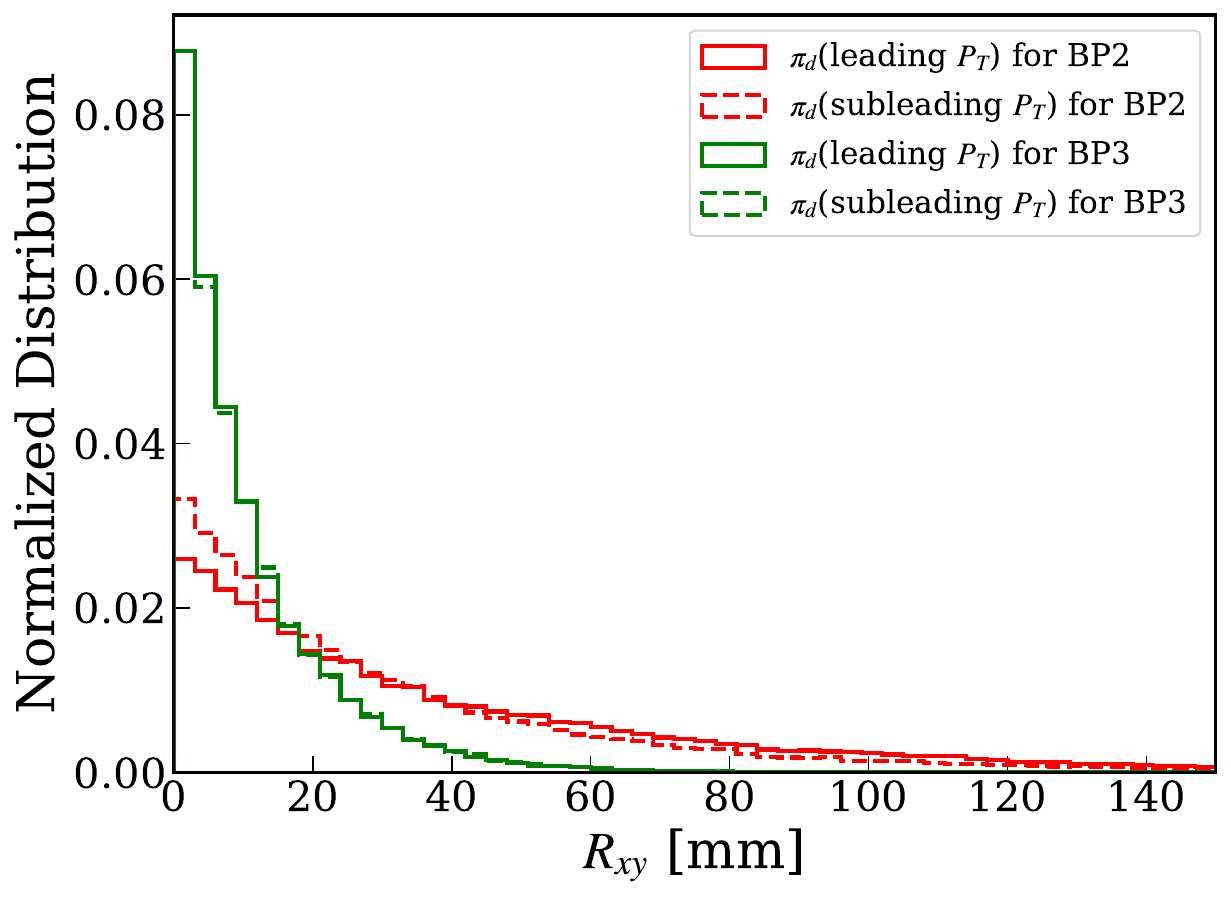}
\hfill
\includegraphics[width=0.45\textwidth]{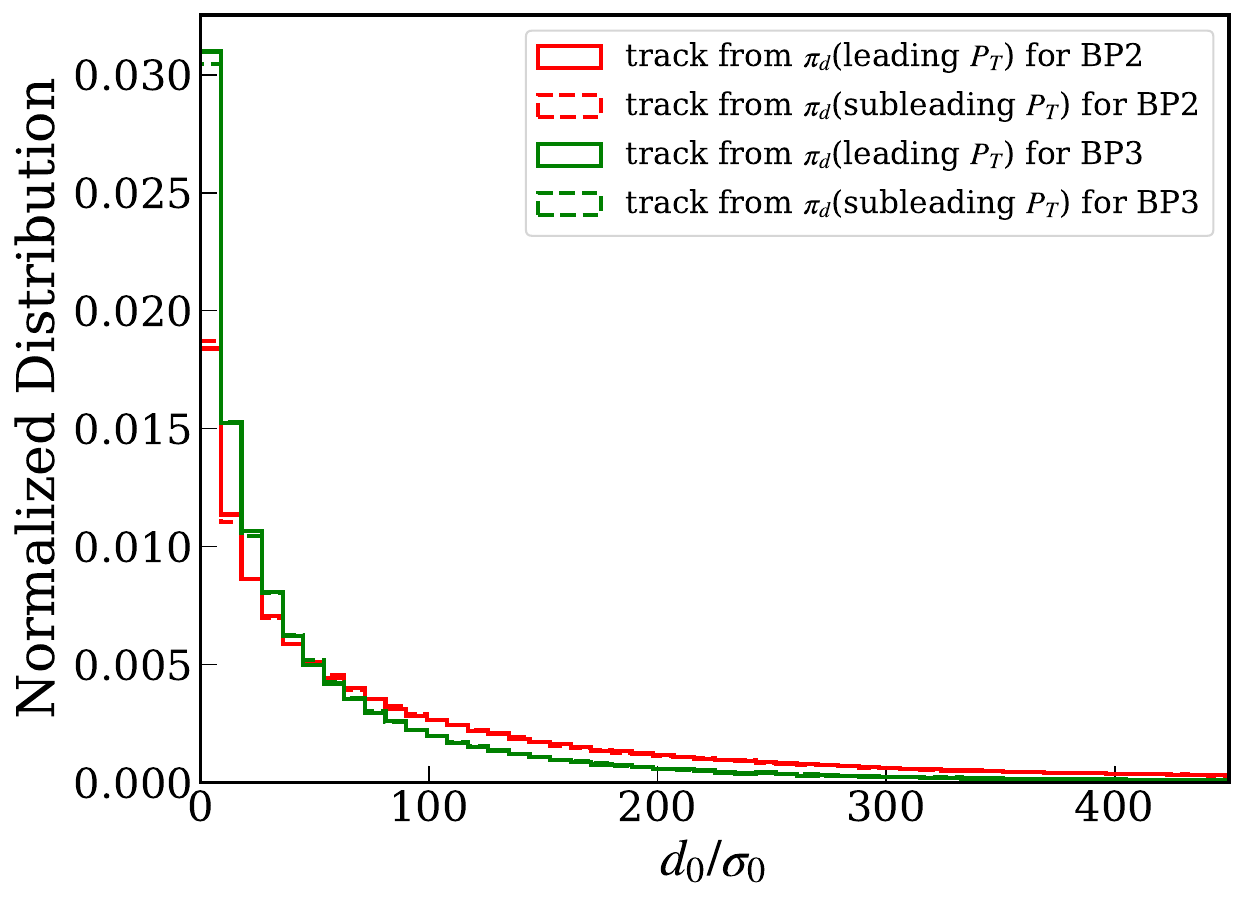}
\caption{
Kinematic distributions for the signal benchmark points BP2 and BP3 at the BaBar. The distributions of the leading $p_T$ and subleading $p_T$ dark pions and their daughter muons are represented by solid and dashed lines, respectively. 
{\color{black}Top left}: transverse momentum $p_T$ of muons from dark pion decays.  
{\color{black}Top right}: invariant mass $m_{\mu^+\mu^-}$ of muon pairs. 
{\color{black}Bottom left}: transverse decay distance $R_{xy}$ of dark pions. 
{\color{black}Bottom-right}: track impact parameter significance $d_0/\sigma_0$. 
}
\label{fig:babar__kinematics}
\end{figure}

Figure~\ref{fig:babar__kinematics} shows the relevant kinematic distributions of these two signal BPs. First, the transverse momentum distribution of the muon, $P_T(\mu)$ ({\color{black}top left}), is generally larger for heavier $\pi_d$ masses. This is because, under a lower total $\pi_d$ yield, each $\pi_d$ carries higher average energy, leading to muons from its decay with larger $p_T$. 
Second, in the $\mu^+\mu^-$ invariant mass distribution, $m_{\mu^+\mu^-}$ ({\color{black}top right}), the position of the reconstructed mass peak directly reflects the mass of $\pi_d$. 
Regarding the transverse decay length, $R_{xy}$ ({\color{black}bottom left}), since $\pi_d$ is long-lived, its decay vertex is significantly displaced from the IP in the transverse plane. Moreover, due to the proper decay length $c\tau_0 \propto 1/m_{\pi_d}^3$, a lighter $\pi_d$ has a longer decay length, resulting in an $R_{xy}$ distribution that extends to larger values and exhibits a more pronounced long tail for BP2 compared to BP3.
Finally, the transverse impact parameter $d_0$ ({\color{black}bottom right}) further characterizes the long-lived nature of $\pi_d$. The value of $d_0$ is significantly larger than the uncertainty $\sigma_0$ which clearly distinguishes $\pi_d$ decays from prompt decays.

\begin{figure}[h]
\centering 
\begin{minipage}{0.48\textwidth}
\includegraphics[width=\textwidth]{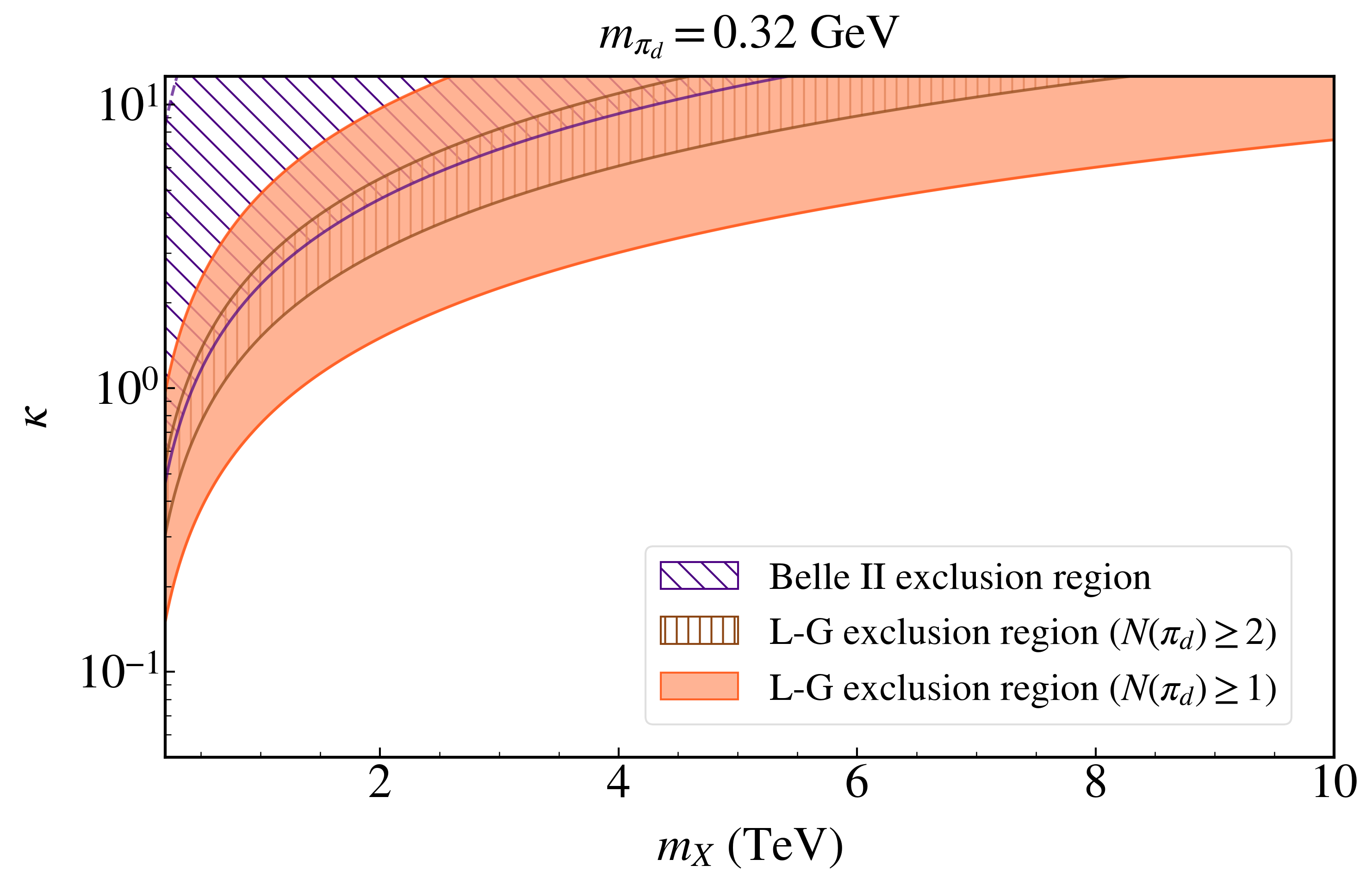}
\end{minipage}
\begin{minipage}{0.48\textwidth}
\includegraphics[width=\textwidth]{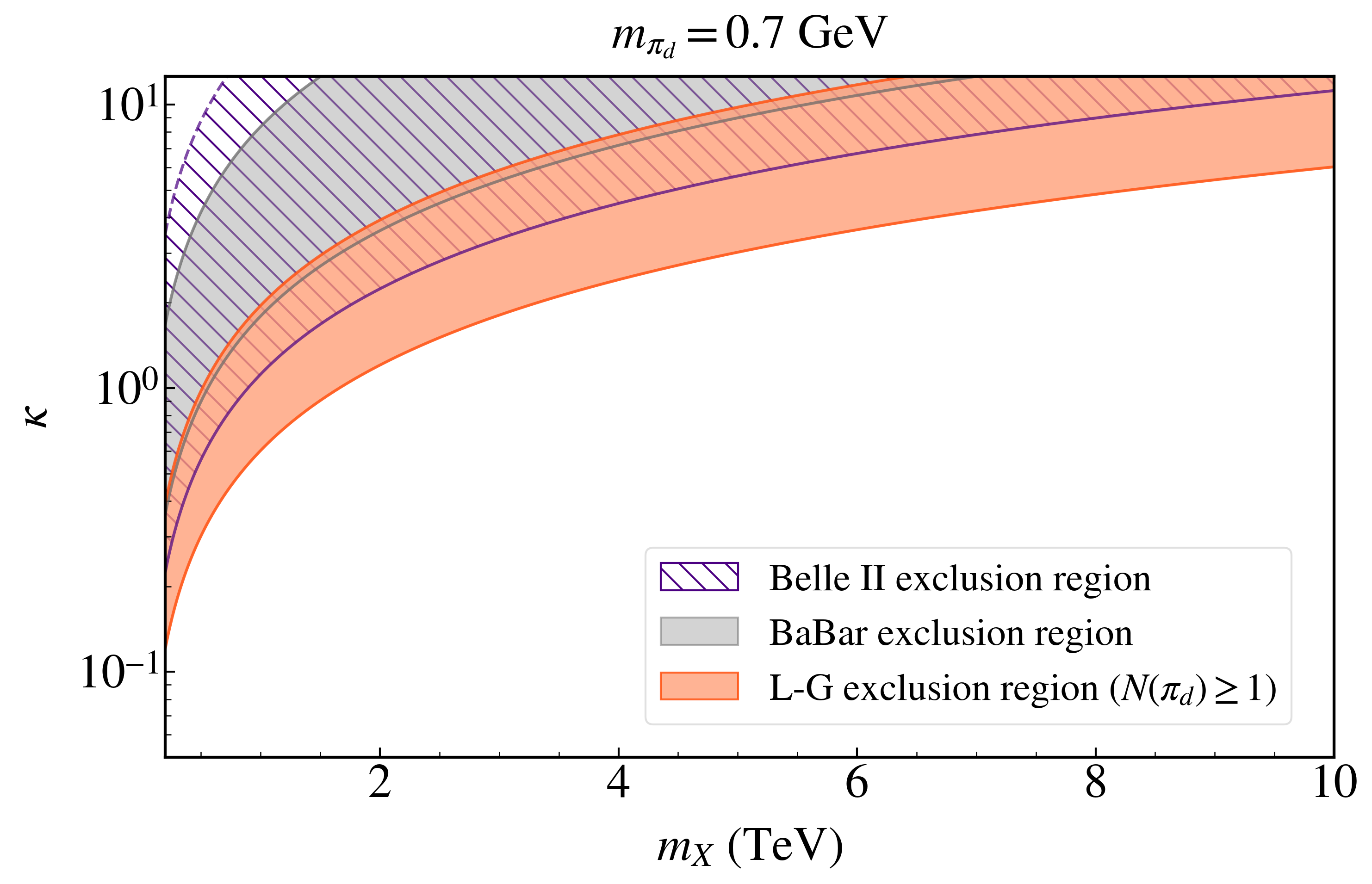}
\end{minipage}
\\
\begin{minipage}{0.48\textwidth}
\centering
\includegraphics[width=\textwidth]{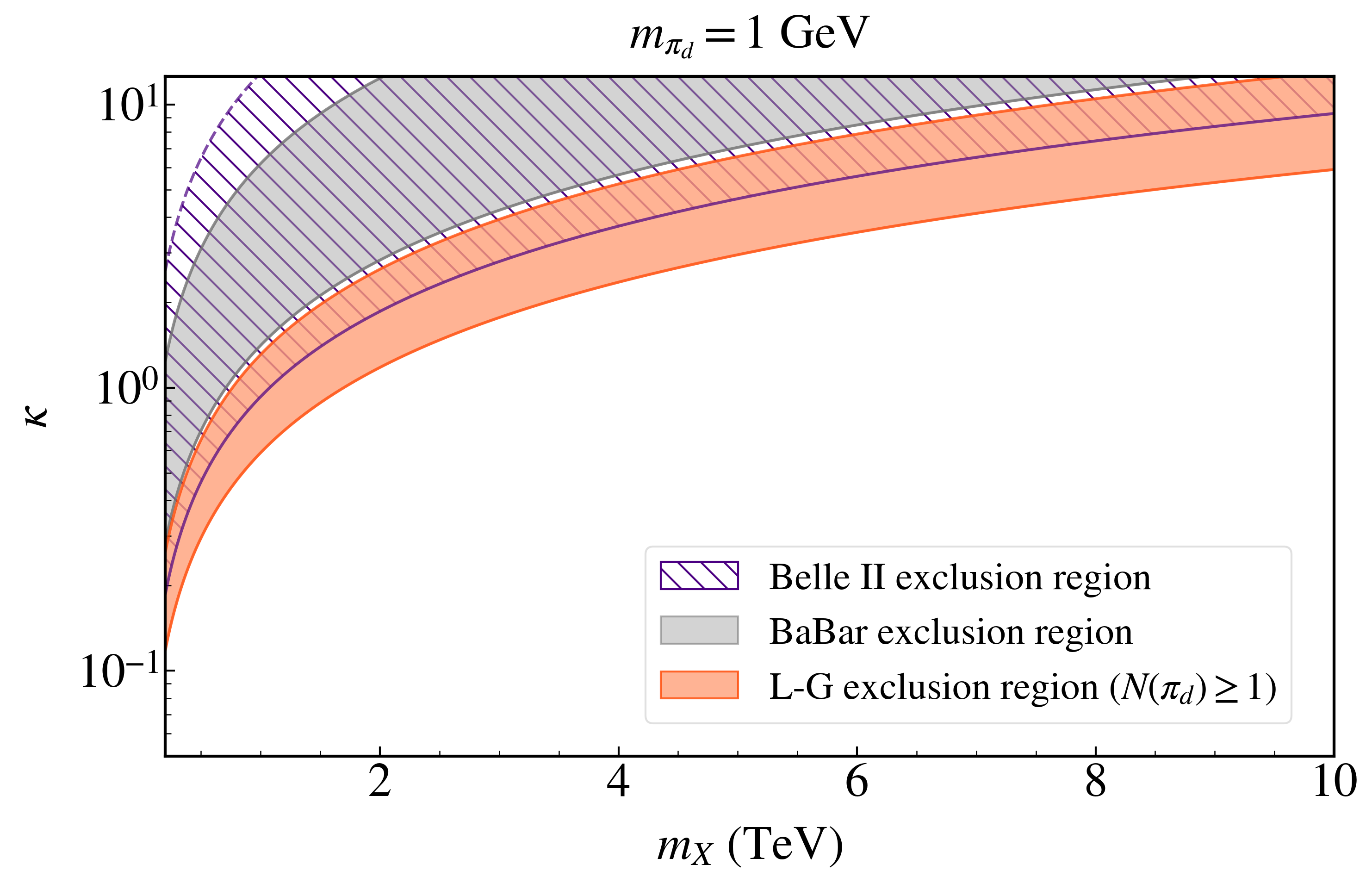}
\end{minipage}
\caption{Exclusion limits for three benchmark points, $m_{\pi_d}=0.32$, $0.7$ and $1$ GeV, for three experiments: BaBar, Belle II and L-GAZELLE.}
\label{fig:belle-gazelle-babar}
\end{figure}

The production {\color{black}cross section} for $e^+ e^- \to \overline{q_d} q_d$ scales as $(\kappa/m_X)^4$, while the proper lifetime of $\pi_d$ is inversely proportional to this factor. We{\color{black}, therefore,} recast BaBar LLP search constraints onto the $(m_X, \kappa)$ parameter plane. Specifically, we extract exclusion limits from the $\mu^+\mu^-$ channel (red curve in Fig.~2 of Ref.~\cite{BaBar:2015jvu}), plotting LLP mass versus $\sigma(e^+e^- \to LX)\mathcal{B}(L\to f) \epsilon(f)$, where $\sigma(e^+e^- \to LX)$ is the LLP production {\color{black}cross section} ($L$ is the LLP candidate, and $X$ are associated particles){\color{black},} $\mathcal{B}(L\to f)$ is the branching fraction to final state of a pair of $f$ particles (a muon pair in our case){\color{black}, and} $\epsilon(f)$ is the detection efficiency to such final state. For $m_{\pi_d} = 0.7$ GeV and $1$ GeV, exclusion limits appear in gray regions of Fig.~\ref{fig:belle-gazelle-babar}. The upper boundaries correspond to short dark pion lifetimes that preclude LLP classification, while lower boundaries indicate long-lived pions escaping the detector's sensitive volume. The inverse relationship of constraints between $m_X$ and $\kappa$ is clearly evident in Fig.~\ref{fig:belle-gazelle-babar}.


\subsection{Detecting dark showers at the Belle II}

The Belle II experiment operates at the SuperKEKB collider, which collides $7\text{ GeV}$ electrons with $4\text{ GeV}$ positrons at a center-of-mass energy of $\sqrt{s} = 10.58\text{ GeV}$. Dark quark pairs are produced via the process $e^+ e^- \to \overline{q_d} q_d$, followed by dark showers and hadronization, as illustrated in the left panel of Fig.~\ref{fig:Feynman}. This leads to the production of several long-lived dark pions, which predominantly decay into muon pairs. As a result, multiple displaced vertices can be identified as the novel signal signature. Three BPs with $m_{\pi_d} = 0.32$, $0.7$, and $1$ GeV are chosen for illustration, corresponding to $\kappa = 1.0$ and $m_X=200\text{ GeV}$.



\begin{table}[h]
\centering
\begin{tabular}{|c|c|c|c|}
\hline
$m_{\pi_d}$ [\text{GeV}] & $\langle{N}_{\pi_d}\rangle$ & $\langle{E} (\pi_d)\rangle$ [\text{GeV}] & $\langle {P_T} (\mu)\rangle$ [\text{GeV}] \\
\hline
0.32 & 6.20 & 1.78 & 0.63 \\
\hline
0.7 & 3.15 & 3.46 & 1.26 \\
\hline
1 & 2.03 & 5.43 & 1.96 \\
\hline
\end{tabular}
\caption{The average dark pion $\pi_d$ multiplicity, the average energy of each $\pi_d$, and the average transverse momentum of each muon in the final state for three BPs with $m_{\pi_d} = 0.32$, $0.7$, and $1$ GeV at the Belle II. }
\label{tab:belle_II_muon_pid_conut_pt}
\end{table}

\begin{table}[h!]
\centering
\begin{tabular}{|p{4cm}|p{10cm}|}
\hline
\textbf{Trigger} & \textbf{Conditions} \\
\hline
Two tracks &Two tracks with \( p_T > 0.3 \) GeV for each and an azimuthal opening angle \( \Delta\phi > 90^\circ \) at the interaction point, as well as the angle between the position vector of decay vertex and the beam axis with \( 38^\circ < \theta_{\text{lab}} < 127^\circ \) in the lab frame \\
\hline
One muon & One muon with \( p_T > 0.9 \) GeV  and \( 38^\circ < \theta_{\text{lab}} < 127^\circ \) \\
\hline
Displaced vertex & At least one displaced vertex in the event with \( 9 \) mm \( < R < 600 \) mm, formed from two tracks with \( p_T > 0.1 \) GeV and \( 38^\circ < \theta_{\text{lab}} < 127^\circ \) \\
\hline
\end{tabular}
\caption{Three Belle II triggers are considered in this analysis: two tracks, one muon, and displaced vertex, adapted from Ref.~\cite{Bernreuther:2022jlj}.}
\label{tab:trigger_summary}
\end{table}

\begin{table}[h!]
\centering
\begin{tabular}{|p{4cm}|p{10cm}|}
\hline
\textbf{Object} & \textbf{Selection Criteria} \\
\hline
Muons & 
\begin{tabular}[t]{@{}l@{}}
\( p_T(\mu^+),~ p_T(\mu^-) > 0.05 \) GeV \\
\( m_{\mu^+\mu^-} < 0.48 \) GeV or \( m_{\mu^+\mu^-} > 0.52 \) GeV
\end{tabular} \\
\hline
Displaced vertices & 
\begin{tabular}[t]{@{}l@{}}
\( 0.2 \) cm \( < R_{xy} < 60 \) cm \\
\( -55 \) cm \( \leq z \leq 140 \) cm \\
\( 17^\circ \leq \theta_{\text{lab}} \leq 150^\circ \)
\end{tabular} \\
\hline
\end{tabular}
\caption{The event selection criteria for muons and displaced vertices in LLP searches at the Belle II~\cite{Bernreuther:2022jlj}.}
\label{tab:llp_selection_summary}
\end{table}

In all BPs, the average number of $\pi_d$ per event exceeds two as shown in Table~\ref{tab:belle_II_muon_pid_conut_pt}. Thus, we require at least two $\pi_d$'s per event for the signal process in subsequent analyses. All events must satisfy the Belle II trigger selections outlined in Table~\ref{tab:trigger_summary} and then pass the event selection criteria outlined in Table~\ref{tab:llp_selection_summary}~\cite{Bernreuther:2022jlj}. Moreover, key kinematic properties for each BP are summarized in Table~\ref{tab:belle_II_muon_pid_conut_pt}, including the average $\pi_d$ multiplicity, the average energy of each $\pi_d$, and the average transverse momentum of each muon in the final state. 

\begin{table}[h]
\centering
\begin{tabular}{|l|*{3}{S[table-format=2.2]|}}
\hline
\textbf{Trigger Type} & {\( m_{\pi_d} = 0.32 \, \text{GeV} \)} & {\( m_{\pi_d} = 0.70 \, \text{GeV} \)} & {\( m_{\pi_d} = 1.00 \, \text{GeV} \)} \\
\hline
Two Tracks & 75.82\% & 58.72\% & 34.16\% \\ 
After Cuts & 17.78\% & 47.48\% & 25.31\% \\ 
\hline
One Muon & 83.76\% & 89.22\% & 84.17\% \\ 
After Cuts & 9.93\% & 65.01\% & 51.62\% \\ 
\hline
Displaced Vertex & 91.24\% & 79.41\% & 46.24\% \\ 
After Cuts & 11.39\% & 60.12\% & 35.16\% \\
\hline
\end{tabular}
\caption{Trigger efficiencies for three different dark pion masses with fixed \( \kappa = 1.0 \,, m_X = 200 \, \text{GeV} \).} 
\label{tab:trigger_efficiencies}
\end{table}

\begin{figure}[t!]
\centering
\includegraphics[width=0.45\textwidth]{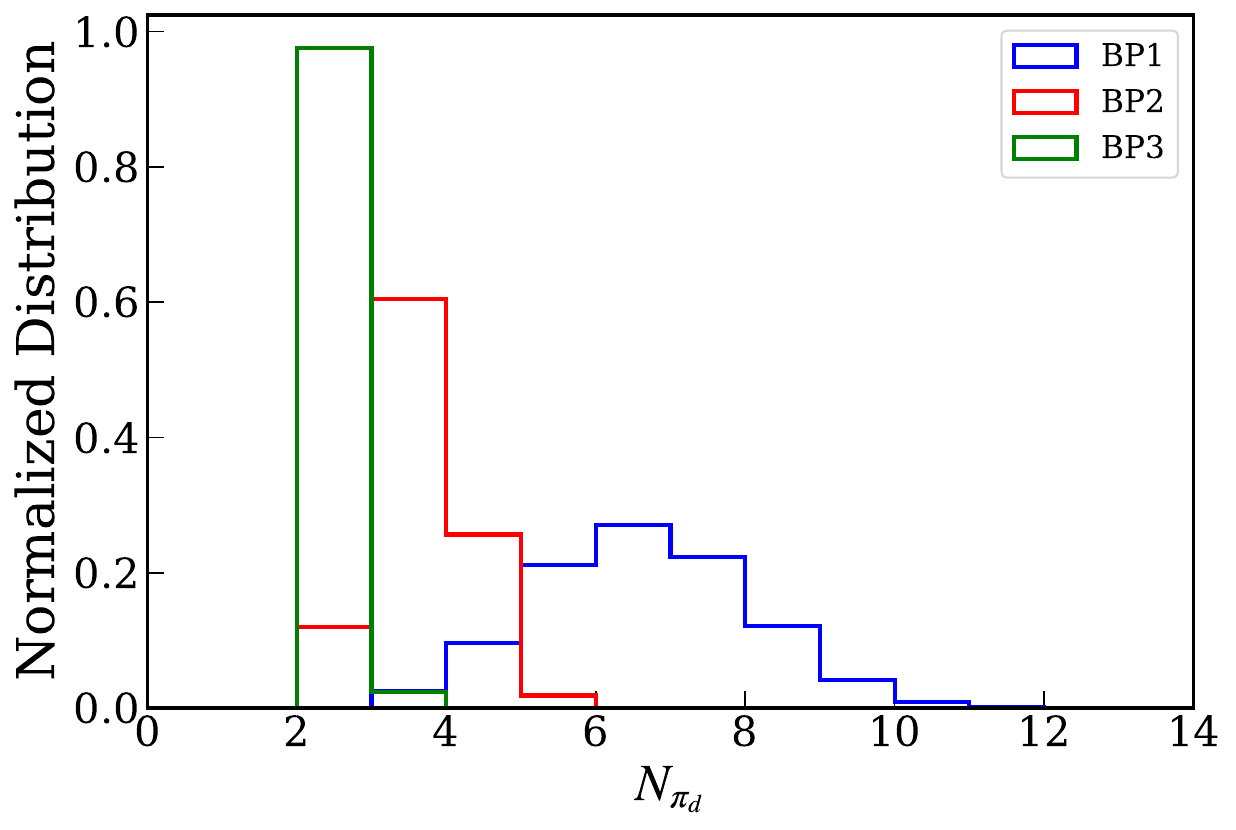}
\hfill
\includegraphics[width=0.45\textwidth]{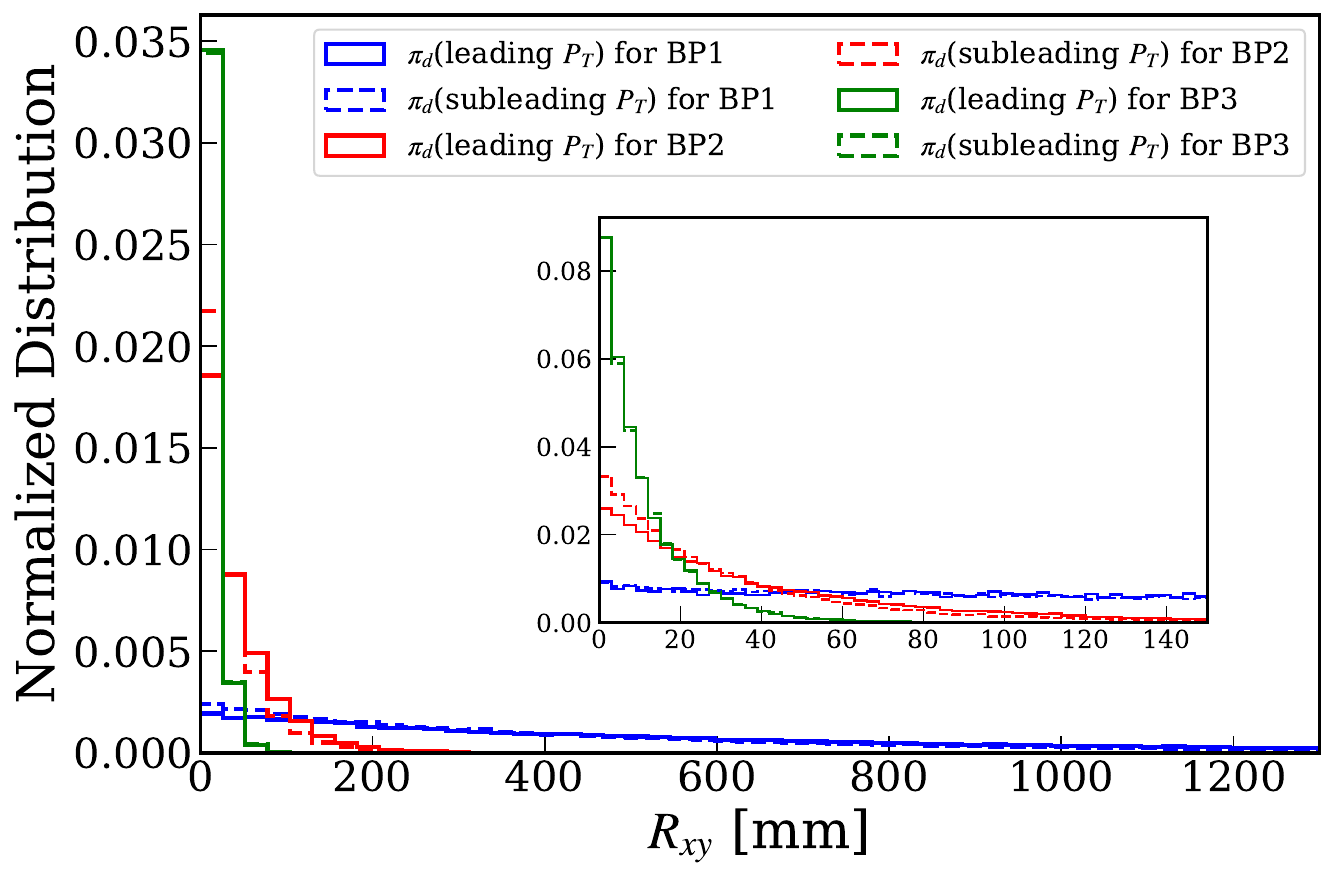}
\vspace{0.5cm}
\includegraphics[width=0.44\textwidth]{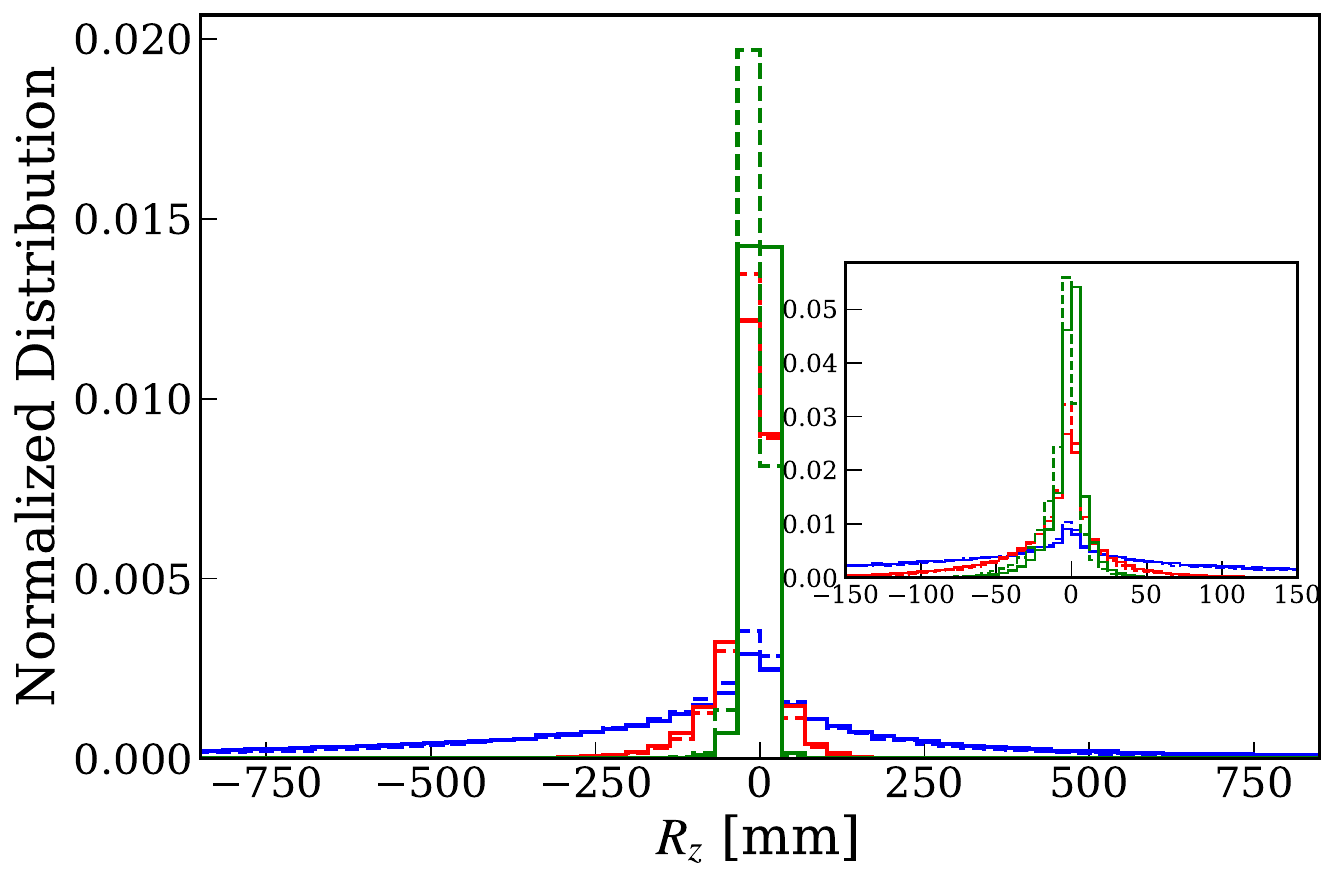}
\hfill
\includegraphics[width=0.45\textwidth]{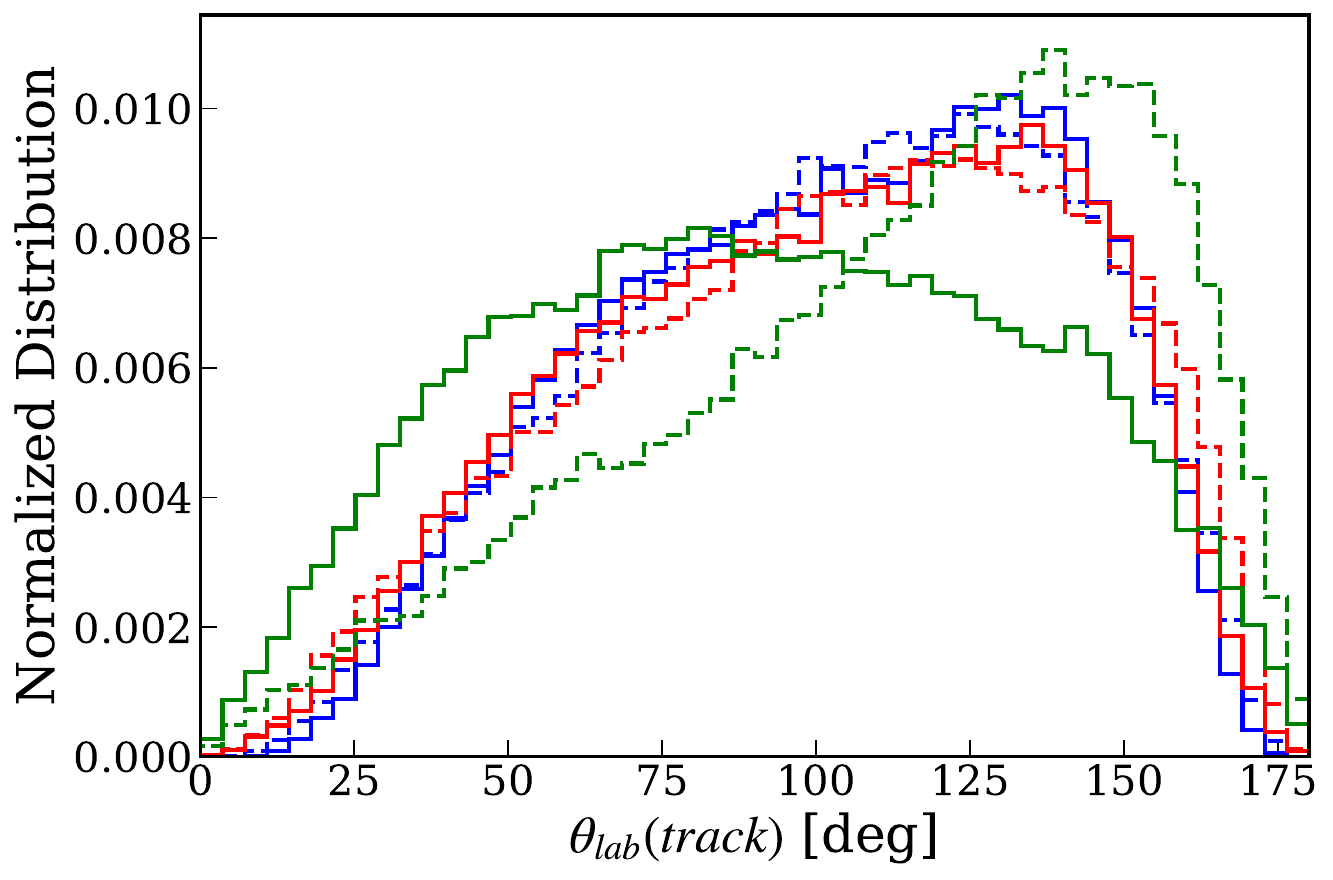}
\caption{Distributions for leading (solid line) and subleading (dashed line) $p_T$ dark pions ($\pi_d$) in BP1, BP2, and BP3 at the Belle II. {\color{black}Top left:} 
{\color{black}number} of dark pions per event, $N_{\pi_d}${\color{black}. Top right: transverse} decay length, $R_{xy}$, of leading and subleading $\pi_d$'s{\color{black}. Bottom left: longitudinal} flight distance, $R_z${\color{black}. Bottom right: angle} between the decay vertex vector and beam axis, $\theta_{\text{lab}}$.}
\label{fig:dark_pion_kinematics}
\end{figure}

In this study, the trigger strategy at the Belle II relies on three categories as follows: 
\begin{itemize}
    \item {\color{black}Two tracks trigger: an} event passes this trigger if it contains two muon tracks, each with \( p_T > 0.3\text{ GeV} \), and the azimuthal opening angle between them at the IP in the lab frame satisfies \( \Delta\phi > 90^\circ \). In addition, the angle between the position vector of decay vertex and the beam axis in the lab frame must satisfy \( 38^\circ < \theta_{\text{lab}} < 127^\circ \).

    \item {\color{black}One muon trigger: an} event passes this trigger if it contains at least one muon with \( p_T > 0.9\text{ GeV} \), and \( 38^\circ < \theta_{\text{lab}} < 127^\circ \).

    \item {\color{black}Displaced vertex trigger: this} trigger is satisfied when the event contains at least one LLP (dark pion) with a transverse decay length \( 9~\text{mm} < R_{xy} < 600~\text{mm} \), which decays into two muons, each with \( p_T > 0.1\text{ GeV} \), and \( 38^\circ < \theta_{\text{lab}} < 127^\circ \).
\end{itemize}

\begin{figure}[t!]
\centering
\includegraphics[width=0.45\textwidth]{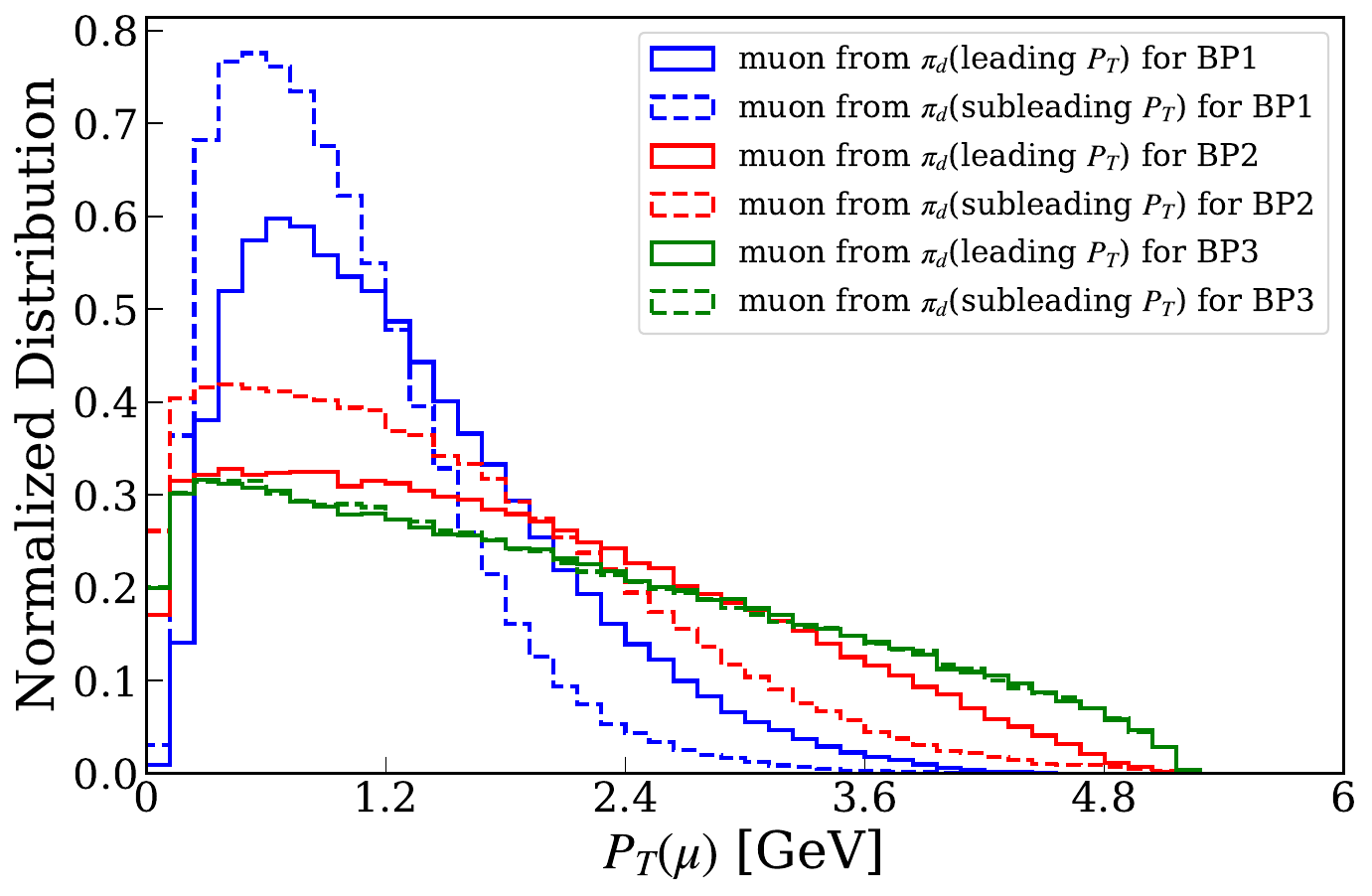}
\hfill
\includegraphics[width=0.45\textwidth]{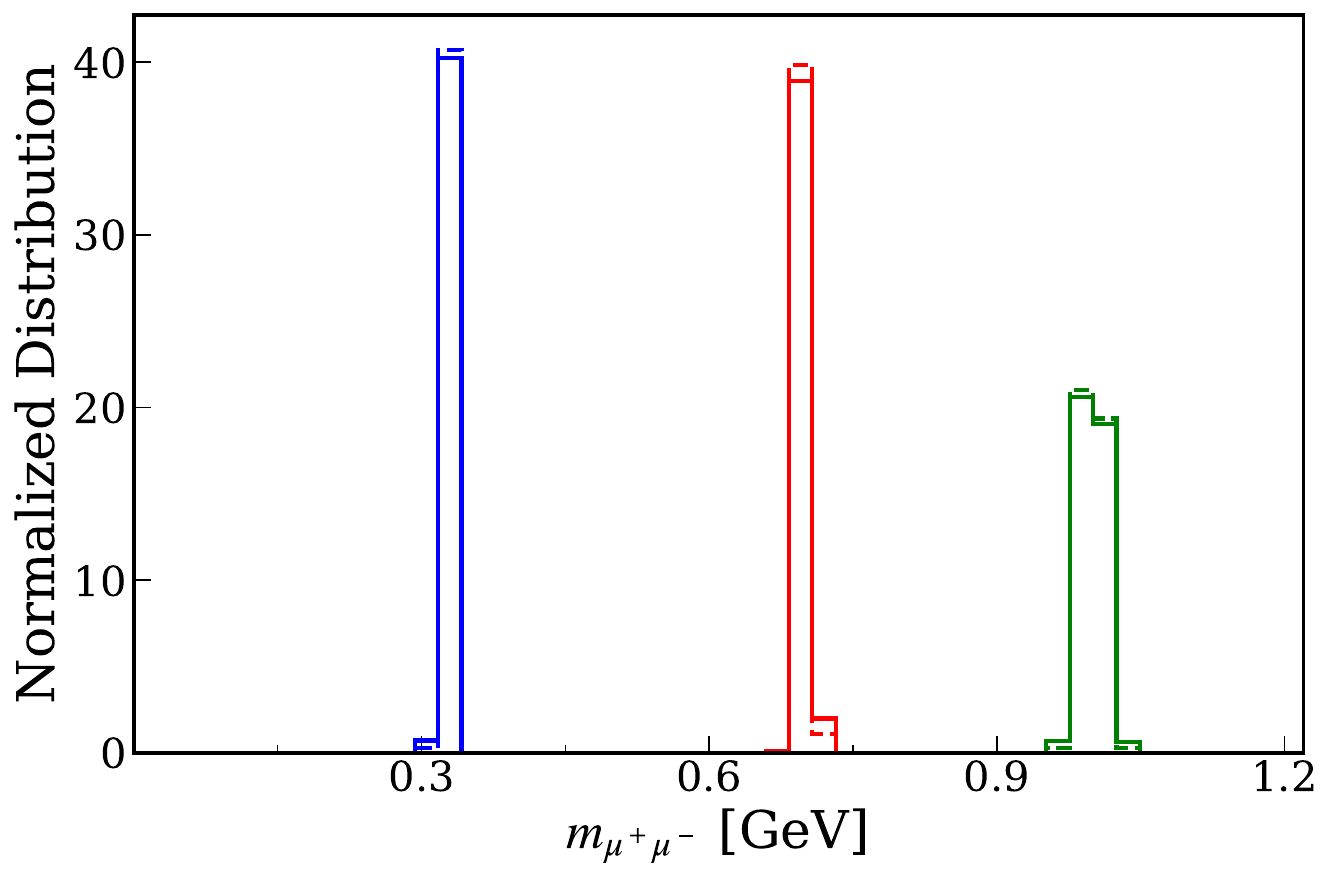}

\vspace{0.5cm}

\includegraphics[width=0.45\textwidth]{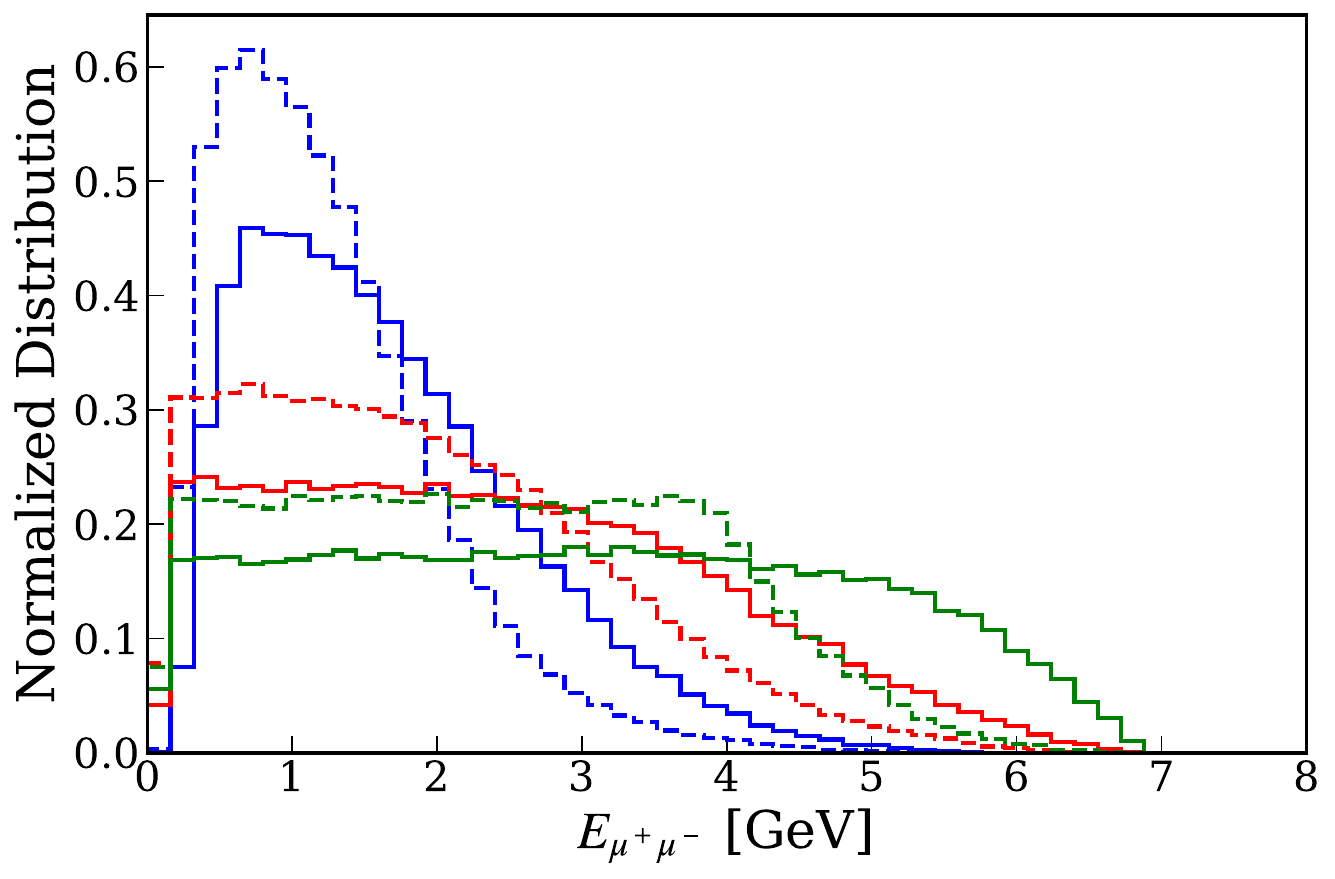}
\hfill
\includegraphics[width=0.45\textwidth]{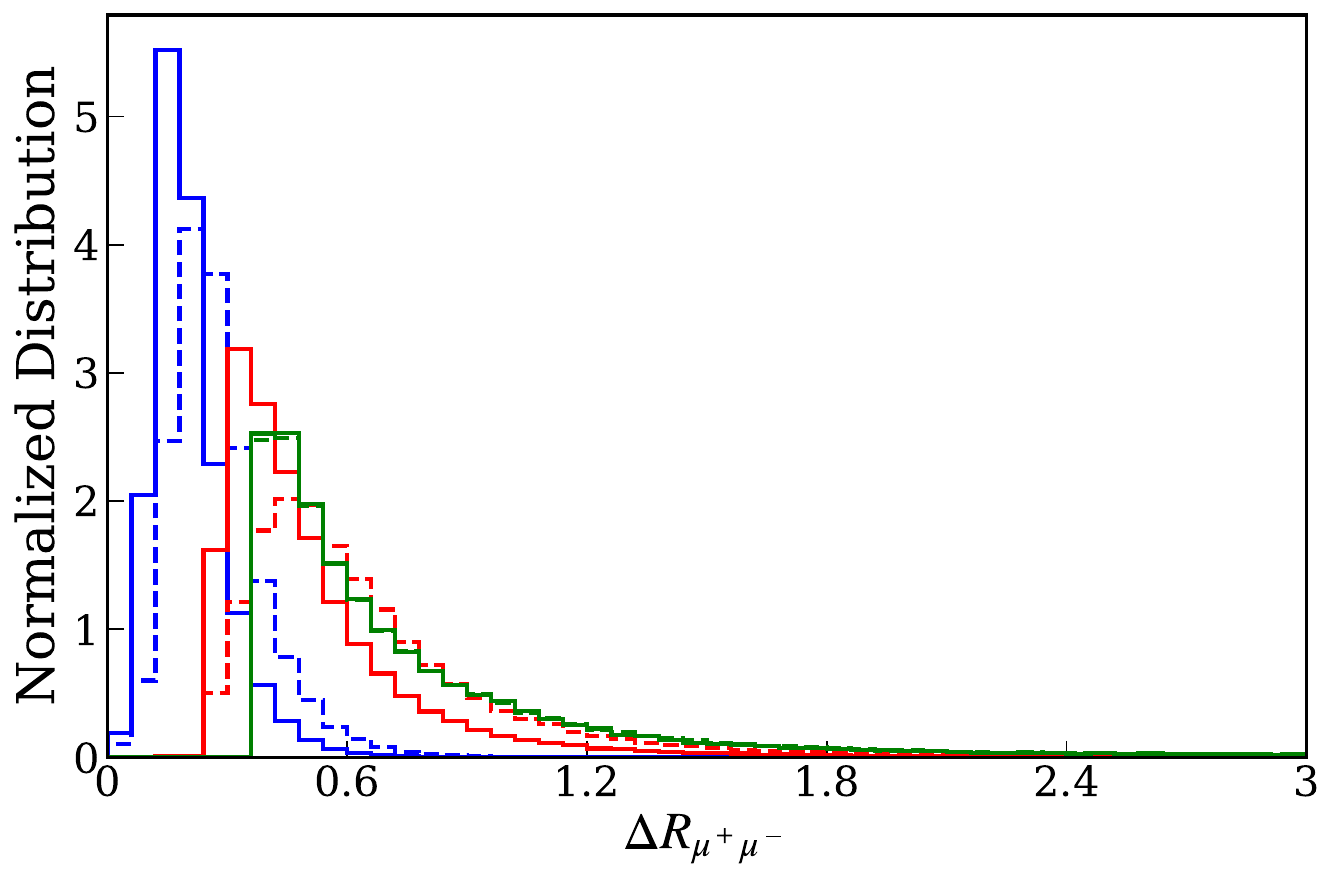}

\caption{The kinematic distributions for muons from leading $p_T$ (solid line) and subleading $p_T$ (dashed line) dark pions in BP1, BP2, and BP3 at the Belle II. {\color{black}Top left: transverse} momentum of muons, $p_T(\mu)${\color{black}. Top right: invariant} mass of muon pairs, $m_{\mu^+\mu^-}${\color{black}. Bottom left: energy} of muon pairs, $E_{\mu^+\mu^-}${\color{black}. Bottom-right: angular} separation between opposite-sign muons, $\Delta R_{\mu^+\mu^-}$.}
\label{fig:muon_kinematics}
\end{figure}

We further apply the event selection criteria summarized in Table~\ref{tab:llp_selection_summary} to 
identify the signal events and suppress the background ones. An event is selected if at least two {\color{black}leading} $p_T$ $\pi_d$ particles satisfy the following conditions:
$0.2~\text{cm} < R_{xy} < 60~\text{cm}$,
$-55~\text{cm} \leq z \leq 140~\text{cm}$, and $17^\circ \leq \theta_{\text{lab}} \leq 150^\circ$. Additionally, their daughter muons must satisfy $p_T > 0.05~\text{GeV}$, and the reconstructed invariant mass of the muon pair must fulfill $m_{\mu^+\mu^-} < 0.48~\text{GeV}$ or $m_{\mu^+\mu^-} > 0.52~\text{GeV}$.  
In this analysis, we require at least two muon pairs with significantly displaced vertices, which effectively suppresses backgrounds from punch-through pions. For decay vertices with $R_{xy} > 0.2~\text{cm}$, the significant displacement from the IP eliminates photon conversion backgrounds. Furthermore, backgrounds originating from detector structures can be effectively rejected for {\color{black}nonelectron} final states. All signal events are required to satisfy at least one of the triggers listed in Table~\ref{tab:trigger_summary} and the selection criteria outlined in Table~\ref{tab:llp_selection_summary}. Under these comprehensive selection criteria, the signal region can be treated as effectively background-free~\cite{Duerr:2020muu,Bernreuther:2022jlj}. 

Table~\ref{tab:trigger_efficiencies} summarizes the efficiencies of three trigger types and after applying event selection criteria for three BPs. Overall, the efficiencies of all three triggers exhibit significant variations with $m_{\pi_d}$, and the relative ordering of efficiencies among the BPs changes considerably after the same selection is applied.
Specifically, the efficiency of {\color{black}the two tracks trigger} decreases as $m_{\pi_d}$ increases, dropping from $75.82\%$ for BP1 to $34.16\%$ for BP3. This behavior can be understood from the distribution of $\theta_{\text{lab}}$. For $m_{\pi_d} = 1~\text{GeV}$, the tracks from the leading and subleading $p_T$ $\pi_d$'s are more dispersed in $\theta_{\text{lab}}$ distribution, with many falling outside the trigger acceptance region of $38^\circ \leq \theta_{\text{lab}} \leq 127^\circ$. In this case, the two selected tracks are very likely to originate from the same dark pion, and their opening angle $\Delta\phi$ rarely exceeds $\pi/2$, leading to the lowest efficiency. In contrast, for $m_{\pi_d} = 0.32$ and $0.7~\text{GeV}$, the $\theta_{\text{lab}}$ distribution is more concentrated, and the higher $\pi_d$ multiplicity results in higher efficiency.
{\color{black}The one muon trigger} maintains high efficiency across all three BPs, peaking at $89.22\%$ for BP2. 
The efficiency of {\color{black}the displaced vertex trigger} decreases with increasing $m_{\pi_d}$. This condition strongly depends on the decay length of $\pi_d$. The proper decay lengths for the three BPs are $c\tau_0 = 116.40~\text{mm}$ (BP1), $8.80~\text{mm}$ (BP2), and $2.94~\text{mm}$ (BP3), respectively. The proper decay length of BP1 is an order of magnitude larger than that of BP2, after multiplying by $\beta\gamma$ in the lab frame, it often exceeds the upper limit of the trigger condition. However, since the proper decay length follows an exponential distribution in the Monte Carlo simulation, the high multiplicity of BP1 enhances the probability that at least one displaced vertex satisfies the trigger condition. In contrast, for $m_{\pi_d} = 1~\text{GeV}$, the $R_{xy}$ distribution of the two leading $p_T$ $\pi_d$'s is concentrated in a very small region, resulting in the fewest events satisfying the trigger condition. 
After applying the cuts in Table~\ref{tab:llp_selection_summary}, the order of efficiencies in these three cases may change. 
From the $R_{xy}$ distribution of $\pi_d$'s, for $m_{\pi_d} = 0.32~\text{GeV}$, many events have decay lengths for both $\pi_d$'s exceeding 60 cm, leading to a significant reduction in efficiency. In comparison, for $m_{\pi_d} = 0.7~\text{GeV}$, most events fall within the required range, resulting in the highest efficiency.

We then analyze relevant signal kinematic distributions for $\pi_d$'s and muons at the Belle II in Figs.~\ref{fig:dark_pion_kinematics} and~\ref{fig:muon_kinematics}. 
First, as the $m_{\pi_d}$ increases from BP1 to BP3, the $N_{\pi_d}$ distribution ({\color{black}top left}) decreases from approximately 6 to 2. 
Regarding the decay length, $R_{xy}$ and $R_z$ represent the transverse and longitudinal decay lengths of the dark pions in the lab frame, respectively. According to Eq.~\eqref{eq:proper_decay_length}, under fixed parameters and with $f_{\pi_d} = m_{\pi_d}$, the proper decay length satisfies $c\tau_0 \propto 1/m_{\pi_d}^3$. Therefore, lighter $\pi_d$'s have longer decay lengths, and their $R_{xy}$ ({\color{black}top right}) and $R_z$ ({\color{black}bottom left}) distributions exhibit larger values and longer tails. 
Finally, the $\theta_{\rm lab}(\text{track})$ distribution ({\color{black}bottom right}) is significantly dispersed for the BP1, while the other two BPs exhibit relatively concentrated distributions. By comparing $E$ and $p_T$ of daughter particles, it can be seen that for BP1 and BP2, the leading and subleading $p_T$ $\pi_d$'s have similar $p_z$ values, leading to small differences in their $\Delta\theta_{\rm lab}$ distributions. In contrast, for the BP3, although its $p_T$ distribution is similar, the leading $\pi_d$ carries higher energy, resulting in a larger $p_z$ difference. The greater difference in boost further leads to a significant separation in their $\theta_{\rm lab}$ distributions.

As shown in Fig.~\ref{fig:muon_kinematics}, the $p_T(\mu)$ distribution ({\color{black}top left}) is similar to that in the BaBar analysis. 
In the $m_{\mu^+\mu^-}$ distribution ({\color{black}top right}), the reconstructed mass peaks clearly appear near the respective $m_{\pi_d}$, $0.32$ GeV, $0.7$ GeV, and $1$ GeV for BP1, BP2, and BP3, respectively, but exhibit different levels of broadening. The width of these distributions is influenced by momentum resolutions of muons~\cite{Adachi:2018qme,Kang:2021oes}, 
\begin{equation}
\frac{\sigma_p}{p_\mu} = 0.0011p_\mu [\text{GeV}] \oplus \frac{0.0025}{\beta}, 
\end{equation} 
the width of the $m_{\mu^+\mu^-}$ distribution increases from BP1 to BP3 as the momentum resolution grows with the $p_T(\mu)$. 
The $E_{\mu^+\mu^-}$ distribution ({\color{black}bottom left}) shows a clear $m_{\pi_d}$ and $p_T(\mu)$ {\color{black}dependence-muon} pairs from the decay of heavier and {\color{black}higher }$p_T$ $\pi_d$'s carry more energy. Finally, in the $\Delta R_{\mu^+\mu^-}$ distribution ({\color{black}bottom right}), the average angular separation between the muons is smaller for higher $\pi_d$ multiplicity, while in BP3, which has fewer $\pi_d$'s, the $\Delta R$ distribution is generally larger. 


Since the coupling $\kappa$ and mediator mass $m_X$ affect both the $e^+e^- \to \overline{q_d}q_d$ production {\color{black}cross section} and $\pi_d$ lifetime, we map exclusion limits in the parameter space of $(m_X, \kappa)$ plane. In particular, $\sigma(e^+e^- \to \overline{q_d}q_d)$ and $\Gamma_{\pi_d}$ both scale as $(\kappa/m_X)^4$. Applying the trigger conditions and event selection criteria (Tables~\ref{tab:trigger_summary} and~\ref{tab:llp_selection_summary}), we obtain the signal efficiency $\epsilon_S$. 
Here an event is considered to pass the trigger selection if at least one of the three trigger conditions is satisfied. 
The expected signal yield is $N_S = \mathcal{L}\times\sigma(e^+e^- \to \overline{q_d}q_d)\times\epsilon_S$, assuming optimistic integrated luminosity $\mathcal{L} = 50~\text{ab}^{-1}$ for the Belle II experiment. With a background-free assumption, $N_S = 2.3$ events establish the exclusion contours of the $90\%$ confidence level (C.L.) which correspond to the lower boundaries with purple solid lines in Fig.~\ref{fig:belle-gazelle-babar}.

As expected, small $\kappa/m_X$ yields a tiny {\color{black}cross section} and excessive $\pi_d$ decay length, reducing $N_S$ below detection ability of the Belle II experiment. Conversely, large $\kappa/m_X$ produces prompt $\pi_d$ decays without the signature of displaced vertices. In this situation, the efficiency drops sharply, necessitating enormous Monte Carlo samples. Therefore, we conservatively set $N_S = 10$ to estimate upper boundaries with purple dashed lines in Fig.~\ref{fig:belle-gazelle-babar}\footnote{Even with this relaxed requirement, more than $10^8$ simulated events are required for this estimation.}. We can observe that the projected Belle II constraints also indicate inverse proportionality between $m_X$ and $\kappa$ for these three BPs. Notably, blank regions in the upper-left corner of $(m_X, \kappa)$ plane could be covered by searching for {\color{black}multimuons} from the prompt decay of dark pions. However, detailed signal-background analysis is beyond the scope of this work. 

Unlike most light DM searches at the Belle II, the {\color{black}monophoton} search strategy lacks sufficient discriminating power between signal and background events in this model. Although undetectable dark pions produce missing energy signatures, their kinematic distributions remain similar to SM backgrounds. 
Because the mediator mass exceeds the center-of-mass energy at the Belle II, this absence of light mediator production results in smooth, continuous photon energy distributions that resemble SM backgrounds~\cite{Bernreuther:2022jlj}. We{\color{black}, therefore, exclude monophoton} constraints from Belle II in this analysis.

\subsection{Long-lived dark pion searches in proposed GAZELLE detector} 

Positioned downstream of the Belle II detector at a considerable distance from the IP, the proposed GAZELLE detector facilitates long-distance tracking and displaced vertex reconstruction which enhances the overall sensitivity for probing LLPs. The GAZELLE detector has three proposed configurations: Baby-GAZELLE, L-GAZELLE, and GODZILLA~\cite{Ko:2025drr}. L-GAZELLE comprises two components, LG-B1 (Region I) and LG-B2 (Region II), which select signal events as detailed later. We chose L-GAZELLE for this analysis due to its superior angular coverage (LG-B1: 0.34 sr [2.7\%]; LG-B2: 0.76 sr [6.0\%]), exceeding Baby-GAZELLE (0.12 sr, 0.95\%) and GODZILLA (0.74 sr, 5.9\%) and thereby improving the detection efficiency. Its moderate detection distances (LG-B1: $x=35$ m, $y=2.3$ m, $z=0$ m; LG-B2: $x=19$ m, $y=2.3$ m, $z=10.5$ m) optimally capture long-lived dark pion decays. This contrasts with Baby-GAZELLE's proximity ($x=10$ m, $y=-3.7$ m, $z=10$ m), which limits the far decay sensitivity, and the GODZILLA distance ($x=-27$ m, $y=18$ m, $z=20$ m), which reduces the near decay sensitivity. Thus, L-GAZELLE offers the optimal balance for our study.

We apply {\color{black}a DDC}~\cite{Domingo:2023dew} for the L-GAZELLE detector simulation.\footnote{The L-GAZELLE detector is approximated with LG-B1 (single-layer approximation within $8.5\%$ error) and LG-B2 (four-layer approximation within $5.0\%$ error).} The decay probability of LLPs within a certain range is calculated as: $P(\ell_1, \ell_2) = \exp\left(-\frac{\ell_1}{\gamma|\vec{v}|\tau_0}\right) - \exp\left(-\frac{\ell_2}{\gamma|\vec{v}|\tau_0}\right)$,
 where $\gamma \equiv (1 - |\vec{v}|^2)^{-1/2}$, $|\vec{v}|$ and $\tau_0$ are the velocity and proper decay length of a LLP with the natural units. The weight for each layer is $\Delta\phi/(2\pi)$, where $\Delta\phi$ is the angular aperture of each layer. The weighted sum of probabilities from all detector layers gives the acceptance $A_i$ for each LLP$_i$, which represents the detection probability of each LLP. 
For event selections, we calculate the probabilities of detecting at least one LLP and at least two LLPs:
\begin{itemize}
\item At least one LLP detected: $\langle P(\geq 1)\rangle = 1 - \langle \prod (1 - A_i)\rangle$. 
\item At least two LLPs detected: $\langle P(\geq 2)\rangle = 1 - \langle \prod (1 - A_i)\rangle - \langle \sum A_i \prod_{j\neq i}(1 - A_j)\rangle$.
\end{itemize}.

For the signal efficiency $\epsilon_S$ obtained for both scenarios, the expected signal yield is given by $N_S = \mathcal{L} \times \sigma(e^+ e^- \to \bar{q}_d q_d) \times \epsilon_S$, assuming an optimistic integrated luminosity $\mathcal{L} = 50~\text{ab}^{-1}$. The $90\%$ C.L. exclusion contours correspond to the lower boundaries, shown as orange-red (at least one LLP detected) and brown (at least two LLPs detected) solid lines in Fig.~\ref{fig:belle-gazelle-babar}. Compared to recast BaBar and projected Belle II constraints, the {\color{black}far detector} L-GAZELLE coverage  extends to smaller $\kappa$ and larger $m_X$ values. 

\begin{table}[htbp]
\centering
\setlength{\tabcolsep}{10pt} 
\renewcommand{\arraystretch}{1.4} 

\begin{tabular}{|c|c|c|c|}
\hline
$m_{\pi_d}$ [GeV] & $\tau_0$ [m] & $\langle P(\geq 1) \rangle$ & $\langle P(\geq 2) \rangle$ \\
\hline
\multirow{3}{*}{0.32}
  & 54.91   & $9.02 \times 10^{-3}$  & $3.71 \times 10^{-5}$ \\
  & 10.98   & $2.16 \times 10^{-2}$  & $1.96 \times 10^{-4}$ \\
  & 5.49    & $2.27 \times 10^{-2}$  & $1.99 \times 10^{-4}$ \\
\hline
\multirow{3}{*}{0.70}
  & 5.25  & $1.22 \times 10^{-2}$  & $2.91 \times 10^{-5}$ \\
  & 1.05  & $2.68 \times 10^{-3}$  & $8.37 \times 10^{-8}$ \\
  & 0.53 & $3.17 \times 10^{-4}$  & $3.33 \times 10^{-11}$ \\
\hline
\multirow{3}{*}{1.00}
  & 1.80    & $4.90 \times 10^{-3}$  & $7.36 \times 10^{-8}$ \\
  & 0.36   & $7.32 \times 10^{-6}$  & $9.96 \times 10^{-19}$ \\
  & 0.18   & $2.02 \times 10^{-9}$  & 0                      \\
\hline
\end{tabular}
\caption{Detection efficiency $\langle P(\geq 1) \rangle$ and $\langle P(\geq 2) \rangle$ for various dark pion proper decay length $\tau_0$ and dark pion mass $m_{\pi_d}$ values. Corresponding production {\color{black}cross sections} are $5.40 \times 10^{-2}$ fb for $m_{\pi_d} = 0.32$ GeV at $\tau_0 = 54.91$ m, $6.59 \times 10^{-2}$ fb for $m_{\pi_d} = 0.7$ GeV at $\tau_0 = 5.25$ m, and $6.32 \times 10^{-2}$ fb for $m_{\pi_d} = 1$ GeV at $\tau_0 = 1.80$ m.}
\label{tab:GAZELLE_efficiencies}
\end{table}

We further evaluated detection efficiencies for different $m_{\pi_d}$ and $\tau_0$ values, as shown in Table~\ref{tab:GAZELLE_efficiencies}. The detection efficiency for at least two LLPs {\color{black}[$\langle P(\geq 2) \rangle$]} is significantly lower than for at least one LLP {\color{black}[{$\langle P(\geq 1) \rangle$]}}. For BP1, $\langle P(\geq 2) \rangle$ is about {\color{black}2} orders of magnitude smaller than $\langle P(\geq 1) \rangle$. As the $m_{\pi_d}$ increases, the efficiency gap widens, particularly at $m_{\pi_d} = 1$ GeV, where events typically produce two LLPs that travel nearly back-to-back, making it highly unlikely for both to be detected within L-GAZELLE. Since $\tau_0$ and the production {\color{black}cross section are anticorrelated}, a decrease in $\tau_0$ enhances the {\color{black}cross section} but causes a more rapid decline in detection efficiency. This efficiency loss prevents meaningful constraints for BP2 and BP3 in the scenario requiring at least two detected LLPs. Consequently, the brown solid line corresponding to BP2 and BP3 is absent in Fig.~\ref{fig:belle-gazelle-babar}. 




\section{Dense dark showers at the CEPC/FCC-ee} 
\label{sec:CEPC}

This section further investigates the detection prospects at high-energy $e^+e^-$ colliders, specifically the proposed {\color{black}CEPC and the FCC-ee}. Both are considered to operate at a center-of-mass energy of $\sqrt{s} = 240~\text{GeV}$, with integrated luminosities of $5.6~\text{ab}^{-1}$ and $5.0~\text{ab}^{-1}$, respectively. Given the expected comparable sensitivity, the analysis in the main text focuses on CEPC, while a direct comparison of the exclusion regions for FCC-ee and CEPC is provided in Appendix~\ref{appendix:A}. We also present exclusion limits for two distinct dark pion mass ranges: $10~\text{MeV} \leq m_{\pi_d} < 2m_\mu$, and $2m_\mu \leq m_{\pi_d} < 2m_\tau$, which correspond to the {\color{black}monophoton} and displaced muon jet (DMJ) signatures, respectively. 

\subsection{The {\color{black}monophoton} signature} 

\begin{figure}[h]
\centering 
\begin{minipage}{0.48\textwidth}
\includegraphics[width=\textwidth]{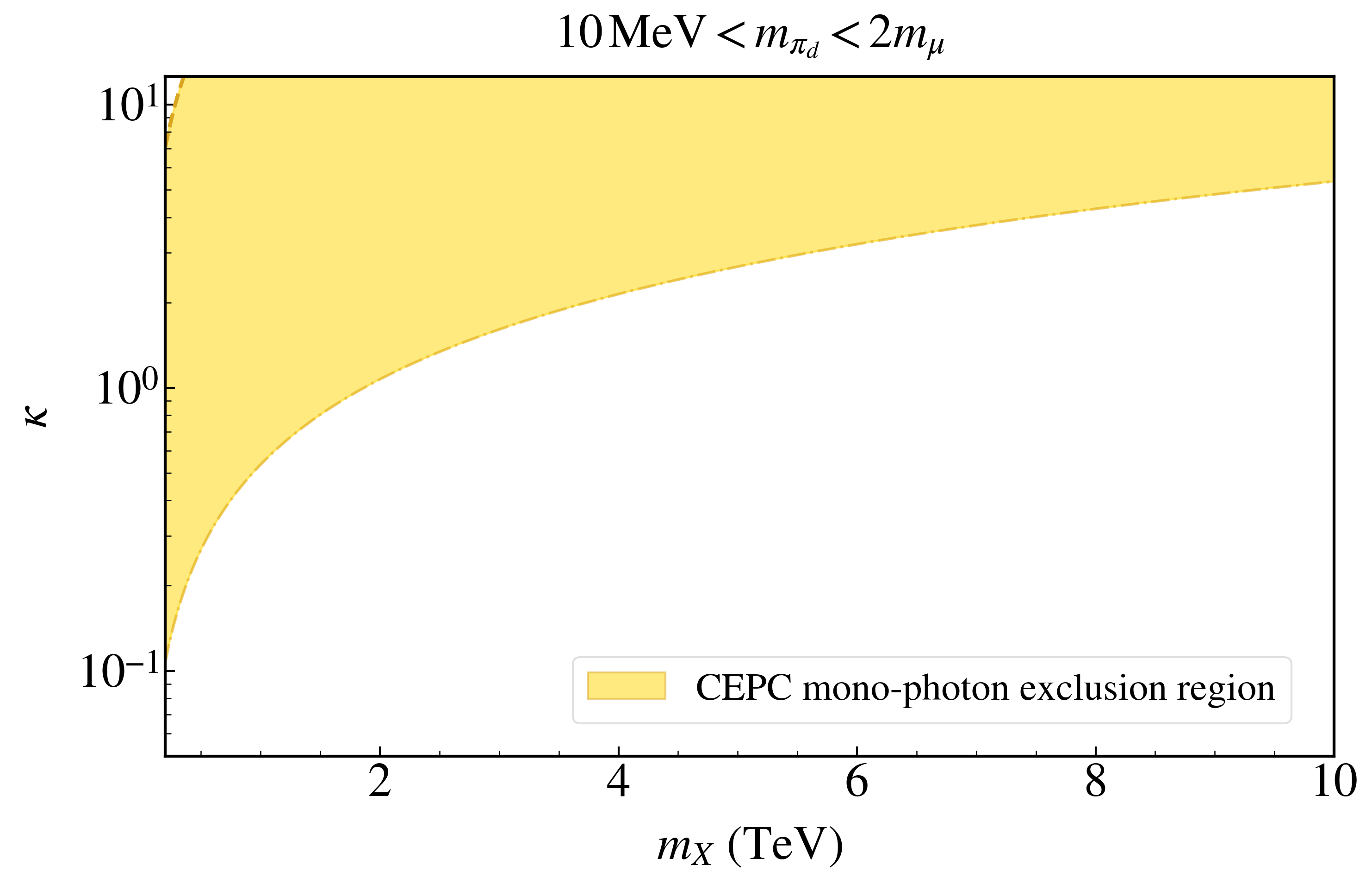}
\end{minipage}
\begin{minipage}{0.48\textwidth}
\includegraphics[width=\textwidth]{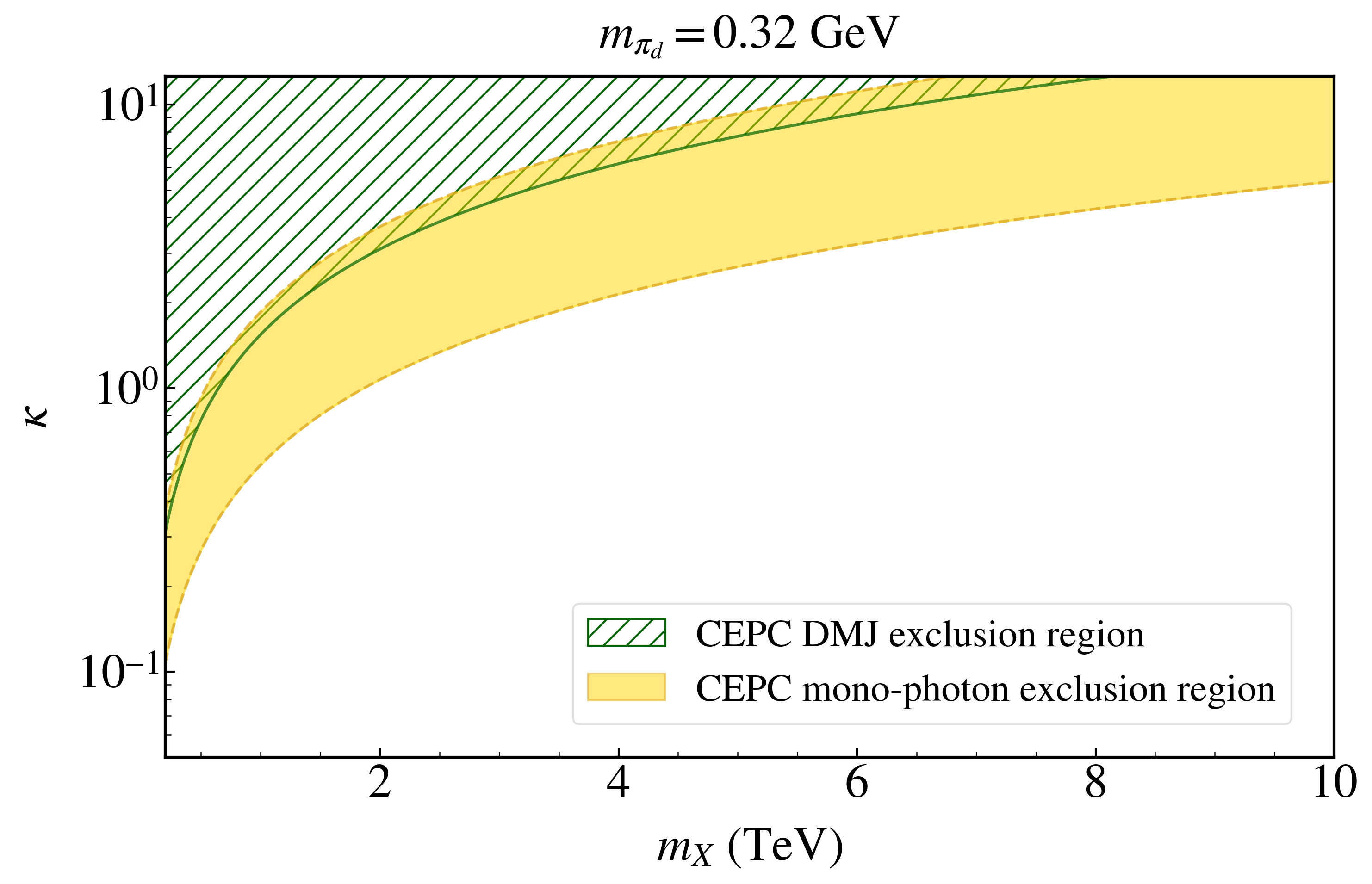}
\end{minipage}
\\
\begin{minipage}{0.48\textwidth}
\includegraphics[width=\textwidth]{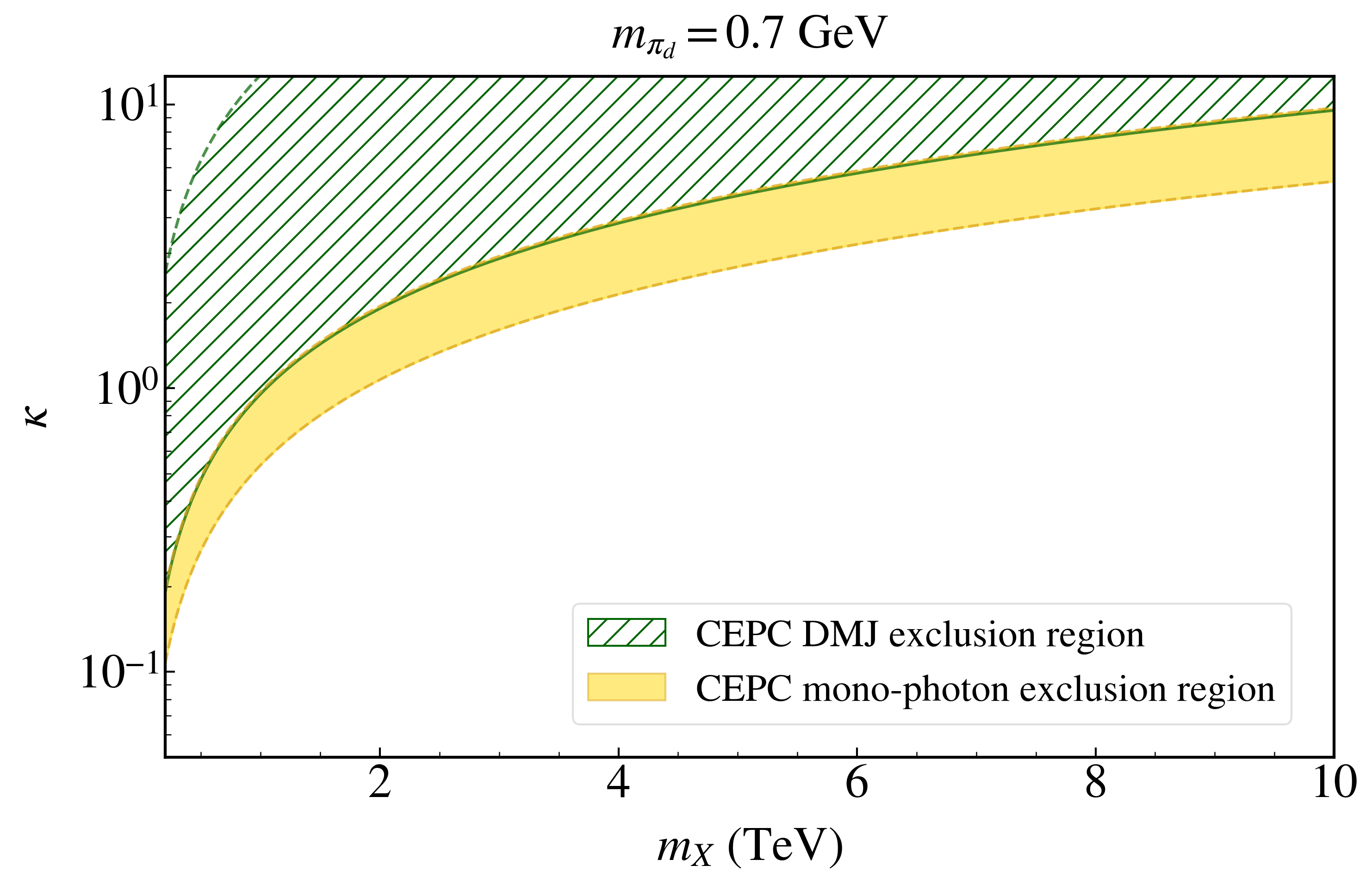}
\end{minipage}
\begin{minipage}{0.48\textwidth}
\centering
\includegraphics[width=\textwidth]{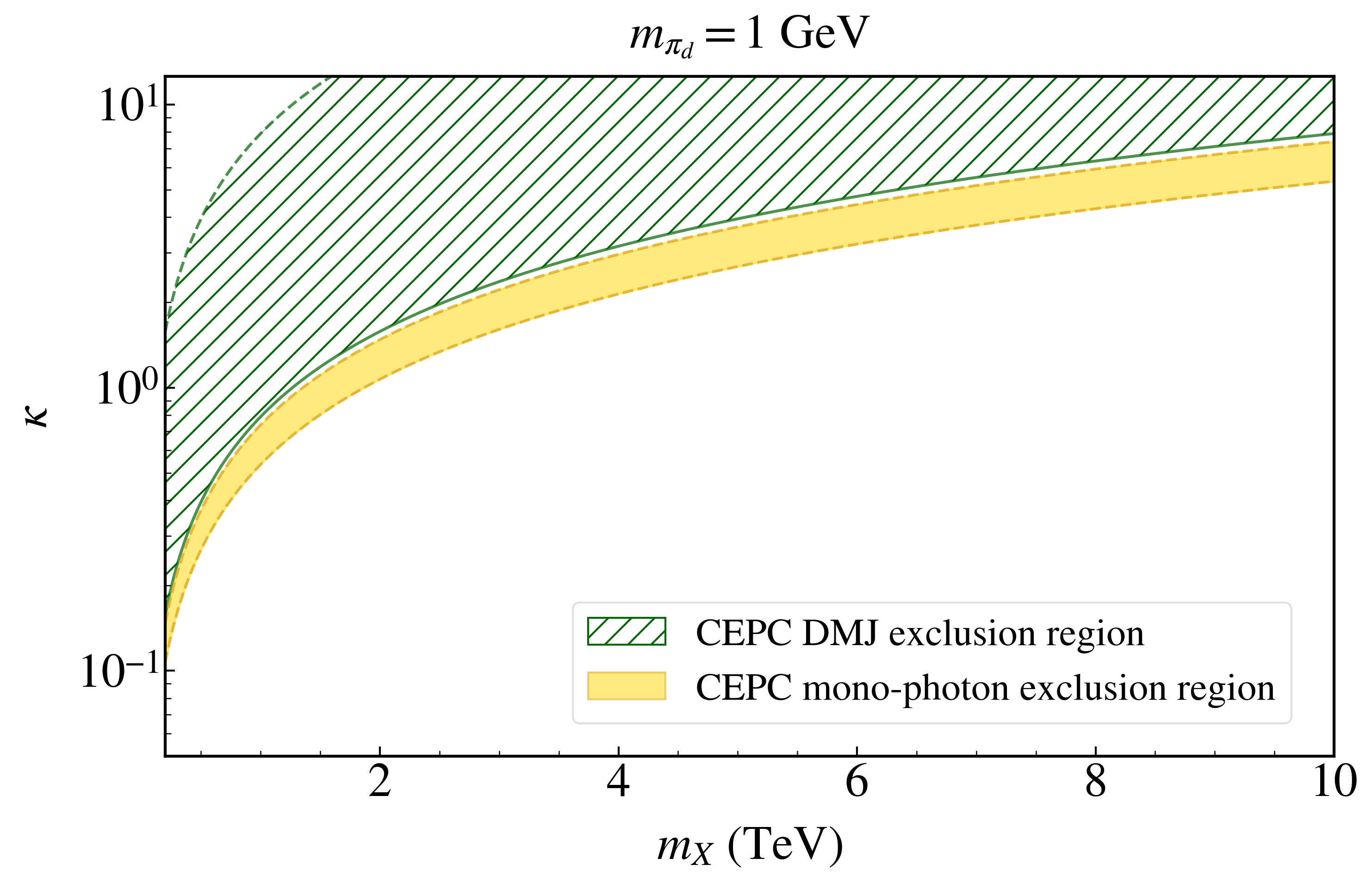}
\end{minipage}
\caption{Exclusion regions for different dark pion mass ranges. The first panel shows the {\color{black}monophoton} signature for $10~\text {MeV} < m_{\pi_d} < 2m_\mu$, while the remaining three panels show benchmark points $m_{\pi_d}=0.32$, $0.7$ and $1~\text {GeV}$ for both CEPC {\color{black}monophoton} and DMJ signatures, covering $2m_\mu < m_{\pi_d} < 2m_\tau$.} 
\label{fig:CEPC-mono-DMJ}
\end{figure}

For the dark pion mass range $10~\text{MeV} \leq m_{\pi_d} < 2m_\mu$, we maintain the conservative setting $m_{\pi_d} = f_{\pi_d} = 0.2~\text{GeV}$ and impose the requirement that the $\pi_d$ proper decay length exceeds $6.08~\text{m}$ to guarantee that all dark mesons decay outside the CEPC detector volume. The resulting upper bounds on $\kappa$, shown as the yellow dashed curve in Fig.~\ref{fig:CEPC-mono-DMJ}, reach values of $\mathcal{O}(1-10)$.
We recast the previous CEPC {\color{black}monophoton} analysis, derived for the $t$-channel scalar mediator effective operator with cutoff scale $\Lambda_t$ {\color{black}[corresponding to the $t$-channel case in Eq.~(2.3)]} and the upper-right panel (H-mode) of Fig.~16 in Ref.~\cite{Liu:2019ogn}, to our leptophilic composite ADM framework through the relation
\begin{equation}
\frac{1}{\Lambda^2} \simeq \frac{1}{(1078~\text{GeV})^2} = 3 \frac{\kappa^2}{m^2_{X}}. 
\end{equation} 
Assuming light dark quarks and determining the lower limits on $\Lambda$, the yellow region in Fig.~\ref{fig:CEPC-mono-DMJ} indicates parameter space where the predicted {\color{black}cross sections} exceed the experimental bounds.

\subsection{The displaced muon jet signature}

For the dark pion mass range $2m_\mu \leq m_{\pi_d} < 2m_\tau$, unlike low-energy $e^+e^-$ colliders, due to the high center-of-mass energy, the initially produced dark quarks undergo dark showering and hadronization processes, yielding a large number of dark pions with the same model parameter settings. These dark pions subsequently decay to muons, resulting in a high multiplicity of displaced muons that collectively form {\color{black}DMJs}. 

\begin{table}[h!]
\centering
\begin{tabular}{|l|c|c|c|}
\hline
\textbf{Detector Component} & $R_{\text{in}}$ & $R_{\text{out}}$ & $\sigma_{xy}$ \\
\hline
Vertex Detector & \SI{16}{mm} & \SI{60}{mm} & \SI{2.80}{\micro\meter}--\SI{6.00}{\micro\meter} \\
Silicon Tracker & \SI{0.15}{m} & \SI{1.81}{m} & \SI{7.20}{\micro\meter} \\
Hadron Calorimeter & \SI{2.30}{m} & \SI{3.34}{m} & \SI{30}{mm} \\
Muon System & \SI{4.40}{m} & \SI{6.08}{m} & \SI{20}{mm} \\
\hline
\end{tabular}
\caption{The CEPC detector geometry and its relevant transverse spatial resolution~\cite{Zhang:2021orr}.}
\label{tab:detector_geometry}
\end{table}

The detailed definition of {\color{black}DMJs} in this analysis is as follows. All candidate muons inside the DMJ are selected by requiring $p_T (\mu) > 1~\text{GeV}$ and $|\eta_{\mu}| < 3.0$, parent particles ($\pi_d$'s) with $R_{xy}$ within one of the four regions defined in Table~\ref{tab:detector_geometry}, and muon pairs with $\Delta\theta > 0.02$ radians. Muons satisfying these criteria are identified as displaced muons. 
We employ the anti-$k_t$ algorithm~\cite{Cacciari:2008gp} as implemented in FASTJET~\cite{Cacciari:2011ma} with a jet radius parameter of $R = 0.5$ to cluster muons into jets, and subsequently identify those containing at least six muons as DMJs.

\begin{figure}[htbp]
    \centering
    \includegraphics[width=0.45\textwidth]{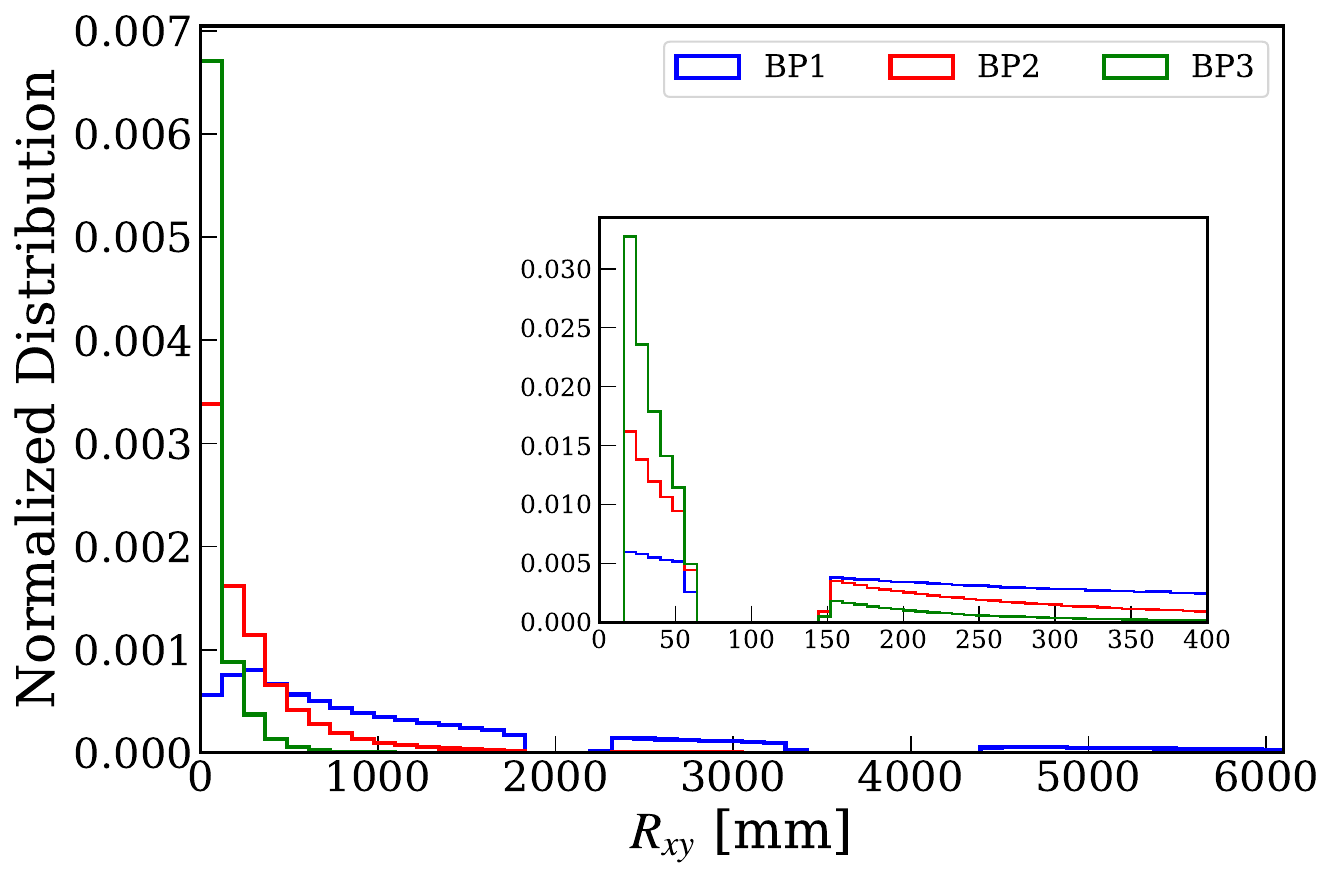}
    \hfill
    \includegraphics[width=0.45\textwidth]{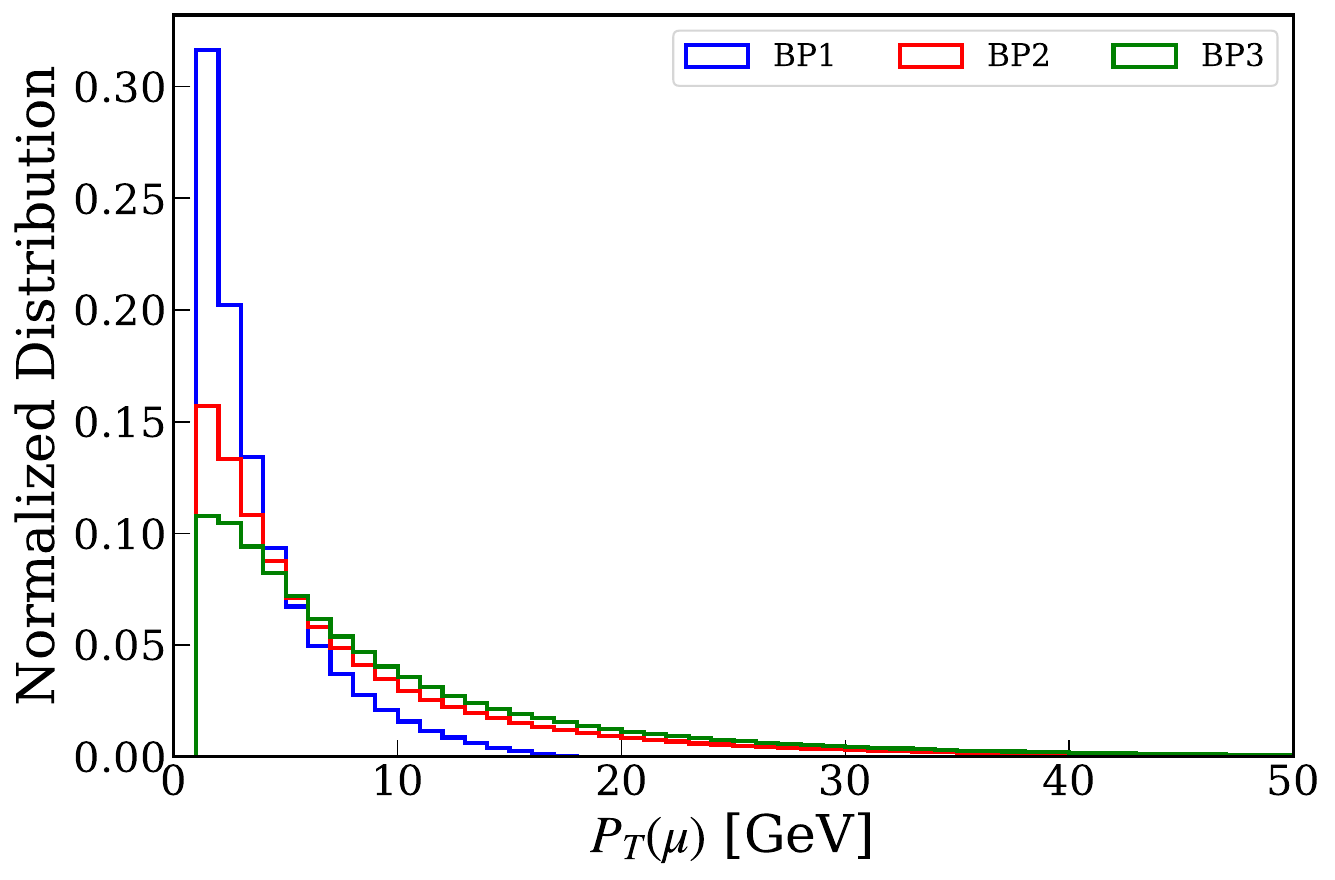}    
    \vspace{0.5cm}
    \includegraphics[width=0.45\textwidth]{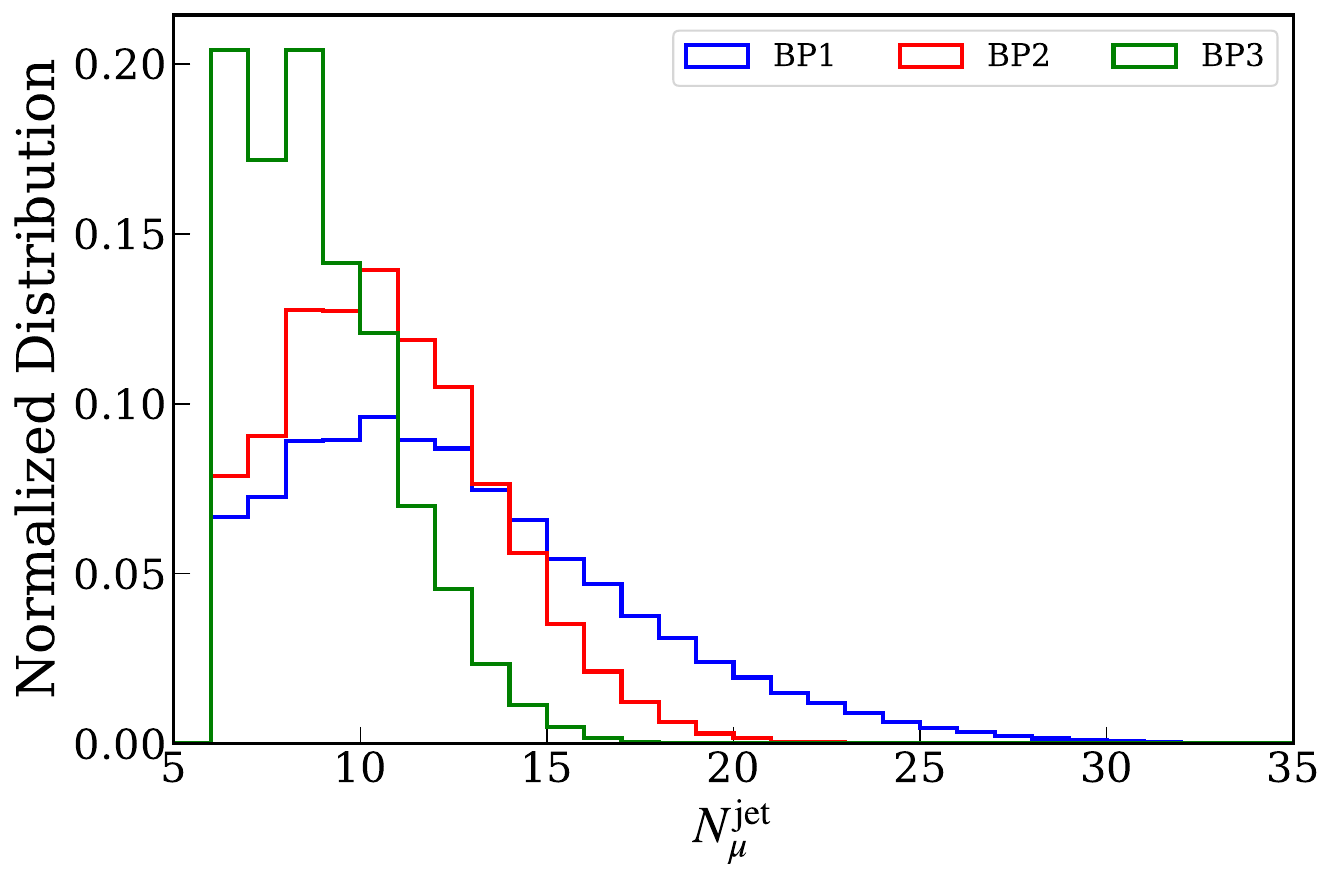}
    \hfill
    \includegraphics[width=0.45\textwidth]{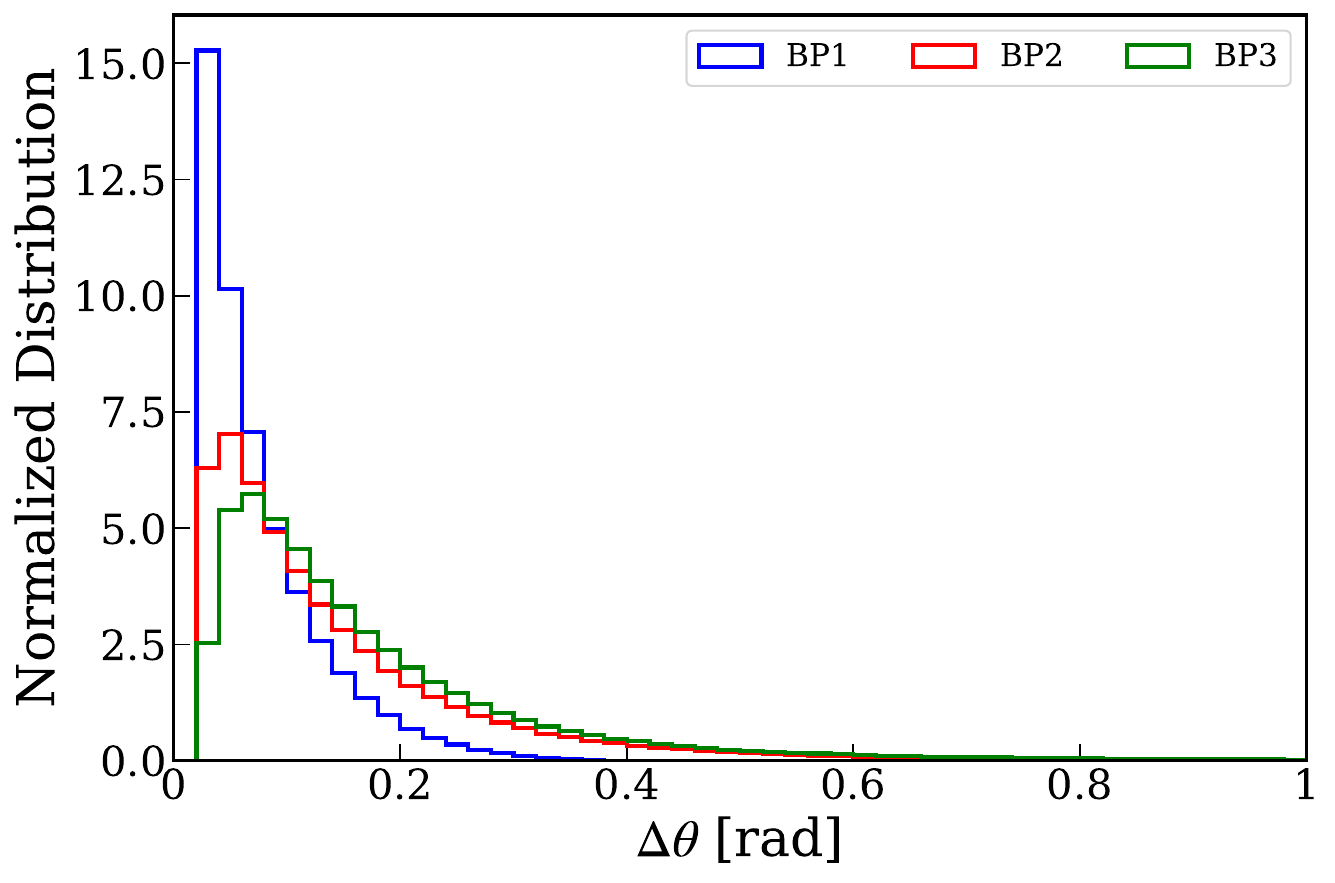}
    \caption{The main kinematic distributions for the DMJ analysis at the CEPC with three benchmark points, $m_{\pi_d} = 0.32$, $0.7$, and $1$ GeV, corresponding to $\kappa = 5$ and $m_X = 1$ TeV. {\color{black}Top left: the} dark pion transverse decay length, $R_{xy}${\color{black}. Top right: muon} transverse momentum, $p_T(\mu)${\color{black}. Bottom left: the} number of muons in each muon jet{\color{black}. Bottom right: the} opening angle $\Delta\theta$ between two muons originating from the same dark pion.}
    \label{fig:cepc_dmj_kinematics}
\end{figure}

We then select the leading $p_T$ DMJ in each event and present in Fig.~\ref{fig:cepc_dmj_kinematics} the main kinematic distributions of its constituent muons and their mother dark pions, using three BPs with $m_{\pi_d} = 0.32~\text{GeV}$, $0.7~\text{GeV}$, and $1~\text{GeV}$, corresponding to $\kappa = 5$ and $m_X = 1~\text{TeV}$. 
The $R_{xy}$ ({\color{black}top left}) and $p_T(\mu)$ ({\color{black}top right}) distributions follow trends similar to those observed at low-energy $e^+ e^-$ colliders, but at the CEPC they extend to significantly higher values due to its much larger collision energy. 
Since the CEPC detector is segmented into four {\color{black}noncontiguous radial regions,} as shown in Table~\ref{tab:detector_geometry}, the $R_{xy}$ distribution reflects this structure{\color{black}. For} three BPs considered here, most dark pions are concentrated in the innermost region, namely the vertex detector.
The calculation of $R_{xy}$ also includes Gaussian smearing effects, with the transverse position resolution for displaced vertices given by~\cite{CEPCStudyGroup:2018ghi}
\begin{equation}
\sigma_{xy} = \left(a \oplus \frac{b}{p_T \cdot \sin^{3/2}\theta}\right) \si{\micro\meter},
\end{equation}
where $a = 5$, $b = \SI{10}{GeV}$, $p_T$ is the average transverse momentum of the track, and $\theta$ is its polar angle.
The distribution of the number of muons inside the leading $p_T$ DMJ ({\color{black}bottom left}) illustrates the multiplicity of displaced muons. Across all three BPs, DMJs exhibit high-multiplicity features, with lighter $\pi_d$'s showing a multiplicity advantage. Such high-multiplicity signatures are distinctive and very difficult for SM backgrounds to mimic~\cite{Zhang:2021orr}.
The opening angle $\Delta\theta$ between two muons originating from the same dark pion ({\color{black}bottom right}) increases from BP1 to BP3. This wider separation for heavier $\pi_d$ results from the larger $p_T$ carried by the muons. However, due to the finite spatial resolution of the detector, muon pairs with excessively small $\Delta\theta$ cannot be resolved and are{\color{black}, therefore,} excluded from the $\pi_d$ reconstruction.

To further identify the signal signature, all events containing at least two DMJs are selected. The CEPC muon system comprises eight sensitive layers alternating with iron absorber plates. With a total iron thickness equivalent to 6.7 interaction lengths, this configuration offers sufficient protection against punch-through backgrounds. The detector achieves excellent muon identification performance while maintaining exceptionally low pion {\color{black}misidentification} rates~\cite{CEPCStudyGroup:2018ghi}. When combined with the signal selection strategy detailed earlier, this methodology guarantees a background-free analysis~\cite{Zhang:2021orr}. The signal efficiency $\epsilon_S$ is then evaluated from the above event selection criteria. The expected signal yield is given by $N_S = \mathcal{L} \times \sigma(e^+e^- \to q_d \bar{q}_d) \times \epsilon_S$. Under the background-free hypothesis, the $2\sigma$ exclusion limit is set at $N_S = 3$, from which we derive the exclusion contour in the $(m_X, \kappa)$ plane, as shown in Fig.~\ref{fig:CEPC-mono-DMJ}.

When determining the upper exclusion boundary, the same challenge as in the Belle II analysis is encountered, and thus conservatively set $N_S = 10$ for the CEPC upper limit estimation, while the lower boundary remains determined by the condition $N_S = 3$. However, for $m_{\pi_d} = 0.32~\text{GeV}$, it remains hopeless to establish an upper boundary under this conservative assumption. We have estimated that more than $10^9$ Monte Carlo events are required for this situation beyond the scope of this work. Therefore, no upper exclusion boundary is presented for $m_{\pi_d} = 0.32~\text{GeV}$.  

\section{Summary and discussion}
\label{sec:conclusion} 

\begin{figure}[h]
\centering 
\begin{minipage}{0.48\textwidth}
\centering
\includegraphics[width=\textwidth]{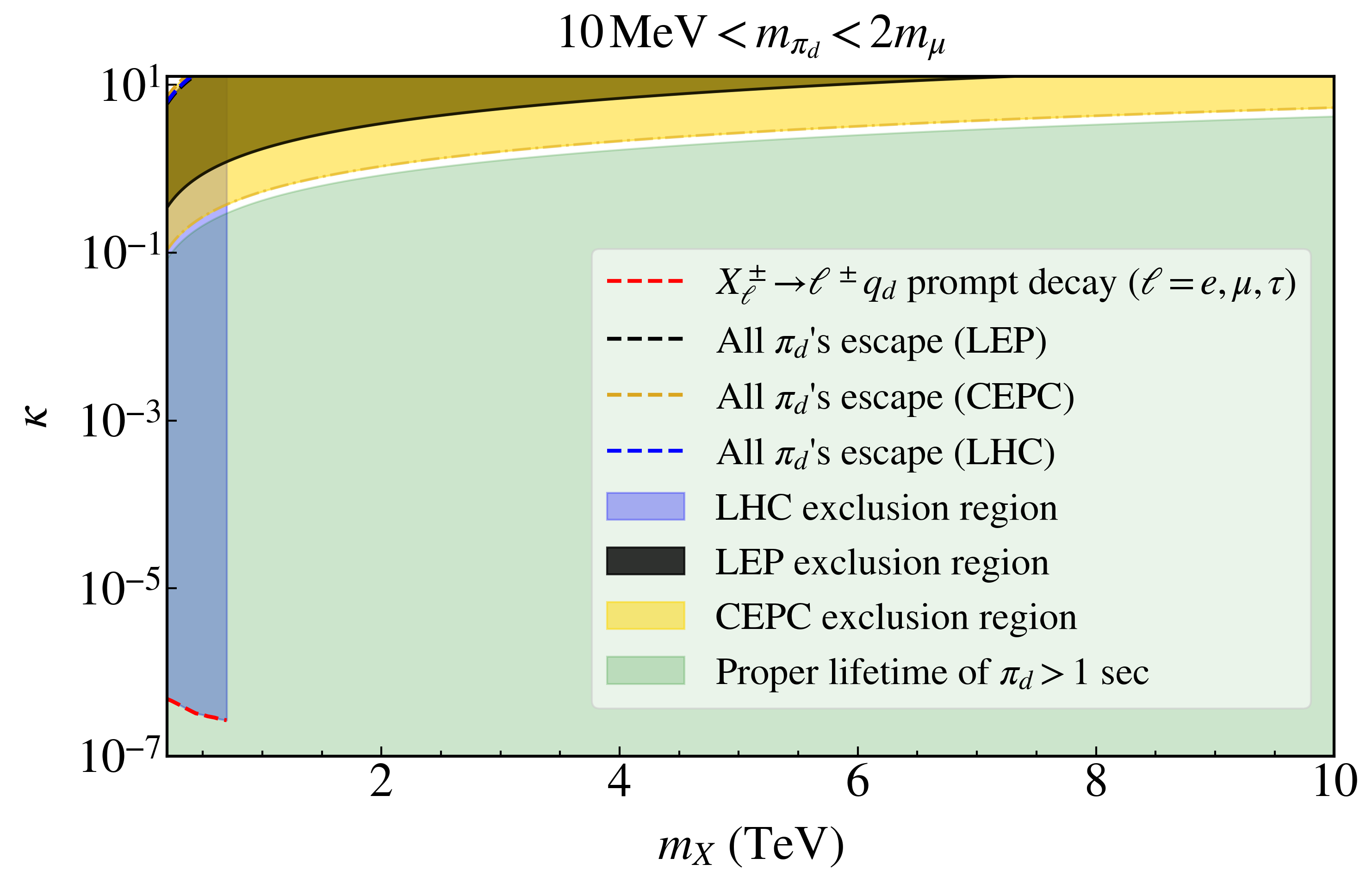}
\end{minipage}
\begin{minipage}{0.48\textwidth}
\centering
\includegraphics[width=\textwidth]{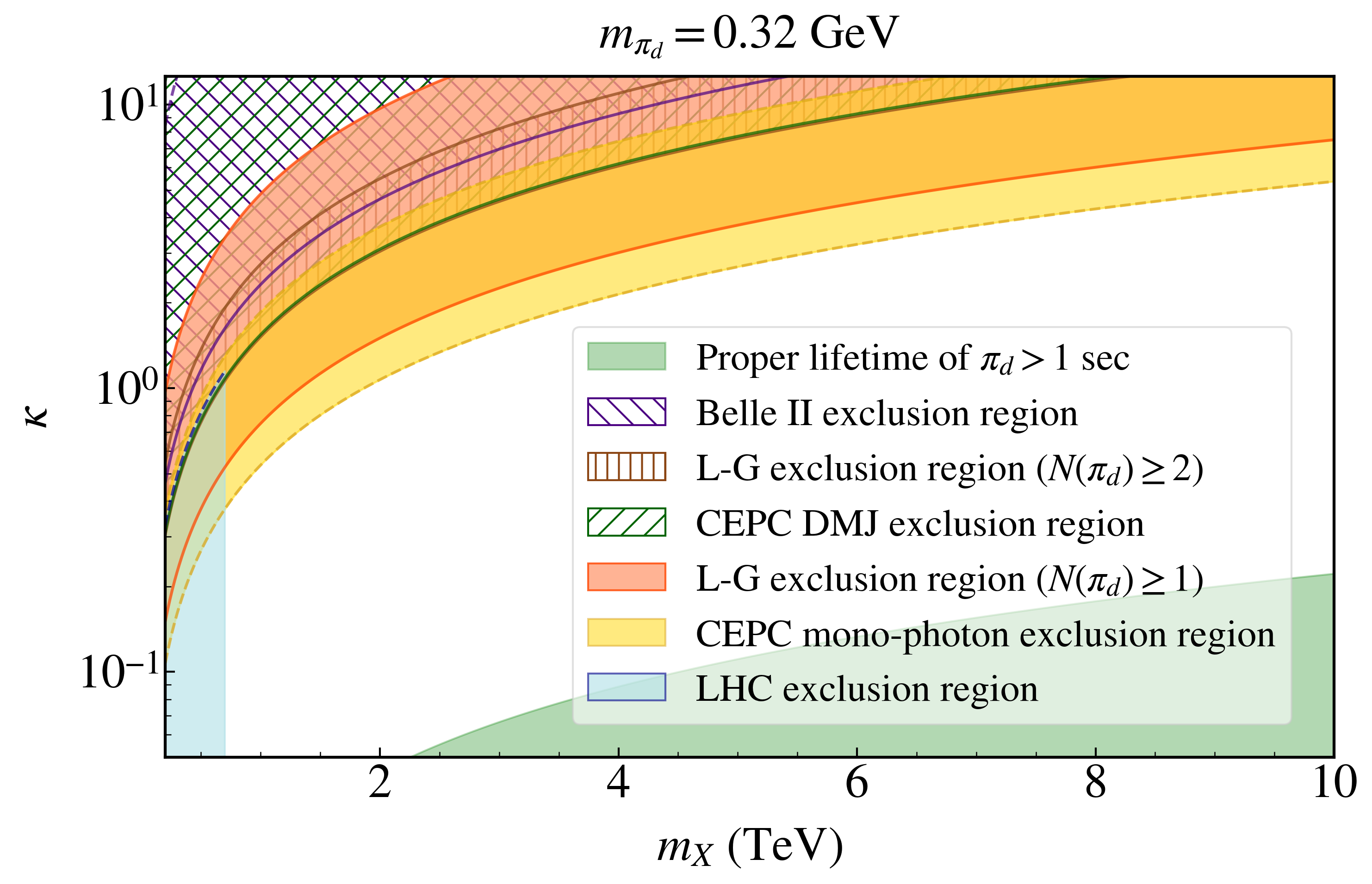}
\end{minipage}
\\
\begin{minipage}{0.48\textwidth}
\centering
\includegraphics[width=\textwidth]{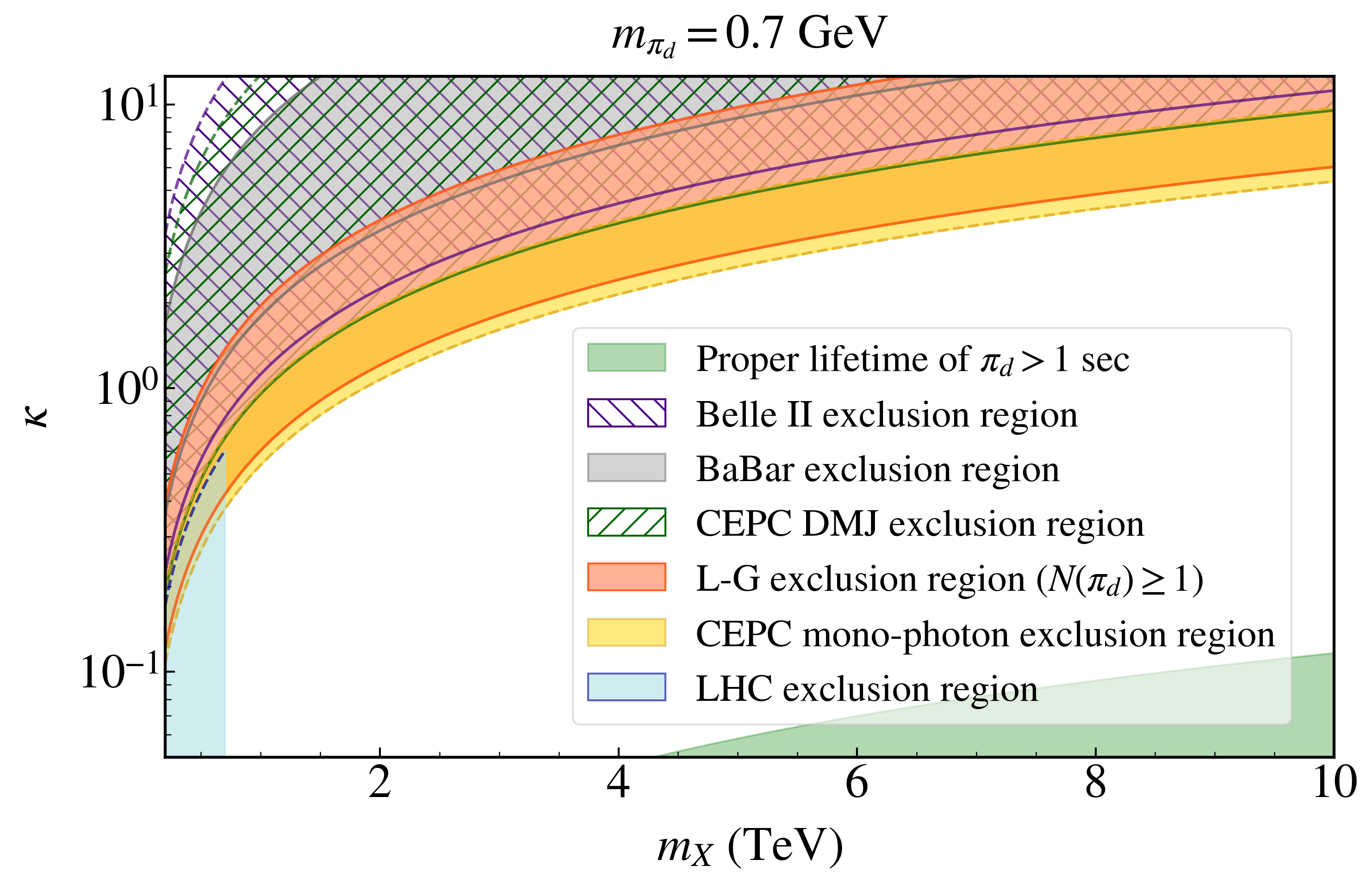}
\end{minipage}
\begin{minipage}{0.48\textwidth}
\centering
\includegraphics[width=\textwidth]{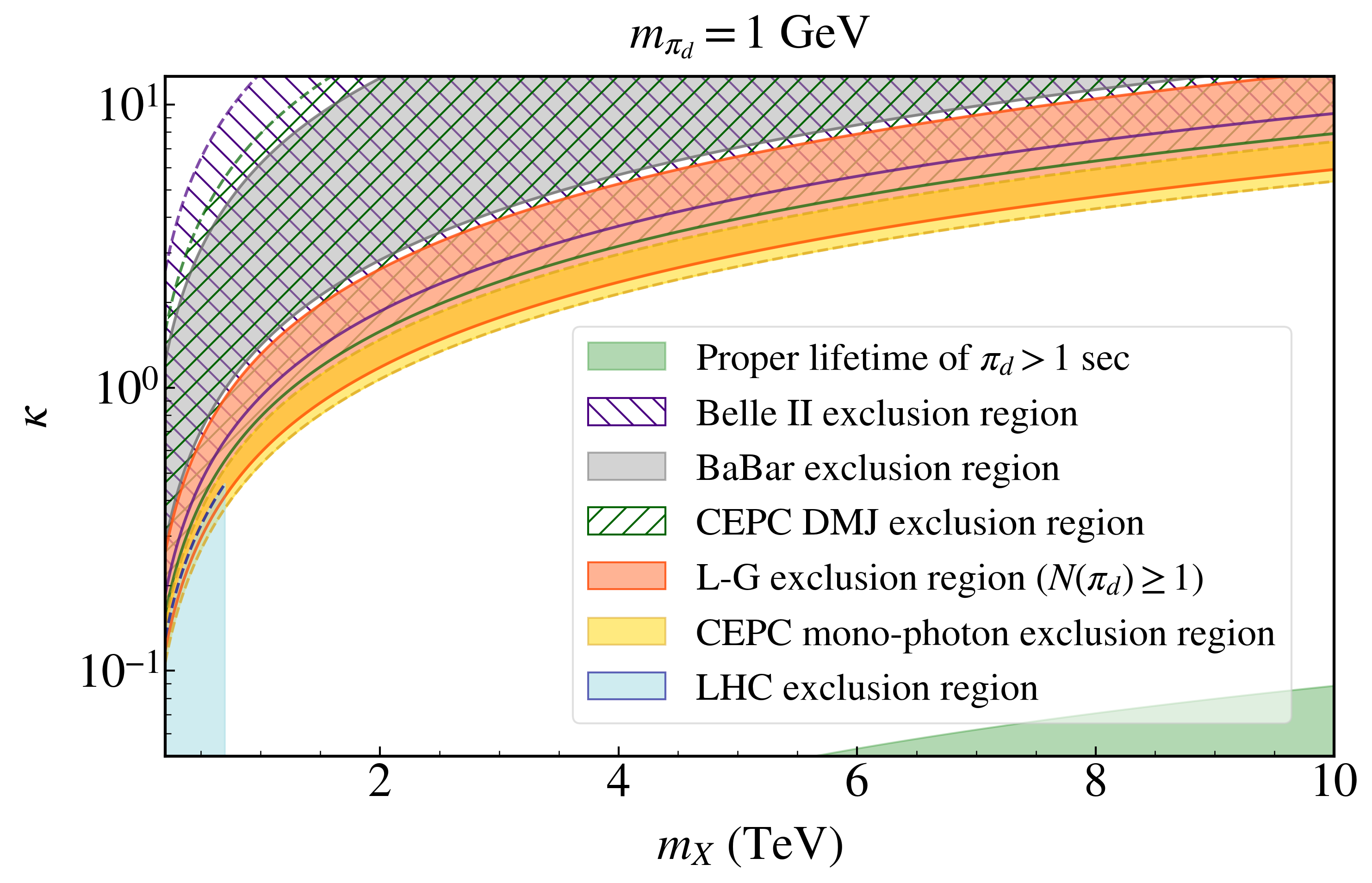}
\end{minipage}
\caption{Combined exclusion limits for different dark pion mass ranges. The first panel shows the exclusion region for $10~\text{MeV} < m_{\pi_d} < 2m_\mu$, while the remaining three panels show benchmark points $m_{\pi_d} = 0.32$, $0.7$ and $1~\text{GeV}$ from LHC, {\color{black}monophoton} searches at LEP and CEPC, displaced vertex searches (BaBar, Belle II, L-GAZELLE) and DMJ search at CEPC. All four panels include the parameter space where the proper lifetime of $\pi_d$ exceeds {\color{black}one second}, conservatively evaluated at $m_{\pi_d} = 0.2~\mathrm{GeV}$ for the light mass range.}
\label{fig:total_limit_compare}
\end{figure}

We explored novel signatures of a leptophilic QCD-like composite asymmetric dark sector at $e^+e^-$ colliders by simulating dark parton showering, hadronization, and detector {\color{black}responses} to map the experimental reach. The study reveals a sharp transition in the dark pion proper decay length at the muon threshold (Fig.~\ref{fig:proper_decay_length}). Recast constraints from {\color{black}monophoton} searches and from slepton/stau reinterpretations limit the low-mass bi-charged scalar mediator region (Figs.~\ref{fig:light} and~\ref{fig:heavy}), while combined projections for BaBar, Belle II, L-GAZELLE, and CEPC produce the characteristic band-shaped sensitivity in the $(m_X,\kappa)$ plane as summarized in Fig.~\ref{fig:total_limit_compare}. After applying cosmological constraints on the $\pi_d$ lifetime, nearly all parameter space for $10~\text{MeV} \leq m_{\pi_d} < 2m_\mu$ and $m_X \lesssim 10~\text{TeV}$ is excluded. Additionally, for $2m_\mu \leq m_{\pi_d} < 2m_\tau$, a significant fraction of parameter space with $\kappa \gtrsim 1$ and $m_X \lesssim 10~\text{TeV}$ is also excluded. 

The primary experimental targets we identify are (i) sparse displaced dimuon vertices at {\color{black}B factories} (kinematics in Figs.~\ref{fig:babar__kinematics},~\ref{fig:dark_pion_kinematics}{\color{black}, and \ref{fig:muon_kinematics}, while} trigger and event selection conditions are summarized in Tables~\ref{tab:trigger_summary} and~\ref{tab:llp_selection_summary}), which Belle II can probe with realistic triggers and event selections, (ii) far detectors (L-GAZELLE) that extend sensitivity to substantially longer lifetimes, and (iii) {\color{black}DMJ} searches at high-energy $e^+e^-$ colliders (main kinematics shown in Fig.~\ref{fig:cepc_dmj_kinematics} and detector assumptions in Table~\ref{tab:detector_geometry}). Although our projections adopt simplifying assumptions, {\color{black}LFU} of the mediator coupling, the benchmark relation $f_{\pi_d}=m_{\pi_d}$, and optimistic/background-free hypotheses in some regions, they nevertheless provide concrete, experimentally accessible targets and strongly motivate a coordinated program across {\color{black}B factories}, far detectors, and Higgs factories.

Finally, we outline promising directions for future work. Advanced data-scouting or trigger-level analyses for low-$p_T$ displaced dimuon pairs at CMS and LHCb could significantly strengthen LHC constraints~\cite{LHCb:2020ysn,CMS:2023hwl,Born:2023vll}. High-energy muon colliders could directly produce the bi-charged scalar mediator via {\color{black}Drell–Yan} processes and thereby provide an independent probe of the parameter space. Furthermore, when the dark pion is heavier than $2m_\tau$, dark jets become $\tau$-rich{\color{black}. This} regime will require dedicated strategies (e.g., displaced and boosted $\tau$-tagging, and optimized reconstruction of hadronic $\tau$ decays) to retain sensitivity.

\appendix 

\section{FCC-ee and CEPC Exclusion Region Comparison}
\label{appendix:A}

This appendix compares the FCC-ee and CEPC experiments in terms of their exclusion regions from {\color{black}monophoton and DMJ} searches. We adopt the same parameter settings and analysis strategies for the {\color{black}monophoton} and DMJ signatures as in the CEPC study. The observed differences are primarily attributed to the distinct detector geometries and the modest variation in integrated luminosity between FCC-ee and CEPC experiments. The geometry and relevant transverse spatial resolutions of the FCC-ee detector are summarized in Table~\ref{tab:FCC-ee_detector_geometry}, following Refs.~\cite{IDEAStudyGroup:2025gbt,Braibant:2021wts}. The transverse spatial resolutions are taken with reference to CEPC. Here we take $m_{\pi_d} = 1.0$ GeV as an example to demonstrate this {\color{black}similarity-other} benchmark points exhibit nearly identical behavior. 

\begin{figure}[htbp]
\centering
\begin{minipage}{0.48\textwidth}
\centering
\includegraphics[width=\textwidth]{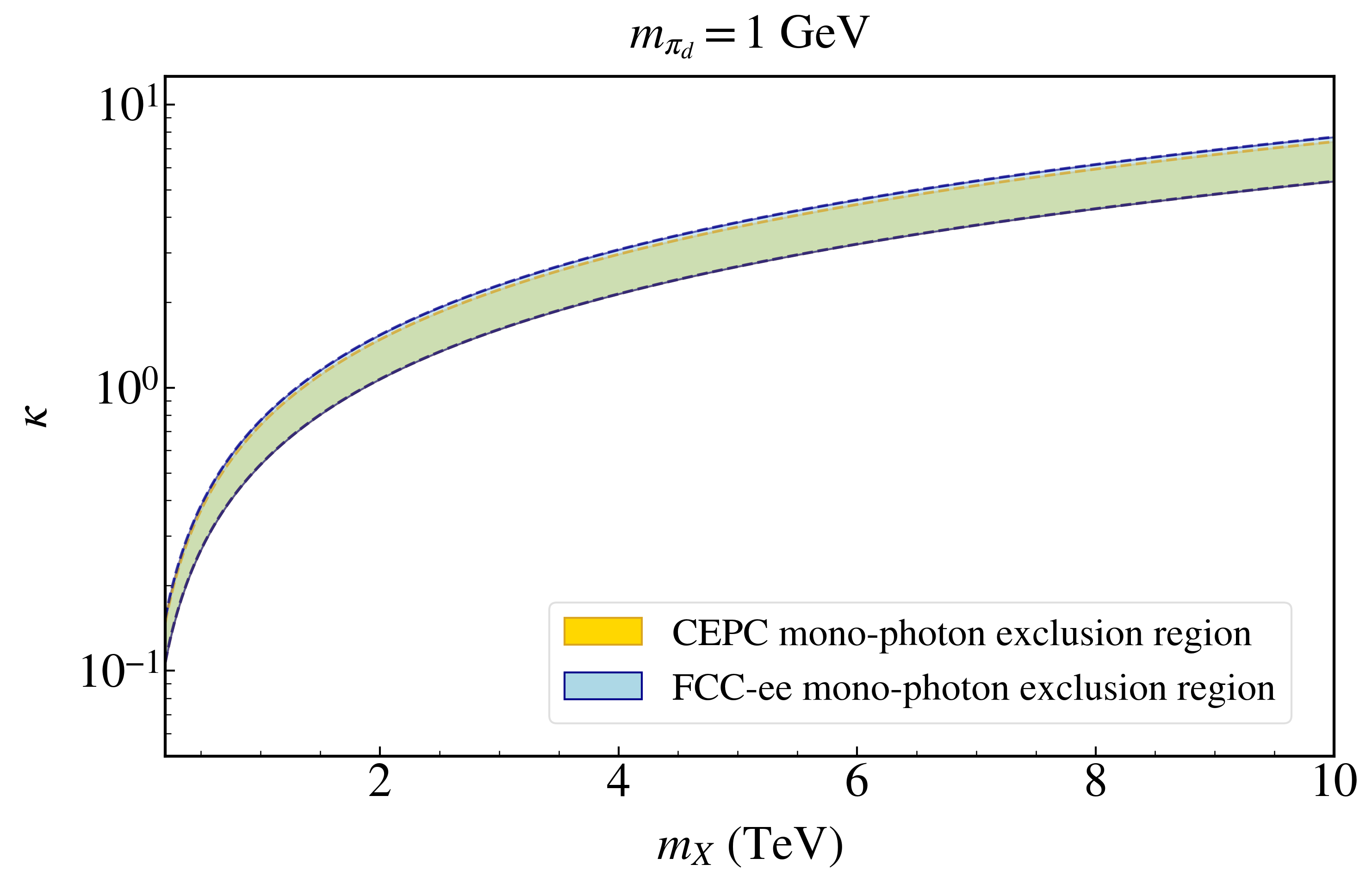}
\end{minipage}
\begin{minipage}{0.48\textwidth}
\centering
\includegraphics[width=\textwidth]{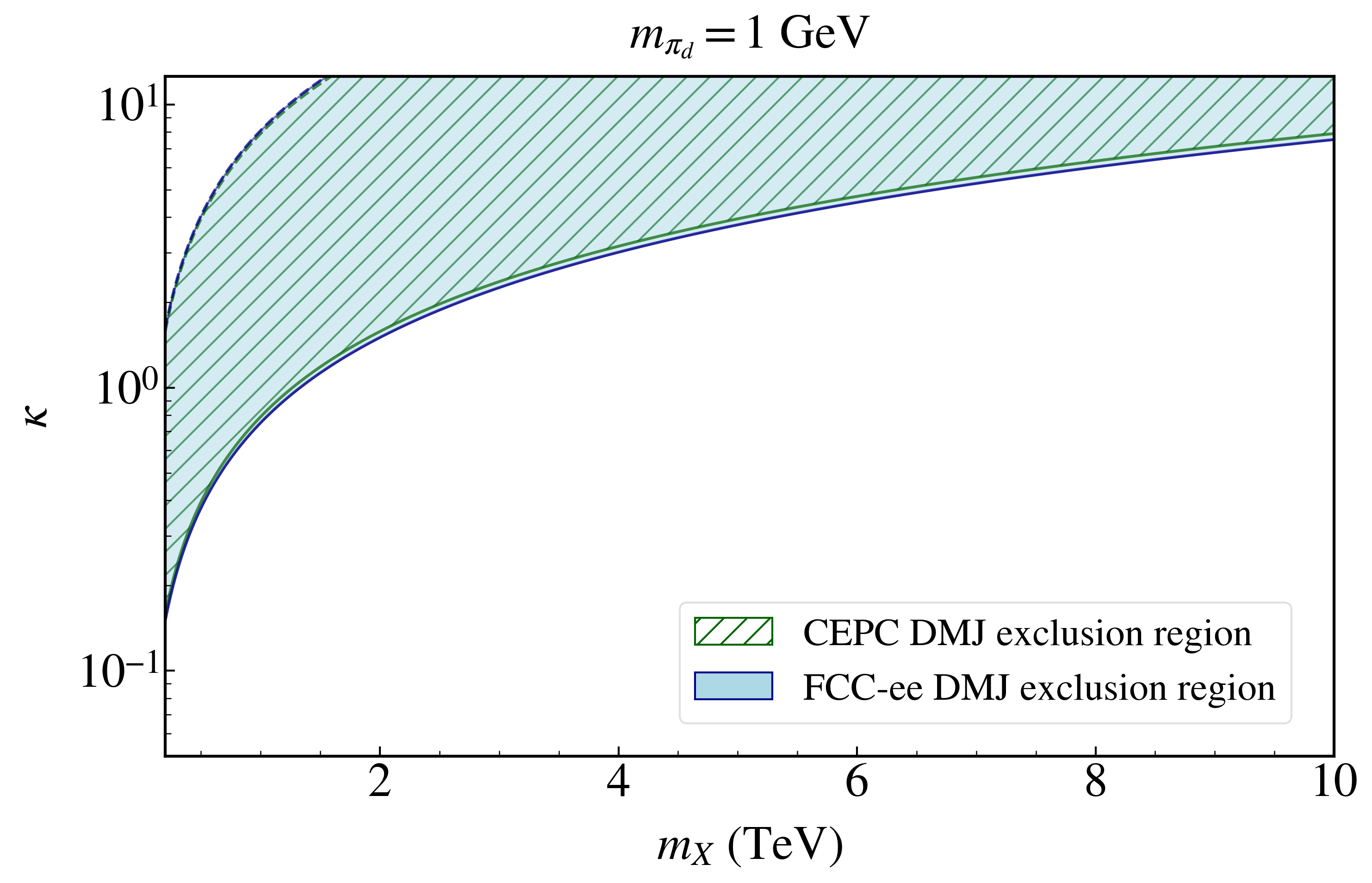}  
\end{minipage}  
\caption{The comparison of exclusion limits from {\color{black}monophoton} (left) and {\color{black}DMJ (}right) searches between the CEPC and FCC-ee experiments, for a dark pion mass of $m_{\pi_d} = 1~\text{GeV}$.}
\label{fig:mpid1_FCC_CEPC}
\end{figure}

As shown in Fig.~\ref{fig:mpid1_FCC_CEPC}, we compare the parameter space exclusion capabilities between FCC-ee and CEPC for $m_{\pi_d} = 1.0$ GeV. Considering the overall coverage of the four detector subsystems, the effective detector length of FCC-ee is slightly larger than that of CEPC, therefore FCC-ee demonstrates a mild advantage. However, this improvement is very small, and the exclusion capabilities of the two experiments are quite comparable. Therefore, for high-energy $e^+e^-$ colliders, we take CEPC as a representative example and perform detailed studies on the three BPs in the main text.

\begin{table}[h!]
\centering
\begin{tabular}{|l|c|c|c|}
\hline
\textbf{Detector Component} & $R_{\text{in}}$ & $R_{\text{out}}$ & $\sigma_{xy}$ \\
\hline
Vertex Detector (Inner) & \SI{13.70}{mm} & \SI{35.60}{mm} & \SI{2.80}{\micro\meter}--\SI{6.00}{\micro\meter} \\
Vertex Detector (Outer) & \SI{0.13}{m} & \SI{0.32}{m} & \SI{2.80}{\micro\meter}--\SI{6.00}{\micro\meter} \\
Drift Chamber & \SI{0.35}{m} & \SI{2.00}{m} & \SI{7.20}{\micro\meter} \\
Hadron Calorimeter & \SI{2.80}{m} & \SI{4.60}{m} & \SI{30}{mm} \\
Muon System & \SI{4.60}{m} & \SI{5.24}{m} & \SI{20}{mm} \\
\hline
\end{tabular}
\caption{The FCC-ee detector geometry and its relevant transverse spatial resolution.}
\label{tab:FCC-ee_detector_geometry}
\end{table}

\section*{Acknowledgments}
We thank Mengchao Zhang and Zeren Simon Wang for helpful discussions. {\color{black}The authors gratefully acknowledge the valuable discussions and insights provided by the members of the China Collaboration of Precision Testing and New Physics. C. T. L. and C. B. X.} are supported by the National Natural Science Foundation of China (NNSFC) under {\color{black}grants} No.~12335005, No.~12575118, and the Special funds for postdoctoral overseas recruitment, Ministry of Education of China. 



\begin{thebibliography}{99} 

\bibitem{Battaglieri:2017aum}
M.~Battaglieri, A.~Belloni, A.~Chou, P.~Cushman, B.~Echenard, R.~Essig, J.~Estrada, J.~L.~Feng, B.~Flaugher and P.~J.~Fox, \textit{et al.}
[arXiv:1707.04591 [hep-ph]].

\bibitem{Lin:2019uvt}
T.~Lin,
PoS \textbf{333}, 009 (2019)
doi:10.22323/1.333.0009
[arXiv:1904.07915 [hep-ph]].

\bibitem{Ferreira:2020fam}
E.~G.~M.~Ferreira,
Astron. Astrophys. Rev. \textbf{29}, no.1, 7 (2021)
doi:10.1007/s00159-021-00135-6
[arXiv:2005.03254 [astro-ph.CO]].

\bibitem{Belenchia:2021rfb}
A.~Belenchia, M.~Carlesso, {\"O}.~Bayraktar, D.~Dequal, I.~Derkach, G.~Gasbarri, W.~Herr, Y.~L.~Li, M.~Rademacher and J.~Sidhu, \textit{et al.}
Phys. Rept. \textbf{951}, 1-70 (2022)
doi:10.1016/j.physrep.2021.11.004
[arXiv:2108.01435 [quant-ph]].

\bibitem{Arcadi:2017kky}
G.~Arcadi, M.~Dutra, P.~Ghosh, M.~Lindner, Y.~Mambrini, M.~Pierre, S.~Profumo and F.~S.~Queiroz,
Eur. Phys. J. C \textbf{78}, no.3, 203 (2018)
doi:10.1140/epjc/s10052-018-5662-y
[arXiv:1703.07364 [hep-ph]].

\bibitem{Arcadi:2024ukq}
G.~Arcadi, D.~Cabo-Almeida, M.~Dutra, P.~Ghosh, M.~Lindner, Y.~Mambrini, J.~P.~Neto, M.~Pierre, S.~Profumo and F.~S.~Queiroz,
Eur. Phys. J. C \textbf{85}, no.2, 152 (2025)
doi:10.1140/epjc/s10052-024-13672-y
[arXiv:2403.15860 [hep-ph]].

\bibitem{Hochberg:2014dra}
Y.~Hochberg, E.~Kuflik, T.~Volansky and J.~G.~Wacker,
Phys. Rev. Lett. \textbf{113}, 171301 (2014)
doi:10.1103/PhysRevLett.113.171301
[arXiv:1402.5143 [hep-ph]].

\bibitem{Feng:2008ya}
J.~L.~Feng and J.~Kumar,
Phys. Rev. Lett. \textbf{101}, 231301 (2008)
doi:10.1103/PhysRevLett.101.231301
[arXiv:0803.4196 [hep-ph]].

\bibitem{Petraki:2013wwa}
K.~Petraki and R.~R.~Volkas,
Int. J. Mod. Phys. A \textbf{28}, 1330028 (2013)
doi:10.1142/S0217751X13300287
[arXiv:1305.4939 [hep-ph]].

\bibitem{Zurek:2013wia}
K.~M.~Zurek,
Phys. Rept. \textbf{537}, 91-121 (2014)
doi:10.1016/j.physrep.2013.12.001
[arXiv:1308.0338 [hep-ph]].

\bibitem{Bernal:2017kxu}
N.~Bernal, M.~Heikinheimo, T.~Tenkanen, K.~Tuominen and V.~Vaskonen,
Int. J. Mod. Phys. A \textbf{32}, no.27, 1730023 (2017)
doi:10.1142/S0217751X1730023X
[arXiv:1706.07442 [hep-ph]].

\bibitem{Baryakhtar:2025jwh}
M.~Baryakhtar, L.~Rosenberg and G.~Rybka,
[arXiv:2504.10607 [hep-ex]].

\bibitem{Villanueva-Domingo:2021spv}
P.~Villanueva-Domingo, O.~Mena and S.~Palomares-Ruiz,
Front. Astron. Space Sci. \textbf{8}, 87 (2021)
doi:10.3389/fspas.2021.681084
[arXiv:2103.12087 [astro-ph.CO]].

\bibitem{Sakharov:1967dj}
A.~D.~Sakharov,
Pisma Zh. Eksp. Teor. Fiz. \textbf{5}, 32-35 (1967)
doi:10.1070/PU1991v034n05ABEH002497

\bibitem{Bodeker:2020ghk}
D.~Bodeker and W.~Buchmuller,
Rev. Mod. Phys. \textbf{93}, no.3, 3 (2021)
doi:10.1103/RevModPhys.93.035004
[arXiv:2009.07294 [hep-ph]].

\bibitem{Davidson:2008bu}
S.~Davidson, E.~Nardi and Y.~Nir,
Phys. Rept. \textbf{466}, 105-177 (2008)
doi:10.1016/j.physrep.2008.06.002
[arXiv:0802.2962 [hep-ph]].

\bibitem{Affleck:1984fy}
I.~Affleck and M.~Dine,
Nucl. Phys. B \textbf{249}, 361-380 (1985)
doi:10.1016/0550-3213(85)90021-5

\bibitem{Blinnikov:1982eh}
S.~I.~Blinnikov and M.~Y.~Khlopov,
Sov. J. Nucl. Phys. \textbf{36}, 472 (1982)
ITEP-11-1982.

\bibitem{Blinnikov:1983gh}
S.~I.~Blinnikov and M.~Khlopov,
Sov. Astron. \textbf{27}, 371-375 (1983)

\bibitem{Khlopov:1989fj}
M.~Y.~Khlopov, G.~M.~Beskin, N.~E.~Bochkarev, L.~A.~Pustylnik and S.~A.~Pustylnik,
Sov. Astron. \textbf{35}, 21 (1991)
FERMILAB-PUB-89-193-A.

\bibitem{Alves:2009nf}
D.~S.~M.~Alves, S.~R.~Behbahani, P.~Schuster and J.~G.~Wacker,
Phys. Lett. B \textbf{692}, 323-326 (2010)
doi:10.1016/j.physletb.2010.08.006
[arXiv:0903.3945 [hep-ph]].

\bibitem{SpierMoreiraAlves:2010err}
D.~Spier Moreira Alves, S.~R.~Behbahani, P.~Schuster and J.~G.~Wacker,
JHEP \textbf{06}, 113 (2010)
doi:10.1007/JHEP06(2010)113
[arXiv:1003.4729 [hep-ph]].

\bibitem{Bai:2013xga}
Y.~Bai and P.~Schwaller,
Phys. Rev. D \textbf{89}, no.6, 063522 (2014)
doi:10.1103/PhysRevD.89.063522
[arXiv:1306.4676 [hep-ph]].

\bibitem{Ibe:2018juk}
M.~Ibe, A.~Kamada, S.~Kobayashi and W.~Nakano,
JHEP \textbf{11}, 203 (2018)
doi:10.1007/JHEP11(2018)203
[arXiv:1805.06876 [hep-ph]].

\bibitem{Ibe:2018tex}
M.~Ibe, A.~Kamada, S.~Kobayashi, T.~Kuwahara and W.~Nakano,
JHEP \textbf{03}, 173 (2019)
doi:10.1007/JHEP03(2019)173
[arXiv:1811.10232 [hep-ph]].

\bibitem{Ibe:2019ena}
M.~Ibe, A.~Kamada, S.~Kobayashi, T.~Kuwahara and W.~Nakano,
Phys. Rev. D \textbf{100}, no.7, 075022 (2019)
doi:10.1103/PhysRevD.100.075022
[arXiv:1907.03404 [hep-ph]].

\bibitem{Hall:2019rld}
E.~Hall, T.~Konstandin, R.~McGehee and H.~Murayama,
Phys. Rev. D \textbf{107}, no.5, 055011 (2023)
doi:10.1103/PhysRevD.107.055011
[arXiv:1911.12342 [hep-ph]].

\bibitem{Zhang:2021orr}
M.~Zhang,
Phys. Rev. D \textbf{104}, no.5, 055008 (2021)
doi:10.1103/PhysRevD.104.055008
[arXiv:2104.06988 [hep-ph]].

\bibitem{Bottaro:2021aal}
S.~Bottaro, M.~Costa and O.~Popov,
JHEP \textbf{11}, 055 (2021)
doi:10.1007/JHEP11(2021)055
[arXiv:2104.14244 [hep-ph]].

\bibitem{Ibe:2021gil}
M.~Ibe, S.~Kobayashi and K.~Watanabe,
JHEP \textbf{07}, 220 (2021)
doi:10.1007/JHEP07(2021)220
[arXiv:2105.07642 [hep-ph]].

\bibitem{Ritter:2022opo}
A.~C.~Ritter and R.~R.~Volkas,
Phys. Rev. D \textbf{107}, no.1, 015029 (2023)
doi:10.1103/PhysRevD.107.015029
[arXiv:2210.11011 [hep-ph]].

\bibitem{Schwaller:2015gea}
P.~Schwaller, D.~Stolarski and A.~Weiler,
JHEP \textbf{05}, 059 (2015)
doi:10.1007/JHEP05(2015)059
[arXiv:1502.05409 [hep-ph]].

\bibitem{Cohen:2015toa}
T.~Cohen, M.~Lisanti and H.~K.~Lou,
Phys. Rev. Lett. \textbf{115}, no.17, 171804 (2015)
doi:10.1103/PhysRevLett.115.171804
[arXiv:1503.00009 [hep-ph]].

\bibitem{Cohen:2017pzm}
T.~Cohen, M.~Lisanti, H.~K.~Lou and S.~Mishra-Sharma,
JHEP \textbf{11}, 196 (2017)
doi:10.1007/JHEP11(2017)196
[arXiv:1707.05326 [hep-ph]].

\bibitem{Park:2017rfb}
M.~Park and M.~Zhang,
Phys. Rev. D \textbf{100}, no.11, 115009 (2019)
doi:10.1103/PhysRevD.100.115009
[arXiv:1712.09279 [hep-ph]].

\bibitem{Renner:2018fhh}
S.~Renner and P.~Schwaller,
JHEP \textbf{08}, 052 (2018)
doi:10.1007/JHEP08(2018)052
[arXiv:1803.08080 [hep-ph]].

\bibitem{Mies:2020mzw}
H.~Mies, C.~Scherb and P.~Schwaller,
JHEP \textbf{04}, 049 (2021)
doi:10.1007/JHEP04(2021)049
[arXiv:2011.13990 [hep-ph]].

\bibitem{Linthorne:2021oiz}
D.~Linthorne and D.~Stolarski,
Phys. Rev. D \textbf{104}, no.3, 035019 (2021)
doi:10.1103/PhysRevD.104.035019
[arXiv:2103.08620 [hep-ph]].

\bibitem{Archer-Smith:2021ntx}
P.~Archer-Smith, D.~Linthorne and D.~Stolarski,
JHEP \textbf{02}, 027 (2022)
doi:10.1007/JHEP02(2022)027
[arXiv:2112.05690 [hep-ph]].

\bibitem{Carrasco:2023loy}
J.~Carrasco and J.~Zurita,
JHEP \textbf{01}, 034 (2024)
doi:10.1007/JHEP01(2024)034
[arXiv:2307.04847 [hep-ph]].

\bibitem{Carmona:2024tkg}
A.~Carmona, F.~Elahi, C.~Scherb and P.~Schwaller,
JHEP \textbf{06}, 198 (2025)
doi:10.1007/JHEP06(2025)198
[arXiv:2411.15073 [hep-ph]].

\bibitem{Liu:2025bbc}
W.~Liu, J.~Lockyer and S.~Kulkarni,
[arXiv:2505.03058 [hep-ph]].

\bibitem{ATLAS:2019lff}
G.~Aad \textit{et al.} [ATLAS],
Eur. Phys. J. C \textbf{80}, no.2, 123 (2020)
doi:10.1140/epjc/s10052-019-7594-6
[arXiv:1908.08215 [hep-ex]].

\bibitem{CMS:2020bfa}
A.~M.~Sirunyan \textit{et al.} [CMS],
JHEP \textbf{04}, 123 (2021)
doi:10.1007/JHEP04(2021)123
[arXiv:2012.08600 [hep-ex]].

\bibitem{CMS:2022syk}
A.~Tumasyan \textit{et al.} [CMS],
Phys. Rev. D \textbf{108}, no.1, 012011 (2023)
doi:10.1103/PhysRevD.108.012011
[arXiv:2207.02254 [hep-ex]].

\bibitem{CMS:2024gyw}
A.~Hayrapetyan \textit{et al.} [CMS],
Phys. Rev. D \textbf{109}, no.11, 112001 (2024)
doi:10.1103/PhysRevD.109.112001
[arXiv:2402.01888 [hep-ex]].

\bibitem{ATLAS:2024fub}
G.~Aad \textit{et al.} [ATLAS],
JHEP \textbf{05}, 150 (2024)
doi:10.1007/JHEP05(2024)150
[arXiv:2402.00603 [hep-ex]].

\bibitem{Fox:2011fx}
P.~J.~Fox, R.~Harnik, J.~Kopp and Y.~Tsai,
Phys. Rev. D \textbf{84}, 014028 (2011)
doi:10.1103/PhysRevD.84.014028
[arXiv:1103.0240 [hep-ph]].

\bibitem{Duerr:2020muu}
M.~Duerr, T.~Ferber, C.~Garcia-Cely, C.~Hearty and K.~Schmidt-Hoberg,
JHEP \textbf{04}, 146 (2021)
doi:10.1007/JHEP04(2021)146
[arXiv:2012.08595 [hep-ph]].

\bibitem{Bernreuther:2022jlj}
E.~Bernreuther, K.~B{\"o}se, T.~Ferber, C.~Hearty, F.~Kahlhoefer, A.~Morandini and K.~Schmidt-Hoberg,
JHEP \textbf{12}, 005 (2022)
doi:10.1007/JHEP12(2022)005
[arXiv:2203.08824 [hep-ph]].

\bibitem{Liu:2019ogn}
Z.~Liu, Y.~H.~Xu and Y.~Zhang,
JHEP \textbf{06}, 009 (2019)
doi:10.1007/JHEP06(2019)009
[arXiv:1903.12114 [hep-ph]].

\bibitem{Alloul:2013bka}
A.~Alloul, N.~D.~Christensen, C.~Degrande, C.~Duhr and B.~Fuks,
Comput. Phys. Commun. \textbf{185}, 2250-2300 (2014)
doi:10.1016/j.cpc.2014.04.012
[arXiv:1310.1921 [hep-ph]].

\bibitem{Alwall:2014hca}
J.~Alwall, R.~Frederix, S.~Frixione, V.~Hirschi, F.~Maltoni, O.~Mattelaer, H.~S.~Shao, T.~Stelzer, P.~Torrielli and M.~Zaro,
JHEP \textbf{07}, 079 (2014)
doi:10.1007/JHEP07(2014)079
[arXiv:1405.0301 [hep-ph]].

\bibitem{Fiaschi:2018xdm}
J.~Fiaschi and M.~Klasen,
JHEP \textbf{03}, 094 (2018)
doi:10.1007/JHEP03(2018)094
[arXiv:1801.10357 [hep-ph]].

\bibitem{Fuks:2013lya}
B.~Fuks, M.~Klasen, D.~R.~Lamprea and M.~Rothering,
JHEP \textbf{01}, 168 (2014)
doi:10.1007/JHEP01(2014)168
[arXiv:1310.2621 [hep-ph]].

\bibitem{ATLAS:2022izj}
G.~Aad \textit{et al.} [ATLAS],
JHEP \textbf{06}, 153 (2023)
doi:10.1007/JHEP06(2023)153
[arXiv:2206.12181 [hep-ex]].

\bibitem{Lindley:1984bg}
D.~Lindley,
Astrophys. J. \textbf{294}, 1-8 (1985)
doi:10.1086/163267

\bibitem{Dev:2025pru}
P.~S.~B.~Dev, Q.~f.~Wu and X.~J.~Xu,
[arXiv:2507.12270 [hep-ph]].

\bibitem{Sjostrand:2014zea}
T.~Sj{\"o}strand, S.~Ask, J.~R.~Christiansen, R.~Corke, N.~Desai, P.~Ilten, S.~Mrenna, S.~Prestel, C.~O.~Rasmussen and P.~Z.~Skands,
Comput. Phys. Commun. \textbf{191}, 159-177 (2015)
doi:10.1016/j.cpc.2015.01.024
[arXiv:1410.3012 [hep-ph]].

\bibitem{Knapen:2021eip}
S.~Knapen, J.~Shelton and D.~Xu,
Phys. Rev. D \textbf{103}, no.11, 115013 (2021)
doi:10.1103/PhysRevD.103.115013
[arXiv:2103.01238 [hep-ph]].

\bibitem{Berlin:2018tvf}
A.~Berlin, N.~Blinov, S.~Gori, P.~Schuster and N.~Toro,
Phys. Rev. D \textbf{97}, no.5, 055033 (2018)
doi:10.1103/PhysRevD.97.055033
[arXiv:1801.05805 [hep-ph]].

\bibitem{Choi:2018iit}
S.~M.~Choi, H.~M.~Lee, P.~Ko and A.~Natale,
Phys. Rev. D \textbf{98}, no.1, 015034 (2018)
doi:10.1103/PhysRevD.98.015034
[arXiv:1801.07726 [hep-ph]].

\bibitem{Kuwahara:2023vfc}
T.~Kuwahara and S.~R.~Yuan,
JHEP \textbf{06}, 208 (2023)
doi:10.1007/JHEP06(2023)208
[arXiv:2303.03736 [hep-ph]].

\bibitem{Bernreuther:2023kcg}
E.~Bernreuther, N.~Hemme, F.~Kahlhoefer and S.~Kulkarni,
Phys. Rev. D \textbf{110}, no.3, 035009 (2024)
doi:10.1103/PhysRevD.110.035009
[arXiv:2311.17157 [hep-ph]].

\bibitem{BaBar:2015jvu}
J.~P.~Lees \textit{et al.} [BaBar],
Phys. Rev. Lett. \textbf{114}, no.17, 171801 (2015)
doi:10.1103/PhysRevLett.114.171801
[arXiv:1502.02580 [hep-ex]].

\bibitem{BaBar:1995bns}
D.~Boutigny \textit{et al.} [BaBar],
SLAC-R-0457.

\bibitem{Adachi:2018qme}
I.~Adachi \textit{et al.} [Belle-II],
Nucl. Instrum. Meth. A \textbf{907}, 46-59 (2018)
doi:10.1016/j.nima.2018.03.068

\bibitem{Kang:2021oes}
D.~W.~Kang, P.~Ko and C.~T.~Lu,
JHEP \textbf{04}, 269 (2021)
doi:10.1007/JHEP04(2021)269
[arXiv:2101.02503 [hep-ph]].

\bibitem{Ko:2025drr}
P.~Ko, Y.~Kwon, C.~T.~Lu and X.~Wei,
JHEP \textbf{08}, 150 (2025)
doi:10.1007/JHEP08(2025)150
[arXiv:2504.19067 [hep-ph]].

\bibitem{Domingo:2023dew}
F.~Domingo, J.~G{\"u}nther, J.~S.~Kim and Z.~S.~Wang,
Eur. Phys. J. C \textbf{84}, no.6, 642 (2024)
doi:10.1140/epjc/s10052-024-13009-9
[arXiv:2308.07371 [hep-ph]].

\bibitem{Cacciari:2008gp}
M.~Cacciari, G.~P.~Salam and G.~Soyez,
JHEP \textbf{04}, 063 (2008)
doi:10.1088/1126-6708/2008/04/063
[arXiv:0802.1189 [hep-ph]].

\bibitem{Cacciari:2011ma}
M.~Cacciari, G.~P.~Salam and G.~Soyez,
Eur. Phys. J. C \textbf{72}, 1896 (2012)
doi:10.1140/epjc/s10052-012-1896-2
[arXiv:1111.6097 [hep-ph]].

\bibitem{CEPCStudyGroup:2018ghi}
J.~B.~Guimar{\~a}es da Costa \textit{et al.} [CEPC Study Group],
[arXiv:1811.10545 [hep-ex]].

\bibitem{LHCb:2020ysn}
R.~Aaij \textit{et al.} [LHCb],
JHEP \textbf{10}, 156 (2020)
doi:10.1007/JHEP10(2020)156
[arXiv:2007.03923 [hep-ex]].

\bibitem{CMS:2023hwl}
A.~Hayrapetyan \textit{et al.} [CMS],
JHEP \textbf{12}, 070 (2023)
doi:10.1007/JHEP12(2023)070
[arXiv:2309.16003 [hep-ex]].

\bibitem{Born:2023vll}
S.~Born, R.~Karur, S.~Knapen and J.~Shelton,
Phys. Rev. D \textbf{108}, no.3, 035034 (2023)
doi:10.1103/PhysRevD.108.035034
[arXiv:2303.04167 [hep-ph]].

\bibitem{IDEAStudyGroup:2025gbt}
M.~Abbrescia \textit{et al.} [IDEA Study Group],
[arXiv:2502.21223 [physics.ins-det]].

\bibitem{Braibant:2021wts}
S.~Braibant and P.~Giacomelli,
Eur. Phys. J. Plus \textbf{136}, no.11, 1143 (2021)
doi:10.1140/epjp/s13360-021-02115-2

\end{thebibliography}
\end{document}